\documentclass[3p]{elsarticle}

\usepackage[dvipsnames,table]{xcolor}
\usepackage{algorithm,algorithmic}
\usepackage{caption}
\usepackage{enumerate}
\usepackage{float,multicol}
\usepackage{hyperref}
\usepackage{graphicx}
\usepackage{listings}
\usepackage{multirow}
\usepackage{newfloat}
\usepackage{smartdiagram}
\usepackage{svg}
\usepackage{tabto}
\usepackage{textcomp}
\usepackage{threeparttable}
\usepackage{tikz}
\usepackage{url}
\usepackage{verbatim}
\DeclareFloatingEnvironment[fileext=diag,placement={!ht},name=Diagram]{diag}

\begin{document}

\begin{frontmatter}

\title{\texttt{can-train-and-test:} A Curated CAN Dataset for Automotive Intrusion Detection}

\author[1]{Brooke Lampe}
\ead{blam@dtu.dk}

\author[1]{Weizhi Meng}
\ead{weme@dtu.dk}

\affiliation[1]{organization={Technical University of Denmark},
addressline={Anker Engelunds Vej 101},
city={Kongens Lyngby},
postcode={2800},
country={Denmark}}



\begin{abstract}
    When it comes to in-vehicle networks (IVNs), the controller area network---CAN---bus dominates the market; automobiles manufactured and sold around the world depend on the CAN bus for safety-critical communications between various components of the vehicle (e.g., the engine, the transmission, the steering column). Unfortunately, the CAN bus is inherently insecure; in fact, it completely lacks controls such as authentication, authorization, and confidentiality (i.e., encryption). Therefore, researchers have travailed to develop automotive security enhancements. The automotive intrusion detection system (IDS) is especially popular in the literature---due to its relatively low cost in terms of money, resource utilization, and implementation effort. That said, developing and evaluating an automotive IDS is often challenging; if researchers do not have access to a test vehicle, then they are forced to depend on publicly available CAN data---which is not without limitations. Lack of access to adequate CAN data, then, becomes a barrier to entry into automotive security research.

    We seek to lower that barrier to entry by introducing a new CAN dataset to facilitate the development and evaluation of automotive IDSs. Our dataset, dubbed \texttt{can-train-and-test}, provides CAN data from four different vehicles produced by two different manufacturers. The attack captures for each vehicle model are equivalent, enabling researchers to assess the ability of a given IDS to generalize to different vehicle models and even different vehicle manufacturers. Our dataset contains replayable \texttt{.log} files as well as labeled and unlabeled \texttt{.csv} files, thereby meeting a variety of development and evaluation needs. Furthermore, \texttt{can-train-and-test} offers nine unique attacks, ranging from denial of service (DoS) to gear spoofing to standstill; as such, researchers can select a subset of the attacks for training and save the remainder for testing in order to assess a given IDS against unseen attacks. Many of our attacks, particularly the spoofing-related attacks, were conducted during live, on-the-road experiments with real vehicles. These attacks have known, \textit{physical} impacts. As a benchmark, we pit a number of machine learning IDSs against our dataset and analyze the results. We present \texttt{can-train-and-test} as a contribution to the existing catalogue of open-access datasets in hopes of filling in the gaps left by those datasets.
\end{abstract}

\begin{keyword}
Automotive security, automotive attacks, controller area network (CAN), in-vehicle network (IVN), intrusion detection system (IDS), logging, spoofing, dataset, benchmark, machine learning
\end{keyword}

\end{frontmatter}

\section{Introduction} \label{sec:introduction}

The modern automobile has shifted from essentially mechanical to markedly electronic. Electronic control units---ECUs---now manage automotive components from the engine to the transmission to the steering column to the airbags. Modern passenger vehicles routinely contain over 80 ECUs; when it comes to luxury vehicles, which often incorporate a number of autonomous and semi-autonomous features, the number of ECUs can be much higher \cite{ecus, autopi}.

\begin{figure}[hbt!]
\centering
\includegraphics[width=1.0\columnwidth]{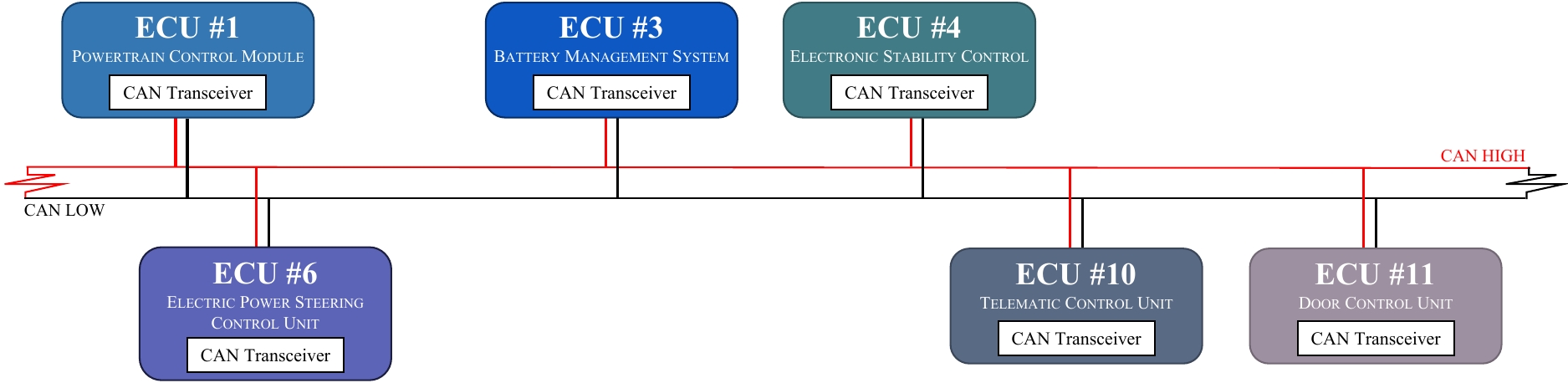}
\caption{Schematic diagram of the CAN bus}
\label{can-diagram}
\end{figure}

ECUs communicate via in-vehicle networks (IVNs); the controller area network (CAN) is by far the most popular. The CAN bus, depicted in Figure \ref{can-diagram}, facilitates reliable multiplex communication across twisted-pair electrical wiring. It leverages differential signaling to minimize electrical interference (i.e., noise). In the absence of an attack, it is a highly robust protocol; however, it is fundamentally insecure. Modern-day security practices---authentication, authorization, confidentiality (i.e., encryption)---have never been incorporated into the CAN bus. While the CAN bus does have integrity controls when it comes to error-handling, it offers no integrity in an attack scenario \cite{can-bus-government-whitepaper, deep-learning-survey}.

Researchers have identified a number of attacks against the CAN bus, which can be organized into six families. These six attack families are enumerated in Table \ref{can-attacks}. Many of the attacks are simple---even trivial---to implement, and the impacts range from irritating to potentially lethal. Multiple CAN-related exploits have been published (see Figure \ref{can-exploits}); fortunately, to our knowledge, no automotive attacks have been conducted in the wild.

The automotive intrusion detection system (IDS) has emerged as a favorite when it comes to CAN bus security. The automotive IDS is a relatively inexpensive solution---in terms of monetary cost, resource cost, and implementation cost. Re-engineering effort is a major roadblock to cryptographic CAN bus security solutions (e.g., authentication, encryption). Generally speaking, automotive intrusion detection systems do not engender a redesign of existing controller area networks \cite{deep-learning-survey}. In fact, automotive IDSs need not be integrated into the vehicle during manufacturing; they can be deployed at any time during the vehicle's service life. Some automotive IDSs can even be deployed by ordinary consumers with no mechanical or technical expertise whatsoever \cite{id-sequence-android}.

\subsection{Motivation}

As mentioned earlier, intrusion detection systems are exceedingly valuable to automotive security and to automotive security research. However, automotive IDSs depend on automotive data---i.e., CAN data---for development, experimentation, and evaluation. If researchers do not have access to a test vehicle, then the lack of sufficient CAN data becomes a barrier to entry into automotive security research. Therefore, we curated a new CAN dataset, \texttt{can-train-and-test}, to address the limitations of existing open-access datasets.

We noted the following limitations of existing publicly available datasets:

\begin{enumerate}
    \item Lack of sufficient attack-free data
    \item Lack of sufficient attack variation
    \item Lack of sufficient vehicle variation
    \item Lack of fidelity
    \item Lack of severity
    \item Lack of modernity
    \item Lack of labels
\end{enumerate}

\textbf{Lack of sufficient attack-free data.} When training a machine learning IDS, especially a \textit{deep} learning IDS, attack-free (``normal") data is crucial. Attack-free data provides a machine learning-based IDS with a much-needed baseline; once the IDS has established a baseline, it can flag deviations from the baseline as potential attacks. Many publicly available datasets provide insufficient attack-free data to train an effective machine learning classifier.

\textbf{Lack of sufficient attack variation.} Generally speaking, automotive IDSs are envisaged as anomaly-based IDSs, meaning that they classify traffic as either (1) ``normal" or (2) ``anomalous." Typically, they build a profile of ``normal" traffic during training. During testing, if they detect a deviation from the ``normal" baseline, then they flag it as ``anomalous." Anomaly-based IDSs are expected to detect previously unseen attacks. If an automotive dataset contains exactly one type of attack, then an anomaly-based IDS cannot be trained on one attack and tested on a second attack. If an automotive dataset contains no attacks at all, then it cannot be used to even train---let alone test---a supervised anomaly-based IDS.

The more attacks a dataset contains, the more options researchers have for training and testing with different combinations of known and unknown attacks. Therefore, when constructing our dataset, we crafted a variety of unique, realistic automotive attacks.

\textbf{Lack of sufficient vehicle variation.} The CAN protocol is ubiquitous, but the implementation is not. Different automotive manufacturers use different CAN messages to communicate different information. As such, an automotive IDS trained and tested against one type of vehicle might not generalize to vehicles produced by a different manufacturer. Such an IDS might not even generalize to different vehicle models produced by the same manufacturer. Thus, an automotive dataset should include the same types of attacks against different vehicles, allowing researchers to assess the detection capabilities of a proposed IDS against multiple vehicles.

\textbf{Lack of fidelity.} A number of the existing publicly available datasets were collected during simulations, from testbed environments, or while the vehicle was stationary. When experimenting with automotive attacks, safety is a major concern, prompting researchers to focus on lower-fidelity options (e.g., simulations and testbeds). However, the goal of automotive security research is to develop solutions that will be deployed on real vehicles---not simulations and testbeds. Therefore, it is prudent to evaluate automotive security solutions with higher-fidelity datasets. If the evaluation dataset is highly realistic, then we can be more confident that a high-performing IDS will perform effectively in a real-world deployment. If, however, the evaluation dataset is highly \textit{un}realistic, then we cannot extrapolate that a high-performing IDS will still perform effectively when faced with a much-different real-world deployment.

\textbf{Lack of severity.} When pondering the possibility of a malicious, in-the-wild automotive attack, few people would expect the adversary to turn on the check engine light and stop there. Instead, we would expect a real-world attack to be a high-severity attack. Perhaps the adversary would turn the steering wheel 180 degrees while the vehicle was cruising down the interstate \cite{advanced-can-injection-whitepaper, blackhat-advanced-can-injection}. Perhaps the adversary would deploy the airbag \cite{airbag}. Perhaps the adversary would spoof ignition messages in order to start the vehicle and drive off with it \cite{theft}. In a CAN dataset, low-severity automotive attacks are very useful; however, it is important to include high-severity attacks as well.

\textbf{Lack of modernity.} Some automotive manufacturers have programmed enhanced security features into the CAN bus and/or the ECUs. For example, when we experimented with a 2016 Chevrolet Silverado, we would be ejected from the CAN bus if we sent too many spoofed CAN frames at too high a frequency. In contrast, the 2011 Chevrolet Impala rarely excluded us from CAN communications. To our knowledge, few of the existing datasets include attacks which have been adapted to subvert the enhanced security features of some modern automobiles. With each publicized automotive exploit, security gains traction with automotive manufacturers; therefore, it is important to build datasets that contain attacks capable of eluding enhanced security features.

\textbf{Lack of labels.} Many IDSs rely on supervised machine learning, which, in turn, relies on labeled datasets. Few of the existing datasets are adequately labeled; as such, they are ill-suited to training and testing supervised machine learning IDSs. Moreover, while unsupervised machine learning does not depend on labeled training data, an unsupervised machine learning IDS would typically be evaluated against labeled testing data, so that researchers could quantify its performance.

In summary, motivated by the limitations of existing datasets, we seek to supply a new CAN dataset which provides (1) ample attack-free data, (2) a variety of different attack types, (3) a variety of different vehicle types---e.g., manufacturer, model, year, (4) high-fidelity data, (5) high-severity attacks, (6) attacks capable of evading modern security enhancements, and (7) labeled data.

While we focus the domain of automotive intrusion detection, we note that our dataset could be used to design and assess automotive security solutions beyond IDSs (e.g., firewall- and filtering-based techniques).

\subsection{Contributions}

    Our contributions can be summarized as follows:
    
    \begin{enumerate}
        \item We present a new open-access CAN dataset for use in automotive intrusion detection. Our dataset includes
        \small
            \begin{itemize}
                \item A high volume of attack-free data
                \item Nine unique attacks
                \item Data from four different vehicles
                \item Attack-free data and attack data that was collected during live, on-the-road experiments
                \item High-severity attacks
                \item Attacks adapted to evade modern security enhancements
                \item Labeled attack-free and attack samples
            \end{itemize}
        \normalsize
        \item We compile the \texttt{can-train-and-test} dataset to help researchers develop and evaluate machine learning IDSs. The dataset is subdivided into four train/test sub-datasets: \texttt{set\_01, set\_02, set\_03, set\_04}. Within each train/test sub-dataset, we provide one training subset and four testing subsets.
        \item We curate the \texttt{can-train-and-test} dataset to optimize its assessment capabilities. The dataset provides testing subsets which challenge an automotive IDS's ability to generalize to different attacks and vehicles:
        \small
            \begin{itemize}
                \item \textbf{train\_01:} Train the model
                \item \textbf{test\_01\_known\_vehicle\_known\_attack:} Test the model against a known vehicle (seen in training) and known attacks (seen in training)
                \item \textbf{test\_02\_unknown\_vehicle\_known\_attack:} Test the model against an unknown vehicle (not seen in training) and known attacks (seen in training)
                \item \textbf{test\_03\_known\_vehicle\_unknown\_attack:} Test the model against a known vehicle (seen in training) and unknown attacks (not seen in training)
                \item \textbf{test\_04\_unknown\_vehicle\_unknown\_attack:} Test the model against an unknown vehicle (not seen in training) and unknown attacks (not seen in training)
            \end{itemize}
        \normalsize
        \item Lastly, we evaluate several machine learning IDSs on our \texttt{can-train-and-test} dataset. In our benchmark, we include both supervised and unsupervised paradigms, as well as traditional machine learning and deep learning models. Our benchmark encompasses seven model families, which contain a total of eighteen models.
    \end{enumerate}

\subsection{Paper Organization}

    The remainder of this paper is organized as follows: Section~\ref{sec:background} encapsulates essential background on automotive security and controller area networks. Section~\ref{sec:related-work} explores a number of related works, in particular, existing publicly available CAN datasets. Section~\ref{sec:methodology} discusses our methodology (e.g., data collection, data pre-processing, etc.), while Section~\ref{sec:datasets} specifies the qualitative and quantitative particulars of our datasets. Section~\ref{sec:usage} provides some usage notes and examples. Our benchmark is detailed in depth in Section~\ref{sec:benchmark}. Section~\ref{sec:future-work} highlights open challenges and opportunities for future work; Section~\ref{sec:conclusion} concludes our work.

\section{Background} \label{sec:background}

\subsection{The Protocol}

The controller area network (CAN) bus is the de facto standard when it comes to in-vehicle networks (IVNs). The CAN protocol was developed by Robert Bosch GmbH (i.e., BOSCH), a German automotive control systems manufacturer, in 1983. In 1993, the protocol was adopted as an ISO standard---no. 11898 \cite{history}.

The CAN protocol's design objectives were (1) low latency, (2) high throughput, and (3) reliability. Controller area networks became popular with automotive manufacturers because they drastically reduced the amount of wiring needed. The CAN bus is a dual-wire serial bus. Two wires---a twisted pair---run from the powertrain control module (PCM) to the battery management system (BMS) to the telematic control unit (TCU) and more. Every electronic control unit (ECU) in the vehicle is physically connected to the CAN bus via these two wires. Wiring is dramatically reduced; instead of directly connecting every ECU to every other ECU (and to the CAN bus) using separate wires, every ECU is connected to the two wires that comprise the CAN bus. The ECUs share the CAN bus and take turns communicating. An arbitration scheme allows the ECUs to take turns without conflict or data loss \cite{history, can-bus-government-whitepaper}.

Figure \ref{can-diagram} illustrates the CAN bus: two wires---\texttt{CAN HIGH} and \texttt{CAN LOW} interconnect various ECUs. As the names suggest, \texttt{CAN HIGH} is the high voltage wire; \texttt{CAN LOW} is the low voltage wire. The difference in voltage---not the voltages themselves---determine the signal that is communicated. If \texttt{CAN HIGH} is less than or equal to \texttt{CAN LOW}, then the recessive logic---1---is transmitted. The recessive logic is the default logic. If, instead, \texttt{CAN HIGH} is greater than \texttt{CAN LOW}, then the dominant logic---0---is transmitted. To communicate, an ECU will drive the CAN bus to the dominant state. When communication ceases, 120 $\Omega$ terminating resistors pull the CAN bus back to the recessive state. Generally, both \texttt{CAN HIGH} and \texttt{CAN LOW} are approximately 2.5 V when the bus is in the recessive state; in the dominant state, \texttt{CAN HIGH} is driven to 5 V, while \texttt{CAN LOW} is pulled to ground \cite{history, deep-learning-survey, future}.

As mentioned above, the difference in voltage between \texttt{CAN HIGH} and \texttt{CAN LOW} determines the signal; this property is a form of \textit{differential signaling.} Differential signaling insulates the CAN bus from electrical interference (i.e., noise). Generally, noise will impact both \texttt{CAN HIGH} and \texttt{CAN LOW}; as such, the voltage difference will not change.

\begin{figure}[hbt!]
\centering
\includegraphics[width=1.0\columnwidth]{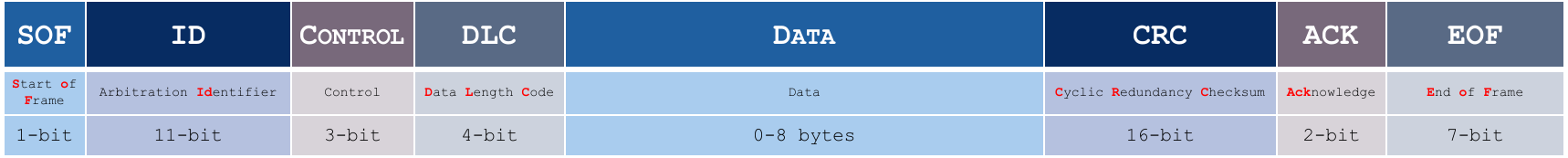}
\caption{CAN frame specification}
\label{can-frame}
\end{figure}

ECUs communicate via packets known as \textit{CAN frames.} The structure of a valid CAN data frame is tightly specified, as shown in Figure \ref{can-frame}. The start of frame---SOF--advertises the start of transmission. The arbitration identifier---arbitration ID---determines the priority of the frame. Typically, the arbitration ID can be used to identify the transmitting ECU. The control field is comprised of several fixed control bits. The data length code---DLC---provides the size of the data field (in bytes). The data field contains the data to be communicated. One or more pieces of information will be encoded in the data field, which can range from zero to eight bytes. The cyclic redundancy checksum---CRC---is a checksum intended to confirm the data's integrity. As the name suggests, the acknowledge---ACK---field is intended for message acknowledgement; any receiving node can use the ACK field to certify that the message was received successfully. Lastly, the end of frame---EOF---demarcates the end of the transmission \cite{deep-learning-survey}.

The CAN protocol itself is standardized across all the automotive manufacturers who have implemented it; however, the implementations are proprietary. Different manufacturers assign different arbitration IDs to different ECUs. Different manufacturers also encode different data into different CAN frames. As such, research work conducted on a particular vehicle often does not generalize to all vehicles fabricated by all manufacturers.

\subsection{The Threat Model}

\begin{figure}[hbt!]
\centering
\includegraphics[width=1.0\columnwidth]{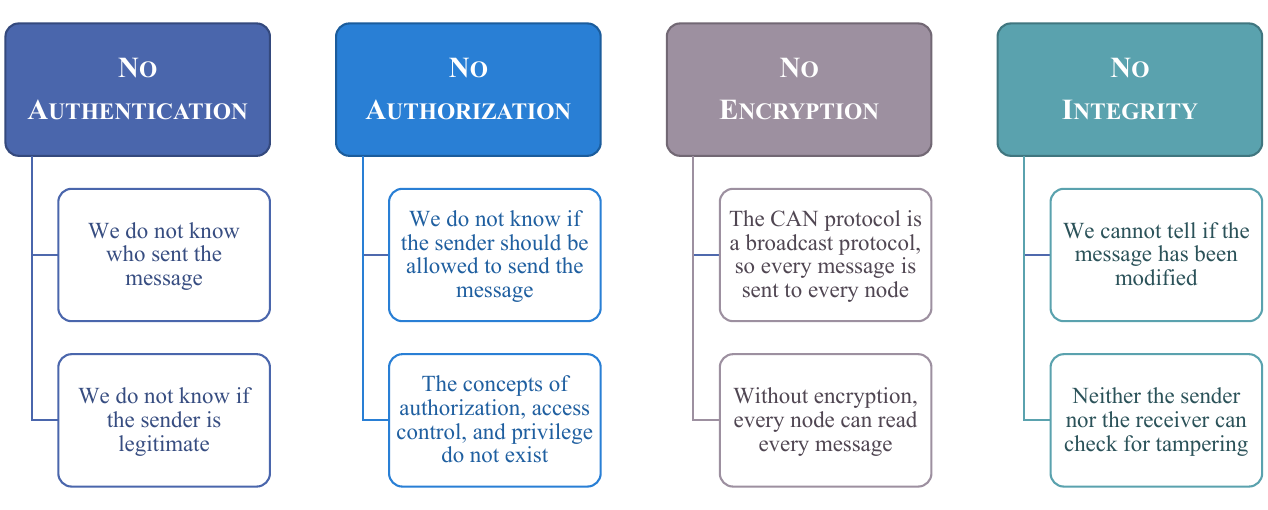}
\caption{Vulnerabilities of the CAN bus}
\label{can-vuln}
\end{figure}

\begin{table*}[hbt!]
\small
\centering

\begin{threeparttable}
\caption{Known CAN Bus Attack Families\tnote{1}} \label{can-attacks}

\begin{tabular}{|p{2cm}|p{5cm}|p{6.5cm}|p{1cm}|} \hline
    \rowcolor{Tan!60}
    \textbf{Attack}           & \textbf{Description}      & \textbf{Impacts}        & \textbf{Figure} \\ \hline \hline
    \rowcolor{Tan!20}
    \textbf{Denial of Service (DoS)}
        & Disable the network---i.e., inhibit legitimate ECUs from communicating over the network---by, e.g., overwhelming it with CAN frames, introducing noise, or manipulating the differential voltage. The lack of network availability constitutes a \textit{denial of service} to legitimate ECUs.
        & Legitimate ECUs will revert to fail-safe parameters.
        & \ref{can-DoS} \\ \hline
    \textbf{Fuzzing}
        & Construct and transmit random or pseudo-random (\textit{fuzzed}) CAN frames. An attacker might randomize both the arbitration ID and the data field, or the attacker might use known, valid arbitration IDs paired with random data fields.
        & Legitimate communication might be disrupted---perhaps by bus errors. Legitimate ECUs generally do not respond to invalid arbitration IDs; however, if valid arbitration IDs are paired with random data fields, then the recipient ECUs might behave erratically. Fuzzing attacks might be used as reconnaissance to prepare for subsequent masquerade or spoofing attacks.
        & \ref{can-fuzz} \\ \hline
    \rowcolor{Tan!20}
    \textbf{Masquerade}
        & \textit{Masquerade} as an authentic ECU in order to, e.g., issue commands to legitimate ECUs. Often, masquerade attacks involve stifling the real ECU in order to eliminate confliction.
        & If deceived, legitimate ECUs could initiate dangerous---even destructive---actions.
        & \ref{can-masquerade} \\ \hline
    \textbf{Replay}
        & Capture and \textit{replay} authentic CAN frames, possibly as the basis for masquerade and spoofing attacks.
        & If deceived, legitimate ECUs could initiate dangerous---even destructive---actions.
        & \ref{can-replay} \\ \hline
    \rowcolor{Tan!20}
    \textbf{Spoofing}
        & Transmit false---i.e., \textit{spoofed}---CAN frames to induce a reaction from legitimate ECUs.
        & If deceived, legitimate ECUs could initiate dangerous---even destructive---actions.
        & \ref{can-spoof} \\ \hline
    \textbf{Suppress}
        & Prevent a legitimate ECU from communicating over the network---i.e., \textit{suppress} CAN frames originating from a particular ECU.
        & Suppressing safety-critical communications (e.g., brake-related communications) could endanger the vehicle, its occupants, and anyone nearby. Suppress attacks are often used to facilitate masquerade attacks.
        & \ref{can-suppress} \\ \hline
\end{tabular}
\footnotesize
\begin{tablenotes}
    \item [1] There is considerable overlap between masquerade, replay, and spoofing attacks. In a replay attack, previously captured CAN frames are simply replayed. However, if an attacker \textit{replays} CAN frames in order to communicate a vehicle status that is no longer true---i.e., \textit{spoofing}---and \textit{masquerade} as a legitimate ECU, then the attack constitutes replay, masquerade, and spoofing. Replay is a relatively simple attack; as such, it is used as a building block for the more complex attacks, masquerade and spoofing.
\end{tablenotes}
\end{threeparttable}

\end{table*}

A number of vulnerabilities afflict the CAN bus, many of which are correlated with four major shortcomings: (1) no authentication, (2) no authorization, (3) no encryption, and (4) no integrity. These vulnerabilities are outlined in Figure \ref{can-vuln}.

\textbf{No authentication.} The CAN bus, as currently implemented, completely lacks authentication. In attack-free conditions, a given arbitration ID will generally map to a particular ECU (or node). However, a malicious node can use a legitimate ECU's arbitration ID to masquerade as the legitimate ECU. If all ECUs were required to authenticate themselves prior to transmitting, then a masquerade attack would not be as simple as using a real ECU's arbitration ID. As it stands, when a CAN frame is received, we cannot confirm that it came from the ECU we expect. We do not know if the CAN frame was sent by a legitimate ECU or a malicious node. Even if we knew that a legitimate ECU sent the CAN frame, we would not be able to determine \textit{which} ECU was the sender.

\textbf{No authorization.} Authorization, as with authentication, is wholly absent from the CAN bus. In an attack-free situation, a given ECU will send CAN frames related to the information it possesses. For example, the door control unit (DCU), as the name suggests, is concerned with the functionalities provided by the doors of modern automobiles (e.g., locking and unlocking doors, rolling windows up and down, child safety locks, etc.). In attack-free conditions, we would expect the DCU to send CAN frames regarding the status of the doors---open or closed, locked or unlocked, etc. The DCU should \textit{not} send CAN frames asserting the vehicle's speed, the temperature of the engine coolant, or the current gear (drive, neutral, reverse, park). If the DCU begins transmitting CAN frames that proclaim ``The vehicle is traveling 15 mph in reverse," then something has gone wrong (perhaps an attacker has compromised the DCU). Without authorization, however, there is nothing to stop an ECU from transmitting information beyond its scope and function.

\textbf{No encryption.} CAN traffic is transparent, unencrypted. While the information contained in the data field of a CAN frame is encoded (e.g., wheel speeds might be converted to hexadecimal and compressed), it is often trivial to decode. Data is encoded in order to maximize throughput, not ensure confidentiality. Moreover, the CAN protocol is a broadcast protocol, meaning that every message is sent to every node. If attackers attach themselves to the CAN bus or compromise even one ECU, they can read every message from every node. With a bit of reconnaissance, attackers can decode the information contained in captured CAN frames. From there, they can craft spoofed CAN frames or simply replay the captured CAN frames. If CAN traffic were encrypted, reconnaissance would be much, much harder, and the compromise of one ECU would not be nearly as catastrophic.

\textbf{No integrity.} As mentioned earlier, the CAN bus does have a cyclic redundancy checksum (CRC) intended to ensure data integrity. However, the CRC is effective only under attack-free conditions. If attackers wished to spoof or modify a CAN frame, they would merely need to recompute the checksum for the modified message. This computation is trivial and is part of the standard---and publicly available---CAN protocol. Therefore, under attack conditions, the CAN bus utterly lacks integrity protections. If integrity controls were incorporated into the CAN bus, we would know if the CAN frame had been subject to tampering, and the origin of the CAN frame would be verifiable.

\begin{figure}[hbt!]
\centering
\includegraphics[width=1.0\columnwidth]{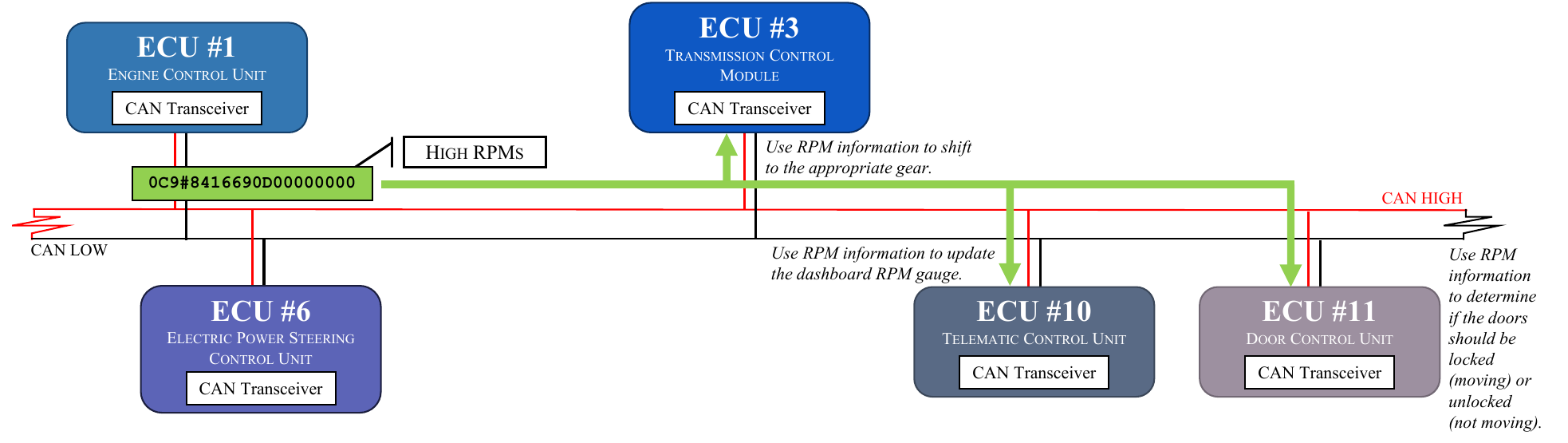}
\caption{Schematic diagram of attack-free CAN communications}
\label{can-attack-free}
\end{figure}

When developing a threat model, the CIA Triad is a popular criterion. The ``C," ``I," and ``A," refer to confidentiality, integrity, and availability, respectively. \textit{Confidentiality} involves protecting sensitive information from unauthorized access. \textit{Integrity} is concerned with tamper-resistance and verification of origin---we should know if the data has been altered, and, if the data has not been altered, then we should know who authored the data. \textit{Availability} ensures that authorized entities can access the data (or service) \cite{cia-triad-can-logging}. The CAN bus falls short in terms of confidentiality, integrity, and availability. The lack of encryption results in a lack of confidentiality. The lack of integrity controls---as well as the lack of authentication---leads to a lack of integrity. Because the CAN bus lacks authentication, a malicious node can overwhelm the network with spoofed CAN frames, preventing legitimate nodes from communicating. As such, the CAN bus also lacks availability.

As we have seen, the CAN bus has a wide array of vulnerabilities; naturally, a wide array of CAN bus attacks have been identified and described. CAN-related attacks can be subdivided into six major attack families:

\begin{enumerate}
    \item Denial of Service (DoS)
    \item Fuzzing
    \item Masquerade
    \item Replay
    \item Spoofing
    \item Suppress
\end{enumerate}

Table \ref{can-attacks} summarizes these known CAN bus attack families.

\begin{figure}[hbt!]
\centering
\includegraphics[width=1.0\columnwidth]{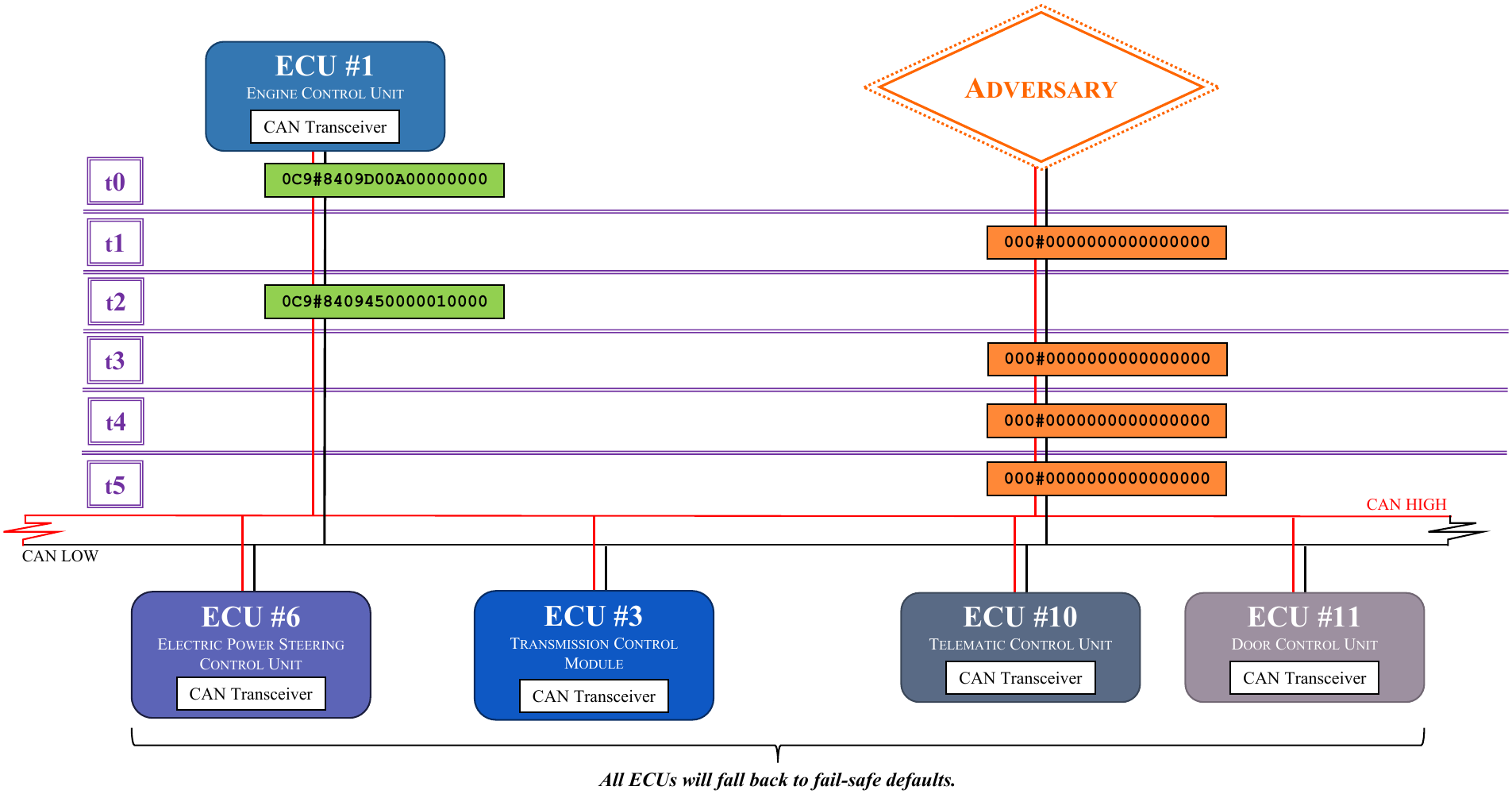}
\caption{Schematic diagram of a denial of service (DoS) attack}
\label{can-DoS}
\end{figure}

\begin{figure}[hbt!]
\centering
\includegraphics[width=1.0\columnwidth]{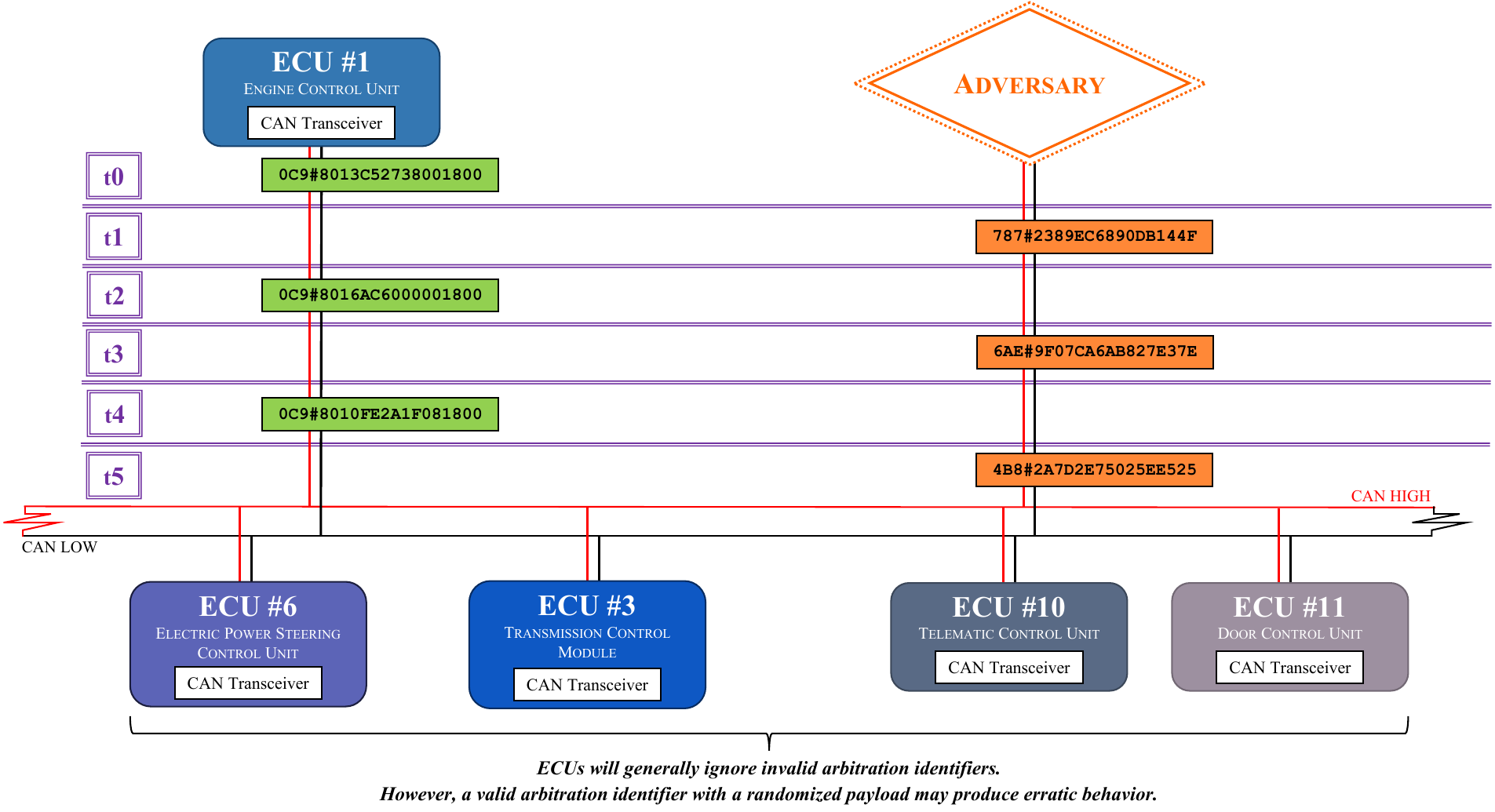}
\caption{Schematic diagram of a fuzzing attack}
\label{can-fuzz}
\end{figure}

\begin{figure}[hbt!]
\centering
\includegraphics[width=1.0\columnwidth]{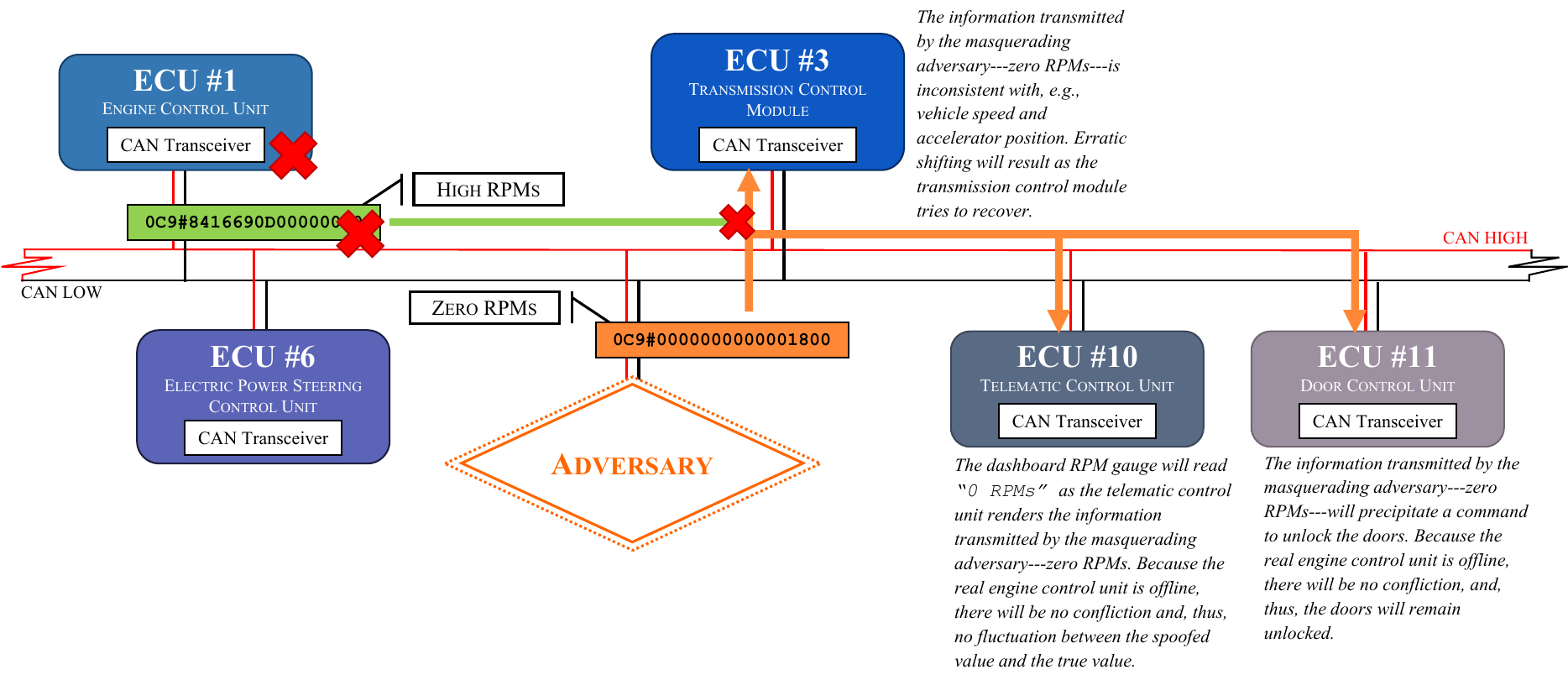}
\caption{Schematic diagram of a masquerade attack}
\label{can-masquerade}
\end{figure}

\begin{figure}[hbt!]
\centering
\includegraphics[width=1.0\columnwidth]{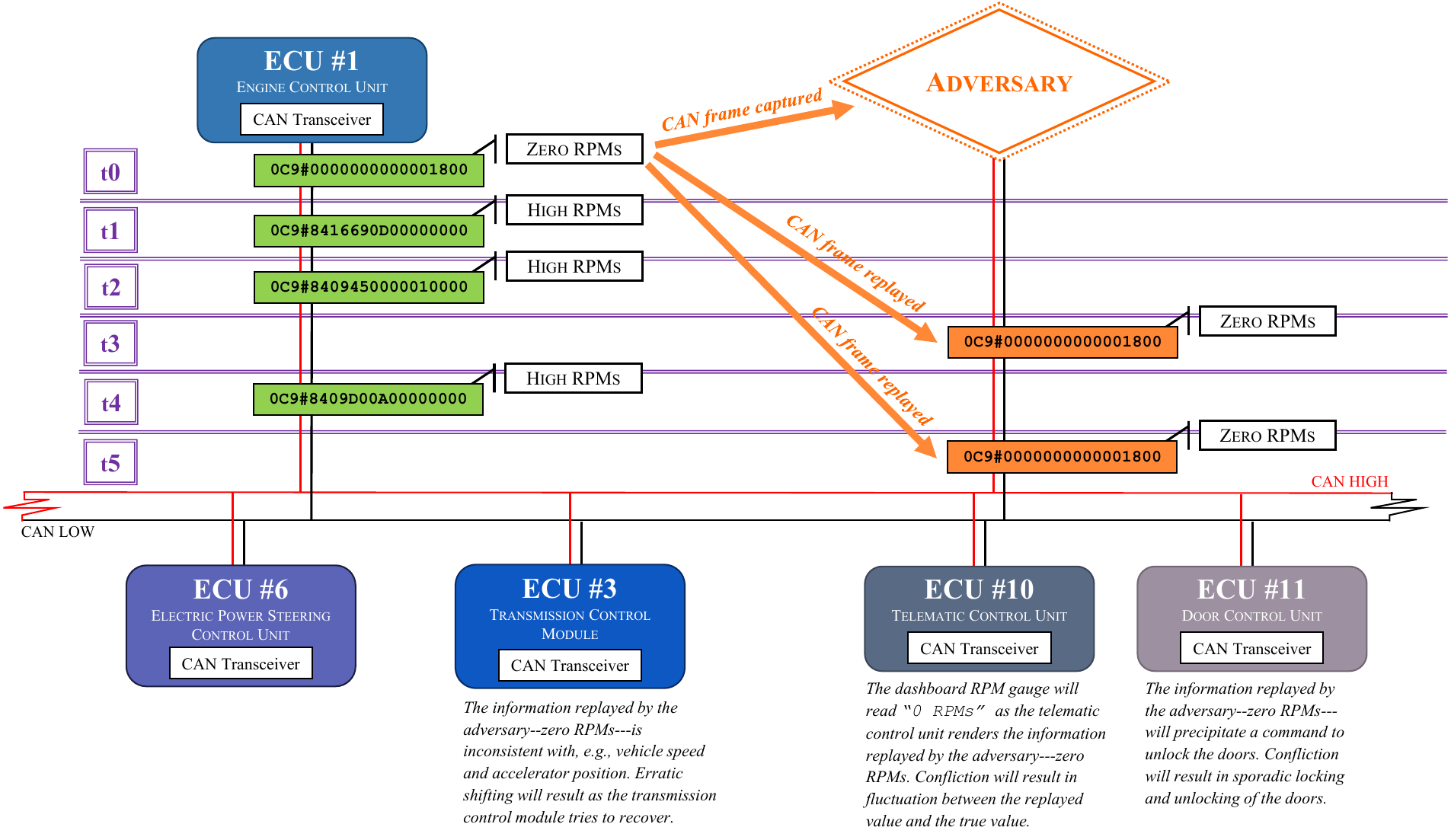}
\caption{Schematic diagram of a replay attack}
\label{can-replay}
\end{figure}

\begin{figure}[hbt!]
\centering
\includegraphics[width=1.0\columnwidth]{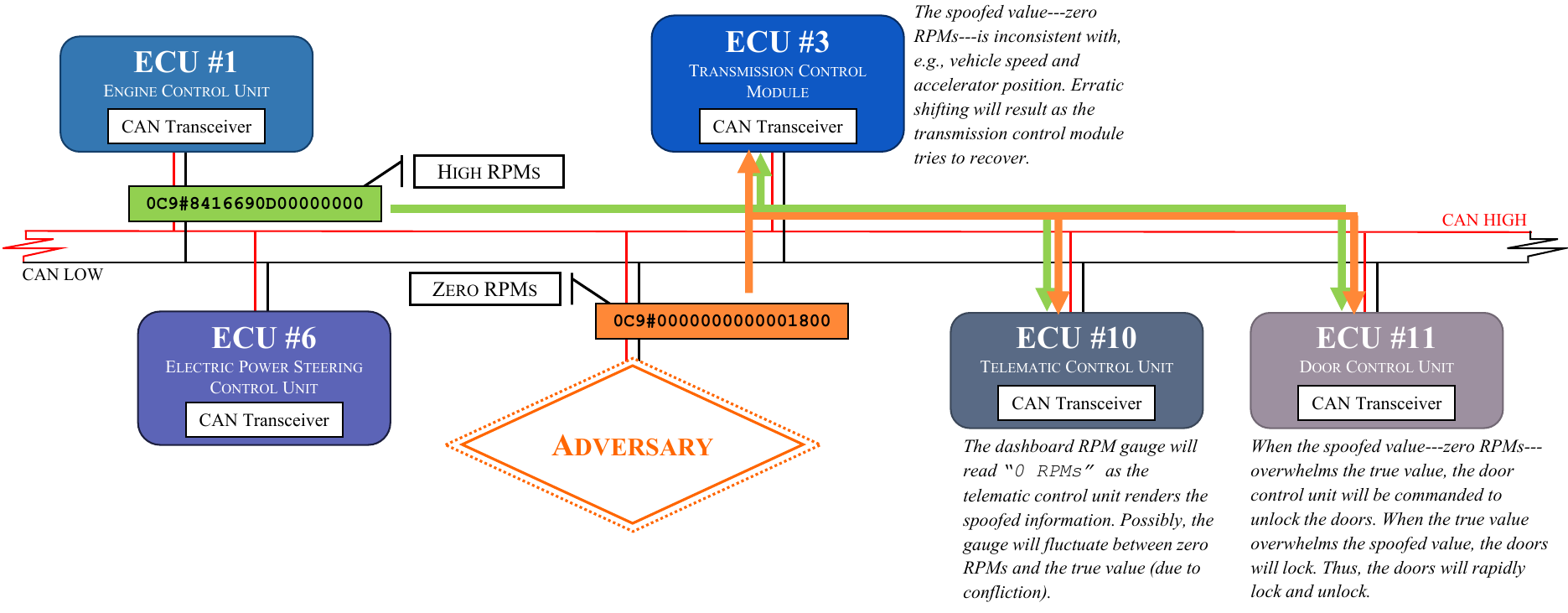}
\caption{Schematic diagram of a spoofing attack}
\label{can-spoof}
\end{figure}

\begin{figure}[hbt!]
\centering
\includegraphics[width=1.0\columnwidth]{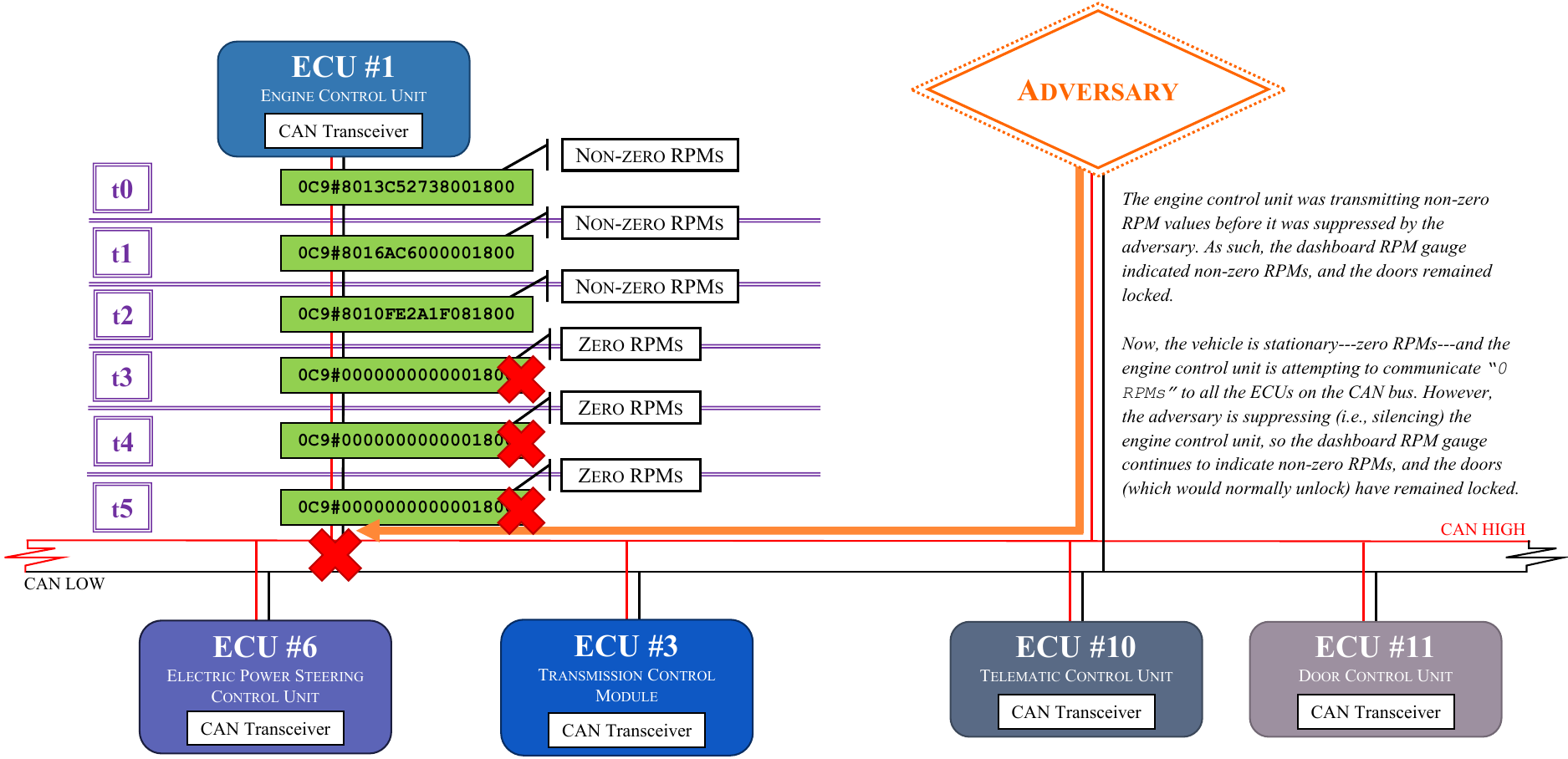}
\caption{Schematic diagram of a suppress attack}
\label{can-suppress}
\end{figure}

Figure \ref{can-attack-free} depicts the CAN bus under attack-free conditions. This diagram provides a ``normal" baseline for CAN communications. We can refer back to this diagram in order to better understand the schematic diagrams of the CAN bus under various attack conditions (Figures \ref{can-DoS}, \ref{can-fuzz}, \ref{can-masquerade}, \ref{can-replay}, \ref{can-spoof}, and \ref{can-suppress}).

Figure \ref{can-DoS} illustrates the CAN bus during a denial of service (DoS) attack. As shown in the figure, the adversary transmits an overwhelming volume of high-priority CAN frames (arbitration ID \texttt{000}). CAN frames from the legitimate ECU are forced to wait on the high-priority traffic. Since the adversary transmits high-priority CAN frames so frequently, the legitimate ECU never has a chance to communicate. Thus, the legitimate ECU experiences a \textit{denial of service.}

Figure \ref{can-fuzz} showcases the CAN bus under fuzzing conditions. The adversary is transmitting randomized CAN frames. In this example, both the arbitration ID and the data field are randomized; as such, most of the fuzzed CAN frames have meaningless arbitration IDs that will be ignored by legitimate ECUs. Legitimate ECUs listen for relevant CAN frames---relevant arbitration IDs---while ignoring CAN frames that are irrelevant.

For example, the door control unit (DCU) will listen for CAN frames bearing the transmission control module (TCM)'s arbitration ID---because the DCU wants to keep the doors locked while driving but unlock them once the vehicle parked. However, the DCU will ignore CAN frames related to fuel injection in the engine, as they are irrelevant to its function.

Therefore, while a randomized fuzzing attack can be disruptive, we would expect it to be less disruptive than a more sophisticated fuzzing attack that uses valid arbitration IDs combined with randomized data fields. This type of sophisticated fuzzing attack might randomly generate (i.e., \textit{fuzz}) data fields that---paired with valid arbitration IDs---cause dangerous and even destructive behavior.

Figure \ref{can-masquerade} demonstrates a particularly sophisticated type of masquerade attack. A sophisticated masquerade attack combines the capabilities of replay, spoofing, and suppress attacks. A malicious node targets a particular ECU (e.g., the engine control unit). To masquerade as the legitimate ECU, the malicious node can either (1) capture and replay the legitimate ECU's CAN frames or (2) spoof CAN frames using the legitimate ECU's arbitration ID.

Of course, the replayed or spoofed CAN frames will clash with the legitimate ECU's CAN frames---this phenomenon is known as ``confliction." To alleviate confliction, the malicious node must prevent the legitimate ECU from communicating, perhaps by intercepting its CAN frames. Miller and Valasek eliminated confliction by putting the legitimate ECU into Bootrom mode (by starting the reprogramming process) \cite{advanced-can-injection-whitepaper, blackhat-advanced-can-injection}.

Once the legitimate ECU is silenced and the malicious node is transmitting CAN frames using the legitimate ECU's arbitration ID, the malicious node is essentially \textit{masquerading} as the legitimate ECU. In our example, the engine control unit has been silenced; as such, the transmission control unit, the telematic control unit, and the door control unit all believe that they are receiving CAN frames from the engine control unit, but, in fact, they are receiving CAN frames from the adversary.

Figure \ref{can-replay} depicts a replay attack against the CAN bus. This type of attack is relatively simple; the adversary simply captures and retransmits (i.e., \textit{replays}) CAN frames originally transmitted by legitimate ECUs. If the adversary collects a sufficient number of legitimate CAN frames, this type of attack can become incredibly powerful. In our example, the engine is generating ``high RPMs" (presumably because the vehicle is in motion), but the adversary has tricked the telematic control unit and the door control unit into believing that the vehicle is stopped by transmitting ``zero RPMs." Of course, due to confliction, the needle in the RPM gauge will waffle between ``zero RPMs"---the spoofed value---and ``high RPMs"---the true value. Similarly, the doors will sporadically lock and unlock.

This type of attack is often a building block for more sophisticated attacks (e.g., masquerade attacks). In a sophisticated masquerade attack, the aforementioned confliction issues---RPM gauge fluctuation, sporadic door locking/unlocking---would be eliminated.

Figure \ref{can-spoof} outlines a CAN bus spoofing attack. A spoofing attack is similar to a replay attack, but it goes a step further. The adversary will still need to eavesdrop on CAN communications. To mount a more effectual attack, the adversary will need to determine which CAN frames should be abused in order to achieve the desired outcome (e.g., disruption, damage, death). The adversary will use his or her reconnaissance to craft tailored (i.e., \textit{spoofed}) CAN frames that achieve the desired outcome.

Figure \ref{can-suppress} depicts the CAN bus during a suppress attack. This attack stifles (i.e., \textit{suppresses}) communication from a target ECU. In our example, the target ECU is the engine control unit. Oftentimes, a suppress attack will be combined with replay or spoofing in order to conduct a sophisticated masquerade attack; however, a suppress attack can be effective on its own. In Figure \ref{can-suppress}, the engine control unit begins by announcing non-zero RPMs (presumably, the vehicle is in motion). Later, the engine control unit wishes to communicate that the RPMs have dropped to zero. The door control unit (DCU) knows that the doors should remain locked while the vehicle is in motion---non-zero RPMs---and unlock only when the RPMs are down to zero (this is a simplified example). As such, even though the vehicle is down to zero RPMs at time \texttt{t3}, the doors will remain locked. In some vehicles, passengers will not be able to manually override the door locks when the vehicle is in motion---or when the DCU believes the vehicle is in motion.

\subsection{The Exploits}

\begin{table*}[hbt!]
\small
\centering

\begin{threeparttable}
\caption{Published CAN Bus Exploits\tnote{1}} \label{can-exploits}

\begin{tabular}{|p{1cm}|p{2.5cm}|p{7.5cm}|p{3cm}|} \hline
    \rowcolor{Tan!60}
    \textbf{Year}       & \textbf{Highlights}       & \textbf{Description}      & \textbf{Source} \\ \hline \hline

    \rowcolor{Tan!20}
    2010 & Manipulating ECUs, embedding malicious code
    & An analysis of the vulnerabilities of internal vehicle networks that uncovered various types of attacks---e.g., bridging between a vehicle's internal subnetworks, manipulating ECUs (selectively engaging and disabling the brakes), and embedding malicious code in the telematics unit, to name a few.
    & Koscher et al. \cite{experimental-hacks} \\ \hline
    
    2014 & Diagnostic commands, ECU reprogramming
    & An analysis of the vulnerabilities of internal vehicle networks that uncovered various types of attacks---e.g., disabling power steering via denial of service attack, sending diagnostic commands (to shut down the engine, to selectively engage and disable the brakes, etc.), and reprogramming ECUs.
    & Miller and Valasek \cite{adventures-whitepaper} \\ \hline

    \rowcolor{Tan!20}
    2015 & Telematics, remote control
    & Exploitation of a popular telematic control unit---plugged into the OBD-II port---to achieve remote control of a vehicle.
    & Foster et al. \cite{telematics-hacks} \\ \hline

    2015 & Remote control of an unaltered private automobile
    & Exploitation of a cellular network vulnerability to achieve arbitrary remote control of an unaltered private automobile---e.g., exfiltrating global positioning system (GPS) coordinates to track the vehicle, locking and unlocking the vehicle, shutting down the engine, etc. 
    & Miller and Valasek \cite{remote-exploitation-whitepaper, blackhat-remote-exploitation} \\ \hline

    \rowcolor{Tan!20}
    2016 & Advanced CAN injection attacks, steering wheel manipulating at speed
    & Eliminate confliction in order to conduct advanced CAN injection attacks, such as manipulating the steering wheel at speed.
    & Miller and Valasek \cite{advanced-can-injection-whitepaper, blackhat-advanced-can-injection} \\ \hline

    2017 & Remote control of an unaltered private automobile
    & A chain of exploits---including browser exploits, local privilege escalation, and ECU reprogramming---was used to compromise and remotely control an unaltered Tesla automobile.
    & Nie, Liu, and Du \cite{keen-tesla} \\ \hline

    \rowcolor{Tan!20}
    2018 & Airbag deployment
    & Penetration testing identified vulnerabilities in a vehicle's pyrotechnic control unit (PCU), which is responsible for airbag deployment. The vulnerabilities were successfully exploited on an automotive testbed, generating the firing impulse that---in a real vehicle---would result in airbag deployment.
    & Dürrwang et al. \cite{airbag} \\ \hline

    2019 & Remote control of an unaltered private automobile, SMS
    & Vulnerabilities in a BMW's infotainment and telematics components were used to remotely compromise the vehicle. Remote services were triggered via short message service (SMS) communications, and cross-domain diagnostic commands were used to access additional ECUs.
    & Cai, Wang, and Zhang \cite{keen-bmw} \\ \hline

    \rowcolor{Tan!20}
    2023 & Grand theft auto
    & Many vehicles provide relatively easy access to the CAN bus from the outside---e.g., via wiring behind a headlight. Thieves can access the CAN bus, inject spoofed CAN frames, suppress conflicting CAN frames, and, ultimately, steal the vehicle.
    & Tindell \cite{theft} \\ \hline
\end{tabular}
\footnotesize
\end{threeparttable}

\end{table*}

Table \ref{can-exploits} highlights a number of published CAN bus exploits, ranging from nuisance attacks to theft to life-threatening remote-controlled attacks.

One of the earliest experimental CAN bus exploits was conducted in 2010 by Koscher et al. \cite{experimental-hacks}. They were able to continuously activate the vehicle's door locks, run the windshield wipers, blare the horn, turn various lights on and off, kill the engine, grind the starter, engage the brakes, disable the brakes, and more.

Miller and Valasek have conducted a number of proof-of-concept CAN bus exploits \cite{adventures-whitepaper, remote-exploitation-whitepaper, blackhat-remote-exploitation, advanced-can-injection-whitepaper, blackhat-advanced-can-injection}. Many of the exploits would be life-threatening---even catastrophic---if conducted by malicious adversaries rather than researchers. For example, the attack against a 2014 Jeep Cherokee \cite{remote-exploitation-whitepaper, blackhat-remote-exploitation} could be generalized to a number of Fiat Chrysler Automotive (FCA) automobiles---including various models from the Chrysler, Dodge, Jeep, and Ram brands. FCA recalled 1.4 million vehicles. If malicious adversaries (e.g., nation-state actors) were to simultaneously compromise those 1.4 million vehicles and disable their brakes, the results could easily be catastrophic.

Miller and Valasek disabled the 2014 Jeep Cherokee's brakes via diagnostic messages, which most of its ECUs would ignore when the vehicle was traveling at speed. Cai, Wang, and Zhang \cite{keen-bmw}, on the other hand, found that most BMW ECUs will respond to diagnostic messages even when the vehicle is driving at normal speeds.

When we zoom out to a ``big picture" view, we can see that there are even bigger problems. Innovations such as advanced driver assistance systems (ADAS), automotive platooning, and smart cities are vulnerable to both \textit{intra-}vehicle and \textit{inter-}vehicle attacks. The ``smarter" the vehicle, the more devastating the potential attack. If a vehicle lacks electronic steering, then an adversary cannot control the steering. A ``smarter" vehicle, equipped with electronic steering (e.g., parking assist, lane-keeping assist), faces the uncomfortable possibility of adversary-controlled steering. Ben Lakhal et al. \cite{future} identified a number of reliability and security issues related to the modern CAN bus that specifically threaten intelligent transportation systems (ITSs). All in all, we can see that the CAN bus is critical to the future of automotive transportation and innovation.

\section{Related Work} \label{sec:related-work}

Constructing a high-fidelity CAN dataset often involves logging---i.e., recording CAN traffic from a real vehicle. Daily and Van \cite{cia-triad-can-logging} describe a cryptographically-secured CAN logging scheme, which uses encryption and hashing to achieve confidentiality and integrity, respectively.

Marx et al. \cite{can-logging-comparison} analyzed and compared different CAN logging techniques in terms of accuracy and file size. They experimented with both frame- and waveform-based logging strategies. Across all strategies, the mean difference between measurements was less than 0.08\%. As such, the authors concluded that, accuracy-wise, both frame- and waveform-based logging techniques would be acceptable, though they noted that waveform-based logging consumes less memory and would be desirable for long-term logging.

Reverse engineering is essential to understanding CAN traffic, particularly when devising CAN bus attacks. The quality of a CAN intrusion detection dataset is heavily dependent on the quality of the attacks. Buscemi et al. \cite{can-reverse-engineering} surveyed CAN bus reverse engineering methodologies. They reviewed data collection---i.e., CAN logging---as well as the popular \texttt{.dbc} file type, which can convert a raw CAN signal into a human-readable, semantic value. Moreover, Buscemi et al. conducted benchmarks of both (1) CAN tokenization algorithms and (2) CAN translation algorithms.

\subsection{Existing Datasets}


\begin{table*}[hbt!]
\footnotesize
\centering

\begin{threeparttable}
\caption{CAN Bus Datasets: \textit{A Summary}} \label{can-datasets}

\begin{tabular}{|p{0.8cm}|p{3cm}|p{1.3cm}|p{2.4cm}|p{1.3cm}|p{1.3cm}|p{0.8cm}|p{2.3cm}|} \hline
    \rowcolor{Tan!60}
    \textbf{Year}               & \textbf{Name}                     & \textbf{Acronym}
    & \textbf{Vehicle(s)}       & \textbf{\# Attacks}               & \textbf{Labeled?}
    & \textbf{Real?}            & \textbf{Source} \\ \hline \hline

    \rowcolor{Tan!20}
    2016 & Simulated CAN Bus dataset                                            & Sim CAN
    & Testbed                       & 1         & No        & No
    & Kang and Kang \cite{sim-can} \\ \hline
    
    2017 & CrySyS Lab CAN dataset                                               & CrySyS CAN
    & Unknown                       & 0         & N/A       & Yes
    & CrySyS Lab \cite{crysys} \\ \hline

    \rowcolor{Tan!20}
    2017 & HCRL CAN dataset                                                     & HCRL CAN
    & Kia Soul                      & 3         & No        & Yes
    & Lee, Jeong, and Kim \cite{hcrl-can-dataset} \\ \hline

    2018 & HCRL Car-Hacking dataset                                             & HCRL CH
    & Hyundai YF Sonata             & 4         & Yes       & Yes
    & Seo, Song, and Kim \cite{hcrl-car-hacking-dataset-1}; Song, Woo, and Kim \cite{hcrl-car-hacking-dataset-2} \\ \hline

    \rowcolor{Tan!20}
    2018 & HCRL Survival Analysis dataset                                       & HCRL SA
    & Chevrolet Spark, Hyundai YF Sonata, Kia Soul
    & 4         & Yes       & Yes
    & Han, Kwak, and Kim \cite{hcrl-survival-analysis-dataset} \\ \hline

    2019 & AEGIS Big Data Project Automotive CAN Bus dataset                    & AEGIS CAN
    & Unknown                       & 0         & N/A       & Yes
    & Kaiser, Stocker, and Festl \cite{aegis} \\ \hline

    \rowcolor{Tan!20}
    2019 & TU Eindhoven CAN bus intrusion dataset v2                            & TU CAN v2
    & Opel Astra, Renault Clio, Testbed
    & 5         & No        & Yes\tnote{1}
    & Dupont et al. \cite{tu-can-v2} \\ \hline

    2020 & HCRL CAN Signal Extraction and Translation dataset                   & HCRL SET
    & Unknown                       & 0         & N/A       & Yes
    & Song and Kim \cite{hcrl-can-signal-extraction-and-translation-dataset} \\ \hline

    \rowcolor{Tan!20}
    2020 & ML350 CAN Bus dataset                                                & ML350 CAN
    & Mercedes ML350                & 2         & Yes       & Yes
    & Sami et al. \cite{ml350-dataset-1, ml350-dataset-2} \\ \hline

    2020 & Real ORNL Automotive Dynamometer CAN intrusion dataset               & ROAD
    & Unknown (mid-2010s)           & 10\tnote{2}           & No        & Yes
    & Verma et al. \cite{road-original}; Verma et al. \cite{road} \\ \hline

    \rowcolor{Tan!20}
    2020 & Synthetic CAN Bus dataset                                            & SynCAN
    & Testbed                                   & 5         & No        & No
    & Hanselmann et al. \cite{syncan} \\ \hline

    2021 & HCRL Attack \& Defense Challenge dataset                             & HCRL A\&D
    & Hyundai Avante CN7                        & 4         & Yes       & Yes
    & Kang et al. \cite{hcrl-attack-and-defense-challenge-dataset} \\ \hline

    \rowcolor{Tan!20}
    2023 & \texttt{can-train-and-test} dataset\tnote{3}                         & \texttt{CT\&T}
    & Chevrolet Impala, Chevrolet Silverado, Chevrolet Traverse, Subaru Forester
    & 9         & Yes       & Yes\tnote{2}
    & Our contribution \\ \hline
\end{tabular}
\footnotesize
\begin{tablenotes}
    \item [1] This dataset contains both real and simulated CAN traffic.
    \item [2] This dataset contains 33 attack traffic captures; however, the attack types are not unique. There are fuzzing attacks, spoofing \& masquerade attacks, and accelerator attacks. Four different signals are spoofed---wheel speed, engine coolant temperature, speedometer value, and reverse light (on/off). For each spoofing attack, Verma et al. provide a masquerade version (legitimate signals are suppressed). We count the fuzzing attack, the four spoofing attacks, the four masquerade attacks, and the accelerator attack as unique attacks, resulting in a total of ten attacks.
    \item [3] \texttt{can-train-and-test} is the name of the repository which contains the final curated dataset (labeled and partitioned). The \texttt{can-train-and-test} repository is linked to the \texttt{can-dataset} and \texttt{can-ml} repositories, which contain raw \texttt{.log} files, additional attack-free traffic, and additional attack traffic. As such, we use the term \texttt{can-train-and-test} to refer to the \texttt{can-train-and-test} repository as well as the linked repositories.
\end{tablenotes}
\end{threeparttable}

\end{table*}


\begin{table}[hbt!]
\footnotesize
\centering

\begin{threeparttable}
\caption{CAN Bus Datasets: \textit{Details}} \label{can-datasets-detail-1}

\begin{tabular}{|p{0.8cm}|p{2.3cm}|p{1.3cm}|p{8.6cm}|p{2cm}|} \hline
    \rowcolor{Tan!60}
    \textbf{Year}               & \textbf{Name}                     & \textbf{Acronym}
    & \textbf{Details\tnote{1}}                                     & \textbf{Source} \\ \hline \hline

    \rowcolor{Tan!20}
    2016 & Simulated CAN Bus dataset                                            & Sim CAN
    & A 3-ECU testbed was constructed to simulate both benign traffic and generalized injection-type attacks. During the simulation, 200,000 packets were generated using the Open Car Testbed and Network Experiments (OCTANE) \cite{octane} packet generator.
    & Kang and Kang \cite{sim-can} \\ \hline
    
    2017 & CrySyS Lab CAN dataset                                               & CrySyS CAN
    & This dataset consists of purely benign data associated with specific driving scenarios (e.g., ``Driving with a speed of 40 km/h, then lane change, then stop"). A ``CAN log infector" is provided to tack on attacks in the post-processing stage.
    & CrySyS Lab \cite{crysys} \\ \hline

    \rowcolor{Tan!20}
    2017 & HCRL CAN dataset                                                     & HCRL CAN
    & This dataset is one of many datasets developed and published by the Hacking and Countermeasure Research Lab (HCRL). It contains benign traffic as well as several types of attacks---DoS, fuzzing, and impersonation (i.e., masquerade). The timestamp, arbitration ID, data length code (DLC), and data field are provided as features. The DoS attack is labeled, but none of the others are.
    & Lee, Jeong, and Kim \cite{hcrl-can-dataset} \\ \hline

    2018 & HCRL Car-Hacking dataset                                             & HCRL CH
    & This dataset contains benign traffic as well as several types of attacks---DoS, fuzzing, RPM spoofing, and gear spoofing. The timestamp, arbitration ID, DLC, data field, and label (``T" for injected, ``R" for normal) are provided as features. During data collection, the vehicle was parked with the engine running. According to Rajapaksha et al. \cite{acm-survey}, this is the most widely used dataset in the literature (to evaluate CAN IDSs).
    & Seo, Song, and Kim \cite{hcrl-car-hacking-dataset-1}; Song, Woo, and Kim \cite{hcrl-car-hacking-dataset-2} \\ \hline

    \rowcolor{Tan!20}
    2018 & HCRL Survival Analysis dataset                                       & HCRL SA
    & Similar to the HCRL Car-Hacking dataset, this dataset contains both benign traffic and attacks---flooding (i.e., DoS) and malfunction. The timestamp, arbitration ID, DLC, data field, and label (``T" for injected, ``R" for normal) are provided as features. An attack-free traffic capture is available for each vehicle; however, the traffic captures are not large enough to train a good classifier \cite{acm-survey}.
    & Han, Kwak, and Kim \cite{hcrl-survival-analysis-dataset} \\ \hline

    2019 & AEGIS Big Data Project Automotive CAN Bus dataset                    & AEGIS CAN
    & This dataset consists of purely benign data---specifically, signal data. It is similar to the dataset curated by Hanselmann et al. \cite{syncan}, though it also contains global positioning system (GPS) data.
    & Kaiser, Stocker, and Festl \cite{aegis} \\ \hline

    \rowcolor{Tan!20}
    2019 & TU Eindhoven CAN bus intrusion dataset v2                            & TU CAN v2
    & This dataset contains benign traffic as well as several types of attacks---DoS, fuzzing, replay, suspension (i.e., suppress), and diagnostic. This dataset contains both real data---from two vehicles---and synthetic data from a testbed. This dataset furnishes the only diagnostic attack publicly available. However, some of the simulated attacks have fidelity shortcomings; for example, the DoS attack was constructed by overwriting ten seconds' worth of traffic---real DoS attacks are much noisier \cite{road}.
    & Dupont et al. \cite{tu-can-v2} \\ \hline
\end{tabular}
\footnotesize
\begin{tablenotes}
    \item [1] The descriptions in the table were aggregated with help from \cite{acm-survey, road, taxonomy-survey}, and \cite{survey-ivn}.
\end{tablenotes}
\end{threeparttable}

\end{table}


\begin{table}[hbt!]
\footnotesize
\centering

\begin{threeparttable}
\caption{CAN Bus Datasets: \textit{Details} (continued)} \label{can-datasets-detail-2}

\begin{tabular}{|p{0.8cm}|p{2.3cm}|p{1.3cm}|p{8.6cm}|p{2cm}|} \hline
    \rowcolor{Tan!60}
    \textbf{Year}               & \textbf{Name}                     & \textbf{Acronym}
    & \textbf{Details\tnote{1}}                                     & \textbf{Source} \\ \hline \hline

    \rowcolor{Tan!20}
    2020 & HCRL CAN Signal Extraction and Translation dataset                   & HCRL SET
    & This dataset consists of purely benign data; it was designed for signal extraction, not intrusion detection. The timestamp, arbitration ID, DLC, and data field are provided as features.
    & Song and Kim \cite{hcrl-can-signal-extraction-and-translation-dataset} \\ \hline

    2020 & ML350 CAN Bus dataset                                                & ML350 CAN
    & This dataset was collected from a real vehicle using a CL2000 CAN bus logger. It contains both benign traffic and attacks---DoS and fuzzing. The label ``R" refers to attack-free messages, while the label ``T" identifies attack messages.
    & Sami et al. \cite{ml350-dataset-1, ml350-dataset-2} \\ \hline

    \rowcolor{Tan!20}
    2020 & Real ORNL Automotive Dynamometer CAN intrusion dataset               & ROAD
    & This dataset contains benign traffic as well as several types of attacks---fuzzing, targeted ID attacks (i.e., spoofing, masquerade), and accelerator attacks. The attacks (except the masquerade attacks) were conducted while the test vehicle was on a dynamometer (a ``rolling road"---the vehicle remains stationary, but the wheels turn as though driving normally \cite{dynamometer}); as such, the attacks are high fidelity. The timestamp, arbitration ID, and data field are provided as features; the dataset is unlabeled.
    & Verma et al. \cite{road-original}; Verma et al. \cite{road} \\ \hline

    2020 & Synthetic CAN Bus dataset                                            & SynCAN
    & This dataset contains benign traffic as well as several types of attacks---flooding (i.e., DoS), playback (i.e., replay), suppress, plateau, and continuous. However, it contains only signal values---no raw CAN data. In this regard, it is similar to the dataset curated by Kaiser, Stocker, and Festl \cite{aegis}. All attacks are simulated; as such, the real impacts---if any---on real vehicles are not known. According to Rajapaksha et al. \cite{acm-survey}, this is the most widely used dataset to evaluate unsupervised payload-based CAN IDSs.
    & Hanselmann et al. \cite{syncan} \\ \hline

    \rowcolor{Tan!20}
    2021 & HCRL Attack \& Defense Challenge dataset                             & HCRL A\&D
    & This dataset contains benign traffic as well as several types of attacks---flooding (i.e., DoS), fuzzing, replay, and spoofing. The timestamp, arbitration ID, DLC, data field, class label (normal or attack), and subclass label (attack type) are provided as features. Generally, the attacks were conducted when the vehicle was stationary---due to safety concerns.
    & Kang et al. \cite{hcrl-attack-and-defense-challenge-dataset} \\ \hline

    2023 & \texttt{can-train\newline-and-test} dataset\tnote{2}                         & \texttt{CT\&T}
    & This dataset contains several simulated attacks as well as live attacks on real vehicles (the attacks were conducted when the vehicle was being driven down rural roads). Both labeled and unlabeled data is provided; for the labeled data, ``0" denotes an attack-free CAN frame, while ``1" identifies an attack frame. The attack type is specified by the file name. The curated dataset---\texttt{can-train-and-test}---enables researchers to assess the capability of a proposed intrusion detection system under several conditions (e.g., known vehicle, known attack; unknown vehicle, known attack; etc.).
    & Our contribution \\ \hline
\end{tabular}
\footnotesize
\begin{tablenotes}
    \item [1] The descriptions in the table were aggregated with help from \cite{acm-survey, road, taxonomy-survey}, and \cite{survey-ivn}.
\end{tablenotes}
\end{threeparttable}

\end{table}

There are a number of existing CAN datasets, each with its own advantages and disadvantages. Table \ref{can-datasets} summarizes the existing datasets, providing information such as vehicle type(s), number of attacks, etc. Meanwhile, Tables \ref{can-datasets-detail-1} and \ref{can-datasets-detail-2} furnish details and highlights of the existing datasets, such as attack types, advantages, and limitations.

A common limitation of existing datasets is the lack of attack data; when developing an intrusion detection system (IDS), attack data is of paramount importance. Insufficient data quantity is a second major issue, especially when it comes to machine learning IDSs, as they require a substantial quantity of data for training (and testing). Quality---that is, fidelity---can also be problematic. Researchers are understandably cautious about conducting CAN bus attacks while driving, but some simulations and testbeds are not the best approximations of a real vehicle's CAN bus.

Several works have conducted in-depth reviews of existing datasets. Wu et al. \cite{survey-ivn} surveyed intrusion detection for in-vehicle networks and included a discussion of datasets and tools used in previous works.

Karopoulos et al. \cite{taxonomy-survey} developed a meta-taxonomy for IVN-based intrusion detection, in which they included a table of publicly available vehicular intrusion detection---VIDS---datasets. They catalogued VIDS datasets for controller area network as well as Ethernet, Wi-Fi, Bluetooth, and others. Karopoulos et al. enumerated the following features:

\begin{enumerate}
    \item Presence of train/test sub-datasets
    \item Number of features
    \item Total number of rows
    \item Number of attacks
    \item Presence of labels
    \item Fidelity (real or simulated)
    \item Number of nodes
\end{enumerate}

Rajapaksha et al. \cite{acm-survey} reviewed AI-based intrusion detection systems and included a section on benchmark datasets. They assessed several datasets released by the Hacking and Countermeasure Research Lab (HCRL) \cite{hcrl-can-dataset, hcrl-car-hacking-dataset-1, hcrl-car-hacking-dataset-2, hcrl-survival-analysis-dataset, hcrl-can-signal-extraction-and-translation-dataset, hcrl-attack-and-defense-challenge-dataset} as well as a number of others. For each dataset, they sought to identify both advantages and limitations. They concluded that the Real ORNL Automotive Dynamometer CAN intrusion dataset (ROAD) \cite{road-original, road} is the most comprehensive CAN intrusion detection dataset available.

Verma et al. \cite{road} focused specifically on open-access CAN intrusion detection datasets. They exhaustively discuss each dataset, including its (1) quality, (2) benefits, (3) drawbacks, and (4) use case. The authors note that their work is the first comprehensive guide to open-access CAN intrusion detection datasets. In addition, they introduced the Real ORNL Automotive Dynamometer CAN intrusion dataset (ROAD) \cite{road-original, road}, which offers a wide variety of attack captures as well as a high volume of attack-free data. Attack-free data was collected both on a dynamometer---a ``rolling road" in which the vehicle remains stationary, but the vehicle operates as though driving normally \cite{dynamometer}---and on the open road. All attacks were conducted while the vehicle was on a dynamometer and actively being driven.

\section{Methodology} \label{sec:methodology}

In this section, we discuss our methodology, including our setup, our attack-free data collection, our attack data collection, and our data pre-processing.

We seek to develop a dataset that is suitable for machine learning---both traditional machine learning and deep learning. As such, during data collection, we emphasize (1) quality and (2) quantity. A machine learning IDS depends on a high volume of data for adequate training and testing. The machine learning model will fit to the training data and will be evaluated on the testing data, so if the data is low quality---i.e., low fidelity---then the model will be ill-fitted to the real problem, and the model's performance during the evaluation will not match its performance when faced with the real problem.

During the data pre-processing phase, we label our dataset so as to facilitate both supervised and unsupervised learning. We provide \texttt{.log} files that can be replayed using \texttt{SocketCAN} \cite{socketcan} and \texttt{can-utils} \cite{can-utils}---no pre-processing needed---as well as \texttt{.csv} files that can be loaded into a program---e.g., the \texttt{pandas} \cite{pandas} Python library.

\subsection{Setup}

    We collected live, on-the-road data from four different vehicles and six different drivers. Our objective was to capture both vehicle and driver idiosyncrasies in order to construct an authoritative, diversified CAN dataset for machine learning.

    \textbf{Vehicles.} For our test vehicles, we varied the manufacturer (Chevrolet, Subaru), the model (Impala, Traverse, Silverado, Forester), and the vehicle type (sedan, sport utility vehicle, pickup truck). Our objective was to collect diverse data, so that a proposed intrusion detection system could be evaluated in terms of its ability to generalize to different vehicles. Our vehicles are enumerated below:
    
    \begin{enumerate}
        \item \textit{2011 Chevrolet Impala:} A four-door sedan
        \item \textit{2011 Chevrolet Traverse:} A full-size sport utility vehicle (SUV)
        \item \textit{2016 Chevrolet Silverado:} A four-door pickup truck
        \item \textit{2017 Subaru Forester:} A compact SUV
    \end{enumerate}
    
    \textbf{Drivers.} For our test drivers, we varied both age and gender, such that our CAN traffic captures would reflect the different driving habits of different people. Our test drivers---in terms of demographics---are enumerated below:
    
    \begin{enumerate}
        \item Male, under 30 years of age
        \item Female, under 30 years of age
        \item Male, 30-60 years of age
        \item Female, 30-60 years of age
        \item Male, over 60 years of age
        \item Female, over 60 years of age
    \end{enumerate}

     Our test drivers are acknowledged in Section \ref{sec:acknowledgments}.
    
\begin{figure}[hbt!]
\centering
\includegraphics[width=.75\linewidth]{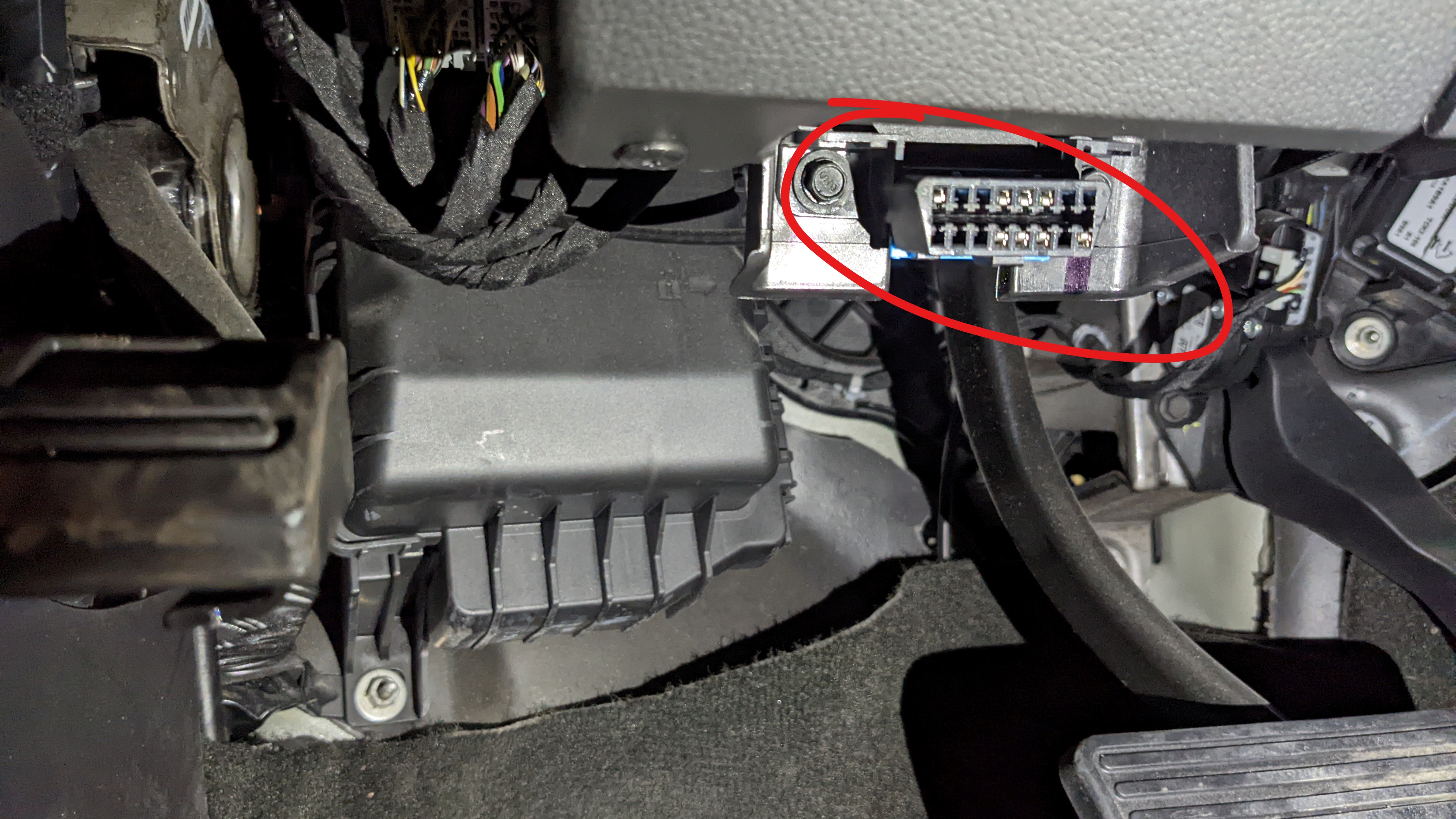}
\caption{The OBD-II port \cite{obdii} of a 2016 Chevrolet Silverado. We accessed this port---via Korlan USB2CAN \cite{8devices}---in order to capture CAN traffic and conduct attacks.}
\label{fig:obd-ii-port}
\end{figure}

    \textbf{Logging.} To log CAN traffic data, we accessed each vehicle's on-board diagnostic port (i.e., OBD-II port) \cite{obdii} (see Figure \ref{fig:obd-ii-port}). U.S. law requires that the OBD-II port be within arm's reach of the driver's seat and accessible without tools \cite{obdii-law}. Similar legislation exists in the European Union. To interface with the OBD-II port, we selected Korlan USB2CAN, an ``8devices" product \cite{8devices}. The cable converts raw OBD-II data into USB data---one end of the cable is plugged directly into the OBD-II port; the opposite end is plugged directly into our laptop. As such, we were able communicate with the OBD-II port via Linux's \texttt{SocketCAN} subsystem \cite{socketcan} and \texttt{can-utils} utilities \cite{can-utils}.

    Our test driver sat in the driver's seat, while our experimenter sat in the front passenger seat, holding the laptop and managing the data collection process.

\subsection{Attack-Free Data}

    We captured attack-free CAN traffic under two conditions:
    
    \begin{enumerate}
        \item Driving mode
        \item Accessory mode
    \end{enumerate}

    While serious (i.e., life-threatening) automotive attacks are generally associated with driving conditions, a number of disconcerting automotive attacks are better suited to stationary settings. For example, an automobile could be tracked while parked. If the automobile connects to home Wi-Fi when parked in the garage, then an attacker could exfiltrate data while the automobile is parked and connected. In addition, Tindell discussed in detail how thieves can access the CAN bus of a parked vehicle in order to steal it \cite{theft} (see Table \ref{can-exploits}).

    Therefore, we collected attack-free traffic in both driving and non-driving settings. With our two types of attack-free data, an IDS can be trained to develop ``normal" baselines for both driving and accessory modes. Since automotive IDSs are typically anomaly-based, high-quality baselines are paramount. Without high-quality baselines, anomaly-based IDSs would be hard-pressed to distinguish between ``normal" traffic and anomalous traffic.

    If an automotive IDS struggles to differentiate attack-free traffic and attack traffic, then it will report a high volume of false positives. False positives are problematic because they dilute the power of a real warning in the event of an actual attack. If drivers are accustomed to false alarms, they may not heed the real warning. If the false alarms are especially frequent and especially irritating, then they may disable the intrusion detection system, leaving the drivers, their passengers, and their vehicles unprotected.

    Since driving mode and accessory mode involve markedly different CAN traffic patterns, we found it prudent to include both types of traffic captures. In accessory mode, many ECUs are either asleep---not running---or silent---not communicating---because they are extraneous when the engine is off and the vehicle is stationary. There are fewer arbitration identifiers and far fewer CAN frames per unit of time. If an IDS were trained exclusively in driving mode and implemented in a vehicle that is occasionally operated in accessory mode, then the loss of many arbitration IDs (and many CAN frames) would probably result in a barrage of false positives. As such, we find it prudent to include both driving and non-driving traffic captures, so that an IDS can be trained for both scenarios.
    
\subsection{Attack Data}

    We conducted and recorded the following types of attacks:
    
    \begin{enumerate}
        \item Denial of Service (DoS)
        \item Combined spoofing
            \begin{enumerate}
                \item Double spoofing
                \item Triple spoofing
            \end{enumerate}
        \item Fuzzing
        \item Gear spoofing
        \item Interval
        \item RPM spoofing
            \begin{enumerate}
                \item Driving mode
                \item Accessory mode
            \end{enumerate}
        \item Speed spoofing
            \begin{enumerate}
                \item Driving mode
                \item Accessory mode
            \end{enumerate}
        \item Standstill
        \item Systematic
    \end{enumerate}

    For two of our attacks---RPM spoofing and speed spoofing---we included both a driving mode and an accessory mode variant. As mentioned earlier, there are stark differences between driving and non-driving CAN traffic captures, and it is important that an IDS be properly trained for both scenarios.

    The denial of service (DoS), fuzzing, and spoofing attack families are common in the literature. However, few existent CAN datasets provide multiple spoofing attacks. Moreover, to our knowledge, none of the existing open-access CAN datasets contain combined spoofing attacks---i.e., spoofing attacks that involve spoofing multiple signals simultaneously. In a double-spoofing attack, for example, an adversary might spoof both the vehicle's gear and RPMs. In some automobiles, the ECU(s) associated with gear shifting will ignore shifting commands when the vehicle's RPMs are non-zero. To conduct a gear shifting attack, an adversary might need to spoof both the vehicle's gear and RPMs simultaneously.

    Many CAN datasets do not---to our knowledge---account for the defenses that automotive manufacturers have begun to incorporate into CAN buses and ECUs. In our spoofing attacks, we encountered such defenses and crafted attacks capable of eluding them. For example, in newer vehicles, our ``malicious" node would be kicked off the bus if we sent too many spoofed messages at too high a frequency. For our attack to succeed, we needed to send enough CAN frames to drown out the legitimate messages coming from the legitimate ECU, but, at the same time, we needed to remain under the threshold that would eject our ``malicious" node from the CAN bus. We experimented with various CAN packet frequencies to optimize our attacks. With the RPM spoofing attacks, for instance, we selected a frequency that would point the tachometer needle to the spoofed value, minimizing fluctuation toward the true value without precipitating our ejection from the CAN bus. The optimal frequency differed by vehicle---the 2011 Chevrolet Impala tolerated extremely high frequencies without kicking us off the CAN bus; the 2016 Chevrolet Silverado would exclude our ``malicious" node from CAN communications within a couple of seconds unless we were very conservative with our frequencies.

    For our RPM spoofing and speed spoofing attacks, we spoofed low, high, and zero values---e.g., 10 miles per hour (mph), 60 mph, and 0 mph for speed spoofing. In addition, we conducted both overt and subtle spoofing attacks. An overt attack, for example, might spoof 60 mph while the vehicle is actually traveling at a sedate 10 mph. In our more subtle attacks, we would spoof something along the lines of 61 or 62 mph while the vehicle was actually traveling at 60 mph. For our gear spoofing attacks, we spoofed ``neutral" (``N") to avoid damaging the vehicle---in particular, the transmission. We conducted successful ``neutral" gear spoofing attacks both when the vehicle was actually in ``drive" (``D") and when the vehicle was actually in ``reverse" (``R"). We believe similar attacks are possible with different combinations of gears; however, if we were to successfully spoof ``reverse" while the vehicle was in ``drive," we would almost certainly damage the transmission (i.e., strip the gears). Such an attack would also be dangerous, especially at speed.

    In addition to the aforementioned attacks, we conducted attacks which we dubbed, ``interval," ``standstill," and ``systematic." Our interval attack leverages a Python script that interfaces with \texttt{SocketCAN} \cite{socketcan} via the \texttt{python-can} package \cite{python-can-package, python-can-docs}. The script watches the CAN bus for a pre-determined target arbitration ID. When the script recognizes the target arbitration ID, it sends one or more spoofed CAN frames with the same arbitration ID. We varied the number of spoofed CAN frames transmitted by the Python script---for some attacks, we sent just one spoofed CAN frame; for others, we sent several spoofed CAN frames. If timed correctly, this type of attack can spoof a vehicle's gear, RPMs, speed, etc. with a minimal number of injected CAN frames. For example, if we send just one spoofed speed message upon receipt of a legitimate speed message, we can keep the speedometer needle hovering near the spoofed value. If, instead, we send spoofed speed messages according to a preset frequency, then we need to send many, many more messages to achieve the same effect.

    Our standstill attack is a spoofing attack, albeit a peculiar one. The effects of our standstill attack are dependent upon the vehicle as well as the current driving conditions. If the attack is conducted in the 2011 Chevrolet Traverse or the 2016 Chevrolet Silverado under the correct conditions, then it will put the vehicle in ``standstill" mode. Essentially, the vehicle behaves as though in neutral: pressing on the accelerator will rev the engine, but the vehicle will not accelerate. The arbitration ID associated with ``standstill" mode is also the arbitration ID associated with RPMs. To trigger ``standstill" mode, an attacker must also spoof zero (or near-zero) RPMs. If there is significant confliction, the attack will fail. This spoofing attack is noteworthy because, normally, in the 2016 Chevrolet Silverado, our ``malicious" node will be ejected from the CAN bus within a few seconds if it sends spoofed CAN frames at even a moderate frequency. However, when spoofing ``standstill" mode, we were able to inject messages at an extremely high frequency without getting kicked off the CAN bus.

    Our systematic attack is similar to our fuzzing attack; however, we do not generate the arbitration identifiers at random. Instead, we generate them \textit{systematically.} As such, we can iterate over all possible 11-bit arbitration IDs. Logically, an attacker might conduct such an attack to determine which arbitration IDs are valid. Similarly, an attacker might use a systematic approach to assess the impact of each arbitration ID. From an IDS's perspective, a systematic attack will look markedly different from a fuzzing attack. Therefore, we feel it is important to include such an attack in order to adequately train and test a machine learning IDS.

\subsection{Pre-processing}

    We provide our raw dataset, \texttt{can-dataset}; a pre-processed and labeled dataset, \texttt{can-ml}; and a curated training and testing dataset, \texttt{can-train-and-test}. Developing \texttt{can-ml} and \texttt{can-train-and-test} necessitated a number of pre-processing steps (\texttt{can-train-and-test} was curated from \texttt{can-ml}). In addition, for all our datasets, we conducted several validation checks to ensure the quality of our data. In this section, we outline six of our pre-processing steps, namely:

    \begin{enumerate}
        \item \texttt{.log} file validation
        \item \texttt{.log} file standardization
        \item \texttt{.log} file conversion (to \texttt{.csv})
        \item \texttt{.csv} file standardization
        \item \texttt{.csv} file labeling
        \item Sub-dataset curation
    \end{enumerate}

    We leveraged the Python programming language to effectuate the bulk of our pre-processing efforts. In particular, we utilized the \texttt{python-can} package \cite{python-can-package, python-can-docs}, which facilitates controller area networking in Python, as well as the \texttt{pandas} package \cite{pandas}, which facilitates data analysis and manipulation.

    \textbf{Validating the \texttt{.log} files.} We should be able to replay a valid \texttt{.log} file using \texttt{canplayer} (one of the Linux \texttt{can-utils} utilities \cite{can-utils}). Playback should be error-free. To validate our \texttt{.log} files, we replayed each one and confirmed that it ran without error.

    \textbf{Standardizing the \texttt{.log} files.} We set up multiple CAN interfaces---both physical and virtual---for our CAN traffic captures. A physical CAN interface, for example, might have been named \texttt{can0}, \texttt{can1}, etc.; similarly, a virtual CAN interface might have been named \texttt{vcan0}, \texttt{vcan1}, etc. During data collection and attack generation, we used multiple interfaces in order to conduct multiple experiments simultaneously. For ease of use, we adopt \texttt{can0} as our standard CAN interface; that is, we replace all CAN interfaces in all \texttt{.log} files with \texttt{can0}. It is much easier to use \texttt{canplayer} to replay CAN \texttt{.log} files when the CAN interface is uniform. In addition, because the CAN interface does not constitute a meaningful feature, it could prove problematic for a machine learning IDS---i.e., the CAN interface could pollute the IDS with noise. By standardizing the CAN interface, we mitigate this issue.

    \textbf{Converting the \texttt{.log} files to \texttt{.csv} files.} Next, we convert the \texttt{.log} files to \texttt{.csv} files. We preserve only meaningful features---i.e., the timestamp, arbitration identifier, and data field (the CAN interface is discarded). During the conversion process, the CAN frame (e.g., \texttt{191\#0670069E067E0000}) is split into two separate features: arbitration ID (e.g., \texttt{191}) and data field (e.g., \texttt{0670069E067E0000}).

    \textbf{Standardizing the \texttt{.csv} files.} A machine learning IDS can fit to arbitrary, ineffectual patterns in a particular dataset's timestamps. For example, if attack-free data was collected on September 12\textsuperscript{th}, and attack data was collected on September 13\textsuperscript{th}, then a machine learning IDS might distinguish attack data from attack-free data by checking the timestamp. This behavior is counterproductive; the IDS is fitting to noise in the data. Therefore, we standardized all of our \texttt{.csv} files to January 1\textsuperscript{st}, 2022. For a given \texttt{.csv} file, the timestamp of the first CAN frame is set to January 1\textsuperscript{st}, and the difference between the original timestamp and the January 1\textsuperscript{st} timestamp is calculated. We use that difference as an offset to update all the remaining timestamps. As such, while the timestamps are changed, the intervals between CAN frames are preserved; thus, we preserve the fidelity of the dataset.

    \textbf{Labeling the \texttt{.csv} files.} We adopted two labeling techniques for our dataset. When labeling a given attack file, we selected the labeling technique best suited to the attack type (i.e., the labeling technique best suited to the characteristics of the injected attack CAN frames). When we simulate an attack, we  begin with an attack-free traffic capture. We replay the attack-free traffic capture, inject attacks, and record the attack-laden traffic. Ultimately, we have an attack-free traffic capture and a corresponding attack-laden traffic capture. By comparing the attack-free capture to the attack-laden capture, we can identify the attack CAN frames. We use this labeling technique for DoS and fuzzing attacks, as the attack CAN frames do not resemble legitimate traffic. For spoofing attacks, however, the attack CAN frames closely resemble---or even match---legitimate traffic. When we replay and recapture CAN traffic, we inadvertently change the timestamps (both the actual dates/times and the intervals), so we cannot distinguish between attack-free and attack CAN frames using the aforementioned comparison technique. Instead, when injecting attack CAN frames, we use a different CAN interface (e.g., \texttt{can1}, \texttt{vcan1}). During the labeling process, we leverage the CAN interface to differentiate attack-free and attack traffic. Once the file is labeled, we standardize the CAN interface to eliminate noise. Ultimately, our labeled \texttt{.csv} files contain the following fields:

    \begin{enumerate}
        \item \texttt{timestamp}
        \item \texttt{arbitration\_id}
        \item \texttt{data\_field}
        \item \texttt{attack} (``0" for attack-free, ``1" for attack)
    \end{enumerate}

    \textbf{Curating the sub-datasets.} We subdivided our dataset into four sub-datasets, each of which contains one ``training" folder and four ``testing" folders. The sub-datasets are intended to be similar in terms of size (e.g., number of samples) and attacks (e.g., attack-free/attack ratio). We planned out our sub-datasets by selecting different vehicle types, attack types, etc. for the various ``training" and ``testing" folders, then we constructed the sub-datasets accordingly. We discuss our sub-datasets in detail in Section \ref{sec:datasets}.

\section{Datasets} \label{sec:datasets}

\begin{table}[hbt!]
\normalsize
\centering

\begin{threeparttable}
\caption{Our Datasets} \label{our-datasets}

\begin{tabular}{|p{4.5cm}|p{2.5cm}|p{2.5cm}|p{2.5cm}|} \hline
    \rowcolor{Tan!60}
    \textbf{Name\tnote{1}}                      & \textbf{\# of Lines\tnote{2}}         & \textbf{Size}
    & \textbf{Format} \\ \hline \hline

    \rowcolor{Tan!20}
    \texttt{can-dataset (can-log)}              & 166,011,012                           & 7.2 GB
    & \texttt{.log} \\ \hline

    \texttt{can-dataset (can-csv)}              & 166,011,310                           & 6.1 GB
    & \texttt{.csv} \\ \hline

    \rowcolor{Tan!20}
    \texttt{can-ml}                             & 315,295,222                           & 13.8 GB
    & \texttt{.log} \\ \hline

    \texttt{can-ml}                             & 630,587,197                           & 23.7 GB
    & \texttt{.csv} \\ \hline

    \rowcolor{Tan!20}
    \texttt{can-train-and-test}\tnote{3}        & 193,241,081                           & 7.5 GB
    & \texttt{.csv} \\ \hline
\end{tabular}
\footnotesize
\begin{tablenotes}
    \item [1] A dataset's ``name" is also the name of its repository.
    \item [2] The number of lines corresponds to the number of samples---or timesteps---in the dataset
    \item [3] \texttt{can-train-and-test} is the name of the repository which contains the final curated dataset (labeled and partitioned). The \texttt{can-train-and-test} repository is linked to the \texttt{can-dataset} and \texttt{can-ml} repositories, which contain raw \texttt{.log} files, additional attack-free traffic, and additional attack traffic. As such, we use the term \texttt{can-train-and-test} to refer to the \texttt{can-train-and-test} repository as well as the linked repositories.
\end{tablenotes}
\end{threeparttable}

\end{table}

\subsection{\texttt{can-dataset}}

    \href{https://bitbucket.org/brooke-lampe/can-dataset/src/master/}{\texttt{can-dataset}} is the name of our raw CAN dataset, available at \url{https://bitbucket.org/brooke-lampe/can-dataset/src/master/}. This dataset contains 596 files (298 \texttt{.log} files and 298 \texttt{.csv} files), which consume 13.3 gigabytes (GB) of space.

    The total number of lines in this dataset is \textbf{\textit{332,022,322.}} Note that this dataset provides raw CAN data in two different formats---\texttt{.log} and \texttt{.csv}. As such, the number of unique samples is approximately half the total number of lines.

    \subsection{\texttt{can-log}}

        \href{https://bitbucket.org/brooke-lampe/can-log/src/master/}{\texttt{can-log}} is a subset of the \texttt{can-dataset}, available at \url{https://bitbucket.org/brooke-lampe/can-log/src/master/}. It contains only the \texttt{.log} files and is available for researchers who do not want to download the full \texttt{can-dataset}, which includes both the \texttt{.log} and \texttt{.csv} files.
    
        \texttt{can-log} contains 298 \texttt{.log} files, which consume 7.2 GB of space. The total number of lines (i.e., samples) in this dataset is \textbf{\textit{166,011,012.}} Each sample corresponds to a particular CAN frame at a particular time.
    
        There are 91,827,504 attack-free samples. Breaking that number down, we have...
    
        \begin{enumerate}
            \item \colorbox{Gray!30}{Chevrolet Impala} \tab \colorbox{Gray!30}{\textbf{14,693,040} attack-free samples}
            \item \colorbox{Gray!5}{Chevrolet Traverse} \tab \colorbox{Gray!5}{\textbf{53,086,095} attack-free samples}
            \item \colorbox{Gray!30}{Chevrolet Silverado} \tab \colorbox{Gray!30}{\textbf{12,867,782} attack-free samples}
            \item \colorbox{Gray!5}{Subaru Forester} \tab \colorbox{Gray!5}{\textbf{11,180,587} attack-free samples}
        \end{enumerate}

    \subsection{\texttt{can-csv}}

        \href{https://bitbucket.org/brooke-lampe/can-csv/src/master/}{\texttt{can-csv}} is a subset of the \texttt{can-dataset}, available at \url{https://bitbucket.org/brooke-lampe/can-csv/src/master/}. It contains only the \texttt{.csv} files. Similar to \texttt{can-log}, \texttt{can-csv} is available for researchers who do not want to download the full \texttt{can-dataset}, which includes both the \texttt{.log} and \texttt{.csv} files.

        \texttt{can-csv} contains 298 \texttt{.csv} files, which consume 6.1 GB of space. The total number of lines (i.e., samples) in this dataset is \textbf{\textit{166,011,310.}} As with \texttt{can-log}, each sample corresponds to a particular CAN frame at a particular time.

\subsection{\texttt{can-ml}}

    We gave the name \href{https://bitbucket.org/brooke-lampe/can-ml/src/master/}{\texttt{can-ml}} to our pre-processed, labeled dataset, as it is intended to support machine learning (i.e., ML)---specifically for intrusion detection. This dataset is available at \url{https://bitbucket.org/brooke-lampe/can-ml/src/master/}.

    \texttt{can-ml} has been subdivided into four directories, as follows:

    \begin{enumerate}
        \item \textit{Pre}-attack unlabeled
        \item \textit{Pre}-attack labeled
        \item \textit{Post}-attack unlabeled
        \item \textit{Post}-attack labeled
    \end{enumerate}

    Essentially, for each traffic capture, we provide both labeled and unlabeled (raw) variants as well as pre-attack (attack-free) and attack variants. The bulk of the files are \texttt{.csv} files, since the comma-separated values (CSV) format is well suited to data labeling---and data loading. Our \texttt{.csv} files can be easily loaded with, e.g., the \texttt{pandas} \cite{pandas} Python library. In the unlabeled pre-attack and post-attack directories, we also include the corresponding raw \texttt{.log} files.

    Discounting the helper files, our dataset contains 1,248 files---\texttt{.csv} and \texttt{.log}---and occupies 37.5 GB of space. There are 312 files per vehicle (Chevrolet Impala, Chevrolet Traverse, etc.). Breaking it down, we have...

    \begin{enumerate}
        \item \textbf{\texttt{.log} files.} There are 416 \texttt{.log} files, totaling 13.8 GB. 104 \texttt{.log} files are associated with each vehicle. Summing up all the samples in all the \texttt{.log} files gives us a total of \textbf{\textit{315,295,222}} samples.
        \item \textbf{\texttt{.csv} files.} There are 832 \texttt{.csv} files, totaling 23.7 GB. 208 \texttt{.csv} files are associated with each vehicle. Summing up all the samples in all the \texttt{.csv} files gives us a total of \textbf{\textit{630,587,197}} samples.
    \end{enumerate}

\subsection{\texttt{can-train-and-test}}

    \href{https://bitbucket.org/brooke-lampe/can-train-and-test/src/master/}{\texttt{can-train-and-test}} is our curated CAN intrusion detection dataset for machine learning IDSs. This dataset is (1) pre-processed, (2) labeled, and (3) organized. \texttt{can-train-and-test} is available at \url{https://bitbucket.org/brooke-lampe/can-train-and-test/src/master/}.

    \texttt{can-train-and-test} contains 236 \texttt{.csv} files and comprises a total of \textbf{\textit{193,241,081}} lines, which corresponds to 7.5 GB of space. It is subdivided into four train/test sub-datasets, as follows:

    \begin{enumerate}
        \item \textbf{\texttt{set\_01}}
            \begin{itemize}
                \item 52 \texttt{.csv} files
                \item 1.7 GB of space
                \item 10,653,152 training samples
                \item 44,659,609 total samples
            \end{itemize}

        \item \textbf{\texttt{set\_02}}
            \begin{itemize}
                \item 56 \texttt{.csv} files
                \item 2.2 GB of space
                \item 17,340,826 training samples
                \item 55,946,808 total samples
            \end{itemize}

        \item \textbf{\texttt{set\_03}}
            \begin{itemize}
                \item 62 \texttt{.csv} files
                \item 1.7 GB of space
                \item 12,025,781 training samples
                \item 43,799,824 total samples
            \end{itemize}

        \item \textbf{\texttt{set\_03}}
            \begin{itemize}
                \item 66 \texttt{.csv} files
                \item 1.9 GB of space
                \item 9,492,819 training samples
                \item 48,834,840 total samples
            \end{itemize}
    \end{enumerate}

    Within each train/test sub-dataset---\texttt{set\_01}, \texttt{set\_02}, \texttt{set\_03}, and \texttt{set\_04}---we provide one training subset and four testing subsets, as follows:

    \begin{enumerate}
        \item \textbf{train\_01:} Train the model
        \item \textbf{test\_01\_known\_vehicle\_known\_attack:} Test the model against a known vehicle (seen in training) and known attacks (seen in training)
        \item \textbf{test\_02\_unknown\_vehicle\_known\_attack:} Test the model against an unknown vehicle (not seen in training) and known attacks (seen in training)
        \item \textbf{test\_03\_known\_vehicle\_unknown\_attack:} Test the model against a known vehicle (seen in training) and unknown attacks (not seen in training)
        \item \textbf{test\_04\_unknown\_vehicle\_unknown\_attack:} Test the model against an unknown vehicle (not seen in training) and unknown attacks (not seen in training)
    \end{enumerate}

\section{Usage} \label{sec:usage}

We provide a number of usage examples for the raw CAN traffic logs---i.e., the \texttt{.log} files. Our usage examples require Linux's \texttt{SocketCAN} subsystem \cite{socketcan} and \texttt{can-utils} utilities \cite{can-utils}. We include the commands to (1) set up a virtual CAN interface and (2) replay the raw CAN traffic logs (using \texttt{canplayer}).

\lstset{
    breaklines=true,
    basicstyle=\small\ttfamily,
    frame=single,
    morecomment=[f][\color{green}][0]{*},
    morecomment=[f][\color{red}][0]{\#},
}

\begin{lstlisting}
# Set up virtual CAN interface
sudo modprobe vcan
sudo ip link add dev vcan0 type vcan
sudo ip link set vcan0 up

# Show the details of the vcan0 link
ip -details -statistics link show vcan0

# Replay a candump log file
canplayer -I your-log-file.log vcan0=can0

# To replay an attack-free log from the can-dataset
canplayer -I can-dataset/2017-subaru-forester/attack-free/attack-free-1.log vcan0=can0

# To replay a DoS log from the can-dataset
canplayer -I can-dataset/2011-chevrolet-impala/DoS-attacks/DoS-1.log vcan0=can0

# Examine packets with the specified arbitration identifier
candump vcan0 | grep " 3E9 "

# Use candump to create a log file that can be replayed using canplayer
candump -l any

# Use candump to create a log file that can be replayed using canplayer,
# specifying the log file name
candump -L any > your-log-file.log

# Use candump to create a log file that can be replayed using canplayer,
# specifying the virtual CAN interface
candump -L vcan0 > DoS-1.log

# Send spoofed packets at set intervals
while true; do cansend vcan0 3E9#1B4C05111B511C69; sleep 0.01; done

# Monitor the bus; send spoofed packets when real packets are detected
candump can0 | grep " 3E9 " | while read line; do cansend can0 3E9#1B4C05111B511C69; done
\end{lstlisting}

\section{Benchmark} \label{sec:benchmark}

\subsection{Setup}

We implemented our benchmark in Python, leveraging the \texttt{pandas} \cite{pandas} Python library to load our dataset and the \texttt{scikit-learn} \cite{scikit-learn} Python library to implement our machine learning intrusion detection systems. We selected several machine learning algorithms---both supervised and unsupervised---as well as two deep learning algorithms from \texttt{scikit-learn}'s offerings.

For supervised traditional machine learning, we experimented with five families of models---Gaussian naive Bayes (1 model), k-nearest neighbor (1 model), regression (2 models), support vector machine (3 models), and tree (5 models). In total, there are twelve supervised machine learning models. For unsupervised machine learning, we explored two families of models---clustering (3 models) and local outlier factor (1 model). We evaluated one supervised machine learning model---a multi-layer perceptron---and one unsupervised deep learning model---a restricted Boltzmann machine.

\begin{itemize}
    \item \textbf{SUPERVISED TRADITIONAL MACHINE LEARNING}
        \begin{itemize}
            \item Gaussian Naive Bayes (NB)
            \item K-Nearest Neighbor (KNN)
            \item Regression
                \begin{itemize}
                    \item \textit{Linear Regression}
                    \item \textit{Logistic Regression}
                \end{itemize}
            \item Support Vector Machine (SVM)
                \begin{itemize}
                    \item \textit{SVM}
                    \item \textit{Linear SVM}
                    \item \textit{One-Class SVM}
                \end{itemize}
            \item Tree
                \begin{itemize}
                    \item \textit{Decision Tree (DT)}
                    \item \textit{Extra Trees (ET)}
                    \item \textit{Gradient Boosting}
                    \item \textit{Isolation Forest (IF)}
                    \item \textit{Random Forest (RF)}
                \end{itemize}
        \end{itemize}
    \item \textbf{UNSUPERVISED TRADITIONAL MACHINE LEARNING}
        \begin{itemize}
            \item Clustering
                \begin{itemize}
                    \item \textit{K-Means Clustering}
                    \item \textit{Mini-Batch K-Means Clustering}
                    \item \textit{Balanced Iterative Reducing and Clustering using Hierarchies (BIRCH)}
                \end{itemize}
            \item Local Outlier Factor (LOF)
        \end{itemize}
    \item \textbf{SUPERVISED DEEP LEARNING}
        \begin{itemize}
            \item  Multi-Layer Perceptron (MLP)
        \end{itemize}
    \item \textbf{UNSUPERVISED DEEP LEARNING}
        \begin{itemize}
            \item Restricted Boltzmann Machine (RBM)
        \end{itemize}
\end{itemize}

We leveraged a compute cluster to conduct our experiments. Each experiment was allocated one CPU and 16 GB of RAM. The time limit was set to 6 days (518400000000000 ns).


\begin{table}[hbt!]
\footnotesize
\centering

\begin{threeparttable}
\caption{Sub-dataset \#1, Testing subset \#1} \label{eval-1-1}

\begin{tabular}{|p{3cm}|p{1.25cm}|p{1.25cm}|p{1.25cm}|p{1.25cm}|p{2.5cm}|p{2.5cm}|} \hline
    \rowcolor{Tan!80}
    \textbf{Model} &
    \textbf{Accuracy} & \textbf{Precision} & \textbf{Recall (TPR)} & \textbf{F1-score} &
    \textbf{Training Time (ns)} & \textbf{Testing Time (ns)} \\ \hline \hline

    \rowcolor{Tan!40}
    Gaussian Naive Bayes &
    0.8820 &        0.9814 &        0.8820 &        0.9273 &
    2071507728 &                    617191930 \\ \hline

    \rowcolor{Tan!20}
    K-Nearest Neighbor &
    0.9855 &        0.3376 &        0.3141 &        0.3254 &
    106205148337753 &               2698926122173 \\ \hline

    \rowcolor{Tan!40}
    Linear Regression &
    0.3540 &        0.9680 &        0.3540 &        0.5042 &
    3544294407 &                    141826800 \\ \hline

    \rowcolor{Tan!20}
    Logistic Regression &
    0.9890 &        0.9871 &        0.9890 &        0.9878 &
    112252658417 &                  380374780 \\ \hline

    \rowcolor{Tan!40}
    Linear Support Vector Machine &
    0.9841 &        0.3018 &        0.3275 &        0.3141 &
    35314797566 &                   344383409 \\ \hline

    \rowcolor{Tan!20}
    One-Class Support Vector Machine\tnote{1} &
    - &             - &             - &             - &
    - &                             - \\ \hline

    \rowcolor{Tan!40}
    Support Vector Machine &
    0.9720 &        0.1505 &        0.3275 &        0.2062 &
    116557173711567 &               20169585048115 \\ \hline

    \rowcolor{Tan!20}
    Decision Tree &
    0.9742 &        0.0724 &        0.1123 &        0.0880 &
    8271963209 &                    401618047 \\ \hline

    \rowcolor{Tan!40}
    Extra Trees &
    0.9915 &        0.8199 &        0.3046 &        0.4442 &
    209531429363 &                  20667658310 \\ \hline

    \rowcolor{Tan!20}
    Gradient Boosting &
    0.6254 &        0.9892 &        0.6254 &        0.7585 &
    2451403662347 &                 8489738261 \\ \hline

    \rowcolor{Tan!40}
    Isolation Forest &
    0.9846 &        0.9779 &        0.9846 &        0.9812 &
    197244480367 &                  88099567128 \\ \hline

    \rowcolor{Tan!20}
    Random Forest &
    0.9879 &        0.0088 &        0.0008 &        0.0015 &
    763511969876 &                  38249637250 \\ \hline

    \rowcolor{Tan!40}
    K-Means Clustering &
    0.8564 &        0.9763 &        0.8564 &        0.9124 &
    97416359356 &                   638434302 \\ \hline

    \rowcolor{Tan!20}
    Mini-Batch K-Means Clustering &
    0.4681 &        0.9669 &        0.4681 &        0.6304 &
    7678285764 &                    3796294688 \\ \hline

    \rowcolor{Tan!40}
    BIRCH &
    0.9889 &        0.9779 &        0.9889 &        0.9834 &
    214197454658 &                  605665184 \\ \hline

    \rowcolor{Tan!20}
    Local Outlier Factor &
    0.9254 &        0.9728 &        0.9254 &        0.9465 &
    53788202512612 &                2439824879120 \\ \hline

    \rowcolor{Tan!40}
    Multi-Layer Perceptron &
    0.9788 &        0.9837 &        0.9788 &        0.9811 &
    3088090156514 &                 11708139937 \\ \hline

    \rowcolor{Tan!20}
    Restricted Boltzmann Machine &
    0.0111 &        0.0001 &        0.0111 &        0.0002 &
    916661700320 &                  9736619899 \\ \hline
\end{tabular}
\footnotesize
\begin{tablenotes}
    \item [1] The one-class support vector machine exceeded our 6-day (518400000000000 ns) time limit
\end{tablenotes}
\end{threeparttable}

\end{table}


\begin{table}[hbt!]
\footnotesize
\centering

\begin{threeparttable}
\caption{Sub-dataset \#1, Testing subset \#2} \label{eval-1-2}

\begin{tabular}{|p{3cm}|p{1.25cm}|p{1.25cm}|p{1.25cm}|p{1.25cm}|p{2.5cm}|p{2.5cm}|} \hline
    \rowcolor{Tan!80}
    \textbf{Model} &
    \textbf{Accuracy} & \textbf{Precision} & \textbf{Recall (TPR)} & \textbf{F1-score} &
    \textbf{Training Time (ns)} & \textbf{Testing Time (ns)} \\ \hline \hline

    \rowcolor{Tan!40}
    Gaussian Naive Bayes &
    0.8652 &        0.9479 &        0.8652 &        0.9044 &
    2071507728 &                    674632590 \\ \hline

    \rowcolor{Tan!20}
    K-Nearest Neighbor &
    0.9714 &        0.0009 &        0.0001 &        0.0002 &
    106205148337753 &               6628601627607 \\ \hline

    \rowcolor{Tan!40}
    Linear Regression &
    0.3112 &        0.9668 &        0.3112 &        0.4499 &
    3544294407 &                    127655570 \\ \hline

    \rowcolor{Tan!20}
    Logistic Regression &
    0.9565 &        0.9514 &        0.9565 &        0.9539 &
    112252658417 &                  199207470 \\ \hline

    \rowcolor{Tan!40}
    Linear Support Vector Machine &
    0.9593 &        0.0538 &        0.0361 &        0.0432 &
    35314797566 &                   354491014 \\ \hline

    \rowcolor{Tan!20}
    One-Class Support Vector Machine\tnote{1} &
    - &             - &             - &             - &
    - &                             - \\ \hline

    \rowcolor{Tan!40}
    Support Vector Machine &
    0.9720 &        0.1505 &        0.3275 &        0.2062 &
    116557173711567 &               20169585048115 \\ \hline

    \rowcolor{Tan!20}
    Decision Tree &
    0.9738 &        0.0000 &        0.0000 &        0.0000 &
    8271963209 &                    471393563 \\ \hline

    \rowcolor{Tan!40}
    Extra Trees &
    0.9740 &        0.0000 &        0.0000 &        0.0000 &
    209531429363 &                  24155191698 \\ \hline

    \rowcolor{Tan!20}
    Gradient Boosting &
    0.6626 &        0.9763 &        0.6626 &        0.7739 &
    2451403662347 &                 8834463042 \\ \hline

    \rowcolor{Tan!40}
    Isolation Forest &
    0.9550 &        0.9492 &        0.9550 &        0.9521 &
    197244480367 &                  100337823607 \\ \hline

    \rowcolor{Tan!20}
    Random Forest &
    0.9720 &        0.0000 &        0.0000 &        0.0000 &
    763511969876 &                  42700298556 \\ \hline

    \rowcolor{Tan!40}
    K-Means Clustering &
    0.8654 &        0.9467 &        0.8654 &        0.9042 &
    97416359356 &                   709549141 \\ \hline

    \rowcolor{Tan!20}
    Mini-Batch K-Means Clustering &
    0.4969 &        0.9271 &        0.4969 &        0.6470 &
    7678285764 &                    4126732218 \\ \hline

    \rowcolor{Tan!40}
    BIRCH &
    0.9745 &        0.9497 &        0.9745 &        0.9620 &
    214197454658 &                  662717530 \\ \hline

    \rowcolor{Tan!20}
    Local Outlier Factor &
    0.1714 &        0.9557 &        0.1714 &        0.2640 &
    53788202512612 &                7698684672802 \\ \hline

    \rowcolor{Tan!40}
    Multi-Layer Perceptron &
    0.9666 &        0.9529 &        0.9666 &        0.9594 &
    3088090156514 &                 13276672102 \\ \hline

    \rowcolor{Tan!20}
    Restricted Boltzmann Machine &
    0.0255 &        0.0006 &        0.0255 &        0.0013 &
    916661700320 &                  10970949396 \\ \hline
\end{tabular}
\footnotesize
\begin{tablenotes}
    \item [1] The one-class support vector machine exceeded our 6-day (518400000000000 ns) time limit
\end{tablenotes}
\end{threeparttable}

\end{table}


\begin{table}[hbt!]
\footnotesize
\centering

\begin{threeparttable}
\caption{Sub-dataset \#1, Testing subset \#3} \label{eval-1-3}

\begin{tabular}{|p{3cm}|p{1.25cm}|p{1.25cm}|p{1.25cm}|p{1.25cm}|p{2.5cm}|p{2.5cm}|} \hline
    \rowcolor{Tan!80}
    \textbf{Model} &
    \textbf{Accuracy} & \textbf{Precision} & \textbf{Recall (TPR)} & \textbf{F1-score} &
    \textbf{Training Time (ns)} & \textbf{Testing Time (ns)} \\ \hline \hline

    \rowcolor{Tan!40}
    Gaussian Naive Bayes &
    0.8839 &        0.9957 &        0.8839 &        0.9361 &
    2071507728 &                    1069338929 \\ \hline

    \rowcolor{Tan!20}
    K-Nearest Neighbor &
    0.9976 &        0.4762 &        0.0005 &        0.0010 &
    106205148337753 &               3326301454816 \\ \hline

    \rowcolor{Tan!40}
    Linear Regression &
    0.3467 &        0.9939 &        0.3467 &        0.5136 &
    3544294407 &                    159399960 \\ \hline

    \rowcolor{Tan!20}
    Logistic Regression &
    0.9976 &        0.9954 &        0.9976 &        0.9964 &
    112252658417 &                  696710769 \\ \hline

    \rowcolor{Tan!40}
    Linear Support Vector Machine &
    0.9975 &        0.0131 &        0.0009 &        0.0017 &
    35314797566 &                   673546732 \\ \hline

    \rowcolor{Tan!20}
    One-Class Support Vector Machine\tnote{1} &
    - &             - &             - &             - &
    - &                             - \\ \hline

    \rowcolor{Tan!40}
    Support Vector Machine &
    0.9976 &        0.0260 &        0.0009 &        0.0018 &
    116557173711567 &               30536179241509 \\ \hline

    \rowcolor{Tan!20}
    Decision Tree &
    0.9821 &        0.0022 &        0.0148 &        0.0039 &
    8271963209 &                    602445631 \\ \hline

    \rowcolor{Tan!40}
    Extra Trees &
    0.9976 &        1.0000 &        0.0001 &        0.0003 &
    209531429363 &  31514357019 \\ \hline

    \rowcolor{Tan!20}
    Gradient Boosting &
    0.9977 &        0.9977 &        0.9977 &        0.9966 &
    2451403662347 &                 11825961968 \\ \hline

    \rowcolor{Tan!40}
    Isolation Forest &
    0.9969 &        0.9957 &        0.9969 &        0.9963 &
    197244480367 &                  134518109020 \\ \hline

    \rowcolor{Tan!20}
    Random Forest &
    0.9976 &        1.0000 &        0.0004 &        0.0009 &
    763511969876 &                  56868215083 \\ \hline

    \rowcolor{Tan!40}
    K-Means Clustering &
    0.8658 &        0.9970 &        0.8658 &        0.9258 &
    97416359356 &                   963140938 \\ \hline

    \rowcolor{Tan!20}
    Mini-Batch K-Means Clustering &
    0.4741 &        0.9969 &        0.4741 &        0.6408 &
    7678285764 &                    5800975603 \\ \hline

    \rowcolor{Tan!40}
    BIRCH &
    0.9976 &        0.9953 &        0.9976 &        0.9965 &
    214197454658 &                  1004912418 \\ \hline

    \rowcolor{Tan!20}
    Local Outlier Factor &
    0.9386 &        0.9965 &        0.9386 &        0.9665 &
    53788202512612 &                2504114371275 \\ \hline

    \rowcolor{Tan!40}
    Multi-Layer Perceptron &
    0.9974 &        0.9953 &        0.9974 &        0.9964 &
    3088090156514 &                 17550210412 \\ \hline

    \rowcolor{Tan!20}
    Restricted Boltzmann Machine &
    0.0024 &        0.0000 &        0.0024 &        0.0000 &
    916661700320 &                  14603996262 \\ \hline
\end{tabular}
\footnotesize
\begin{tablenotes}
    \item [1] The one-class support vector machine exceeded our 6-day (518400000000000 ns) time limit
\end{tablenotes}
\end{threeparttable}

\end{table}


\begin{table}[hbt!]
\footnotesize
\centering

\begin{threeparttable}
\caption{Sub-dataset \#1, Testing subset \#4} \label{eval-1-4}

\begin{tabular}{|p{3cm}|p{1.25cm}|p{1.25cm}|p{1.25cm}|p{1.25cm}|p{2.5cm}|p{2.5cm}|} \hline
    \rowcolor{Tan!80}
    \textbf{Model} &
    \textbf{Accuracy} & \textbf{Precision} & \textbf{Recall (TPR)} & \textbf{F1-score} &
    \textbf{Training Time (ns)} & \textbf{Testing Time (ns)} \\ \hline \hline

    \rowcolor{Tan!40}
    Gaussian Naive Bayes &
    0.8906 &        0.9979 &        0.8906 &        0.9411 &
    2071507728 &                    1671649398 \\ \hline

    \rowcolor{Tan!20}
    K-Nearest Neighbor &
    0.9982 &        0.0190 &        0.0148 &        0.0166 &
    106205148337753 &               11972712642811 \\ \hline

    \rowcolor{Tan!40}
    Linear Regression &
    0.3341 &        0.9758 &        0.3341 &        0.4757 &
    3544294407 &                    137564940 \\ \hline

    \rowcolor{Tan!20}
    Logistic Regression &
    0.9976 &        0.9979 &        0.9976 &        0.9977 &
    112252658417 &                  984385909 \\ \hline

    \rowcolor{Tan!40}
    Linear Support Vector Machine &
    0.9942 &        0.0002 &        0.0009 &        0.0003 &
    35314797566 &                   661366602 \\ \hline

    \rowcolor{Tan!20}
    One-Class Support Vector Machine\tnote{1} &
    - &             - &             - &             - &
    - &                             - \\ \hline

    \rowcolor{Tan!40}
    Support Vector Machine &
    0.9981 &        0.0007 &        0.0006 &        0.0007 &
    116557173711567 &               46633303631277 \\ \hline

    \rowcolor{Tan!20}
    Decision Tree &
    0.9827 &        0.0018 &        0.0276 &        0.0033 &
    8271963209 &                    997735598 \\ \hline

    \rowcolor{Tan!40}
    Extra Trees &
    0.9988 &        0.0000 &        0.0000 &        0.0000 &
    209531429363 &                  49439825750 \\ \hline

    \rowcolor{Tan!20}
    Gradient Boosting &
    0.9984 &        0.9980 &        0.9984 &        0.9982 &
    2451403662347 &                 18095133741 \\ \hline

    \rowcolor{Tan!40}
    Isolation Forest &
    0.9881 &        0.9980 &        0.9881 &        0.9930 &
    197244480367 &                  202233230770 \\ \hline

    \rowcolor{Tan!20}
    Random Forest &
    0.9988 &        0.0011 &        0.0001 &        0.0003 &
    763511969876 &                  88901090110 \\ \hline

    \rowcolor{Tan!40}
    K-Means Clustering &
    0.8872 &        0.9985 &        0.8872 &        0.9392 &
    97416359356 &                   1463804724 \\ \hline

    \rowcolor{Tan!20}
    Mini-Batch K-Means Clustering &
    0.4748 &        0.9984 &        0.4748 &        0.6429 &
    7678285764 &                    9274790788 \\ \hline

    \rowcolor{Tan!40}
    BIRCH &
    0.9990 &        0.9979 &        0.9990 &        0.9984 &
    214197454658 &                  1171693676 \\ \hline

    \rowcolor{Tan!20}
    Local Outlier Factor &
    0.1661 &        0.8882 &        0.1661 &        0.2726 &
    53788202512612 &                7995192332965 \\ \hline

    \rowcolor{Tan!40}
    Multi-Layer Perceptron &
    0.9973 &        0.9979 &        0.9973 &        0.9976 &
    3088090156514 &                 27715472855 \\ \hline

    \rowcolor{Tan!20}
    Restricted Boltzmann Machine &
    0.0010 &        0.0000 &        0.0010 &        0.0000 &
    916661700320 &                  22041127093 \\ \hline
\end{tabular}
\footnotesize
\begin{tablenotes}
    \item [1] The one-class support vector machine exceeded our 6-day (518400000000000 ns) time limit
\end{tablenotes}
\end{threeparttable}

\end{table}


\begin{table}[hbt!]
\footnotesize
\centering

\begin{threeparttable}
\caption{Sub-dataset \#4, Testing subset \#1} \label{eval-4-1}

\begin{tabular}{|p{3cm}|p{1.25cm}|p{1.25cm}|p{1.25cm}|p{1.25cm}|p{2.5cm}|p{2.5cm}|} \hline
    \rowcolor{Tan!80}
    \textbf{Model} &
    \textbf{Accuracy} & \textbf{Precision} & \textbf{Recall (TPR)} & \textbf{F1-score} &
    \textbf{Training Time (ns)} & \textbf{Testing Time (ns)} \\ \hline \hline

    \rowcolor{Tan!40}
    Gaussian Naive Bayes &
    0.9755 &        0.9596 &        0.9755 &        0.9668 &
    1711150808 &                    716855819 \\ \hline

    \rowcolor{Tan!20}
    K-Nearest Neighbor &
    0.9825 &        0.9638 &        0.1975 &        0.3279 &
    68650714770712 &                1751927202489 \\ \hline

    \rowcolor{Tan!40}
    Linear Regression &
    0.3243 &        0.9985 &        0.3243 &        0.4886 &
    4962548165 &                    255239910 \\ \hline

    \rowcolor{Tan!20}
    Logistic Regression &
    0.9784 &        0.9788 &        0.9784 &        0.9677 &
    35900328378 &                   199674490 \\ \hline

    \rowcolor{Tan!40}
    Linear Support Vector Machine &
    0.9784 &        0.0000 &        0.0000 &        0.0000 &
    21097322961 &                   415427420 \\ \hline

    \rowcolor{Tan!20}
    One-Class Support Vector Machine\tnote{1} &
    - &             - &             - &             - &
    - &                             - \\ \hline

    \rowcolor{Tan!40}
    Support Vector Machine &
    0.9810 &        0.9913 &        0.1254 &        0.2226 &
    89048060616553 &                7497330386451 \\ \hline

    \rowcolor{Tan!20}
    Decision Tree &
    0.9880 &        0.9705 &        0.4586 &        0.6229 &
    23308577039 &                   687957287 \\ \hline

    \rowcolor{Tan!40}
    Extra Trees &
    0.9837 &        0.9972 &        0.2467 &        0.3956 &
    486225104765 &                  36751033345 \\ \hline

    \rowcolor{Tan!20}
    Gradient Boosting &
    0.9800 &        0.9771 &        0.9800 &        0.9722 &
    2246450290838 &                 7599878712 \\ \hline

    \rowcolor{Tan!40}
    Isolation Forest &
    0.9774 &        0.9578 &        0.9774 &        0.9672 &
    171923906654 &                  103616521318 \\ \hline

    \rowcolor{Tan!20}
    Random Forest &
    0.9848 &        0.9977 &        0.2961 &        0.4566 &
    1658659682691 &                 59969033775 \\ \hline

    \rowcolor{Tan!40}
    K-Means Clustering &
    0.7698 &        0.9530 &        0.7698 &        0.8511 &
    49276596864 &                   757799028 \\ \hline

    \rowcolor{Tan!20}
    Mini-Batch K-Means Clustering &
    0.7597 &        0.9526 &        0.7597 &        0.8448 &
    7006960680 &                    4772947871 \\ \hline

    \rowcolor{Tan!40}
    BIRCH &
    0.9784 &        0.9572 &        0.9784 &        0.9677 &
    161254066683 &                  678632660 \\ \hline

    \rowcolor{Tan!20}
    Local Outlier Factor &
    0.9552 &        0.9981 &        0.9552 &        0.9760 &
    100295280221578 &               4241296731211 \\ \hline

    \rowcolor{Tan!40}
    Multi-Layer Perceptron &
    0.9799 &        0.9787 &        0.9799 &        0.9715 &
    2655845001046 &                 13873431844 \\ \hline

    \rowcolor{Tan!20}
    Restricted Boltzmann Machine &
    0.1171 &        0.9575 &        0.1171 &        0.1781 &
    761974625735 &                  12918669375 \\ \hline
\end{tabular}
\footnotesize
\begin{tablenotes}
    \item [1] The one-class support vector machine exceeded our 6-day (518400000000000 ns) time limit
\end{tablenotes}
\end{threeparttable}

\end{table}

\subsection{Discussion of Results}

In this section, we discuss the results of our evaluation. We selected sub-dataset \#1 to examine in detail; additional results have been made available in the Appendix (Section \ref{sec:appendix}) as well as the \href{https://bitbucket.org/brooke-lampe/can-benchmark/src/master/}{\texttt{can-benchmark}} repository.

\textbf{Testing subsets.} Tables \ref{eval-1-1}, \ref{eval-1-2}, \ref{eval-1-3}, \ref{eval-1-4} showcase our experimental results for sub-dataset \#1, testing subsets \#1, \#2, \#3, and \#4, respectively. For each model, the experiment with the optimal parameter set---in terms of maximum F1-score---is shown.

Recall that \texttt{test\_01} is designed to evaluate a given model against a \textit{known} vehicle and \textit{known} attacks. Next, \texttt{test\_02} evaluates the model against an \textit{unknown} vehicle and \textit{known} attacks. Inversely, \texttt{test\_03} evaluates the model against a \textit{known} vehicle and \textit{unknown} attacks. Lastly, \texttt{test\_04} evaluates the model against an \textit{unknown} vehicle and \textit{unknown} attacks.

Looking at testing subset \#1 (Table \ref{eval-1-1}), we can see that the logistic regression model achieved the highest F1-score: 0.9878. The logistic regression model performed very well across the board; its accuracy, precision, and recall metrics were 0.9890, 0.9871, and 0.9890, respectively. Training time was approximately 112.3 seconds; testing time was approximately 0.38 seconds.

For testing subset \#2 (Table \ref{eval-1-2}), the BIRCH model earned the highest F1-score: 0.9620---somewhat lower than the highest F1-score for the previous testing subset. The BIRCH model also attained an accuracy of 0.9745, a precision of 0.9497, and a recall of 0.9745. The BIRCH model spent approximately 214.2 seconds on training and 0.66 seconds on testing.

Next, we examine testing subset \#3 (Table \ref{eval-1-3}). We can see that the gradient boosting model attained the highest F1-score: 0.9966. The gradient boosting model attained an accuracy of 0.9977, a precision of 0.9977, and a recall of 0.9977. Training time was approximately 2451.4 seconds ($\approx$41 minutes); testing time was approximately 11.8 seconds.

Lastly, looking at testing subset \#4 (Table \ref{eval-1-4}), we can see that, once again, the BIRCH model earned the highest F1-score: 0.9984. For accuracy, precision, and recall, the BIRCH model attained values of 0.9990, 0.9979, and 0.9990, respectively. As before, the training time was approximately 214.2 seconds, though the testing time was marginally longer---approximately 1.2 seconds.

If we zoom out to the full dataset---which includes all experiments with all parameters---we can see a trend in terms of the highest-performing models. Generally, the following models achieve the top F1-scores:

\begin{enumerate}
    \item Multi-Layer Perceptron
    \item Gradient Boosting
    \item Isolation Forest
    \item BIRCH
    \item Logistic Regression
\end{enumerate}

On the flip side, the following models earned F1-scores of \textit{zero} for one or more experiments:

\begin{itemize}
    \item Restricted Boltzmann Machine
    \item Decision Tree
    \item Extra Trees
    \item K-Nearest Neighbor
    \item Linear Support Vector Machine
    \item Support Vector Machine
    \item Random Forest
\end{itemize}

We do not have an F1-score for the one-class support vector machine---as it ran out of time---but, given the performance of the linear support vector machine and the traditional support vector machine, we do not believe that it would have performed well.

When evaluating the performance of a machine learning model; training time is also a very important metric. From a training time perspective, the Gaussian naive Bayes model is the clear victor. Looking at sub-dataset \#4, testing subset \#2, not only did the Gaussian naive Bayes model achieve the shortest training time---1711150808 nanoseconds ($\approx$1.7 seconds)---but it also earned an impressive 0.9955 F1-score. Moreover, the model achieved an accuracy of 0.9961, a precision of 0.9951, and a recall of 0.9961. The worst performer, time-wise, would be the one-class support vector machine, as it failed to terminate within the 6-day time limit.

As mentioned above, the testing subsets were developed to evaluate different facets of a machine learning model. Some subsets evaluate the given model against \textit{knowns} (i.e., known vehicles, known attacks), while others evaluate the given model against \textit{unknowns} (i.e., unknown vehicles, unknown attacks). The average F1-scores for our testing subsets are given below:

\begin{enumerate}
    \item \colorbox{Gray!30}{Testing subset \#1} \tab \colorbox{Gray!30}{\textbf{F1-score:} 0.49833623}
    \item \colorbox{Gray!5}{Testing subset \#2} \tab \colorbox{Gray!5}{\textbf{F1-score:} 0.42135362}
    \item \colorbox{Gray!30}{Testing subset \#3} \tab \colorbox{Gray!30}{\textbf{F1-score:} 0.45809565}
    \item \colorbox{Gray!5}{Testing subset \#4} \tab \colorbox{Gray!5}{\textbf{F1-score:} 0.44716087}
    \item \colorbox{Gray!30}{All testing subsets} \tab \colorbox{Gray!30}{\textbf{F1-score:} 0.45623659}
\end{enumerate}

The average F1-score varied across testing subsets \#1 through \#4. The average F1-score for testing subset \#1 was $\approx$0.4983. For testing subset \#2, the average F1-score dropped to $\approx$0.4214. For testing subsets \#3 and \#4, the average F1-scores were $\approx$0.4581 and $\approx$0.4472, respectively. The average F1-score across all testing subsets was $\approx$0.4562.

Unsurprisingly, the average F1-score was highest for testing subset \#1, which contains only known vehicles and known attacks. We expected testing subset \#4, which contains both unknown vehicles and unknown attacks, to attain the lowest average F1-score. However, testing subset \#2 (unknown vehicles, known attacks) saw the lowest average F1-score. It would seem that unknown vehicles might be more disruptive to machine learning IDSs than unknown attacks. If that is indeed the case, then it is all the more important to train a machine learning IDS against as many different vehicles (type, manufacturer, model) as possible.

\textbf{Sub-datasets.} With our four testing subsets, our objective was to evaluate different aspects of a machine learning model. Thus, the testing subsets were carefully designed to be \textit{different.} With our four sub-datasets, our objective was \textit{similarity}---we sought to construct sub-datasets that were consistent in terms of size, complexity, difficulty, attack types, etc. Sub-datasets provide supplementary training and testing data; for example, if sub-dataset \#1 is too small, it can be augmented with data from sub-dataset \#2. In addition, sub-datasets could be used to ensure that a machine learning model does not suffer from overfitting; if the model performs well on sub-dataset \#1 but performs terribly on sub-datasets \#2, \#3, and \#4, then perhaps the model might have overfit to sub-dataset \#1.

Therefore, when we compare sub-datasets, we hope to see similar metrics for the same model when evaluated against the same testing subset. Let us look at the BIRCH model when evaluated against testing subset \#1 for both sub-dataset \#1 and sub-dataset \#4. For sub-dataset \#1, accuracy, precision, recall, and F1-score were 0.9889, 0.9779, 0.9889, and 0.9834, respectively. Meanwhile, for sub-dataset \#4, accuracy, precision, recall, and F1-score were 0.9784, 0.9572, 0.9784, and 0.9677, respectively. Across the board, the values are somewhat lower for sub-dataset \#4 than for sub-dataset \#1. That said, the BIRCH model performed exceptionally well on both sub-datasets. The training times were 214197454658 ns ($\approx$214.2 seconds) and 161254066683 ns ($\approx$161.3 seconds) for sub-dataset \#1 and sub-dataset \#4, respectively. Overall, the two sub-datasets are sufficiently similar.

On the opposite end of the spectrum, the k-nearest neighbor model performed poorly when pitted against sub-dataset \#1, testing subset \#1, achieving an F1-score of only 0.3254. Similarly, the k-nearest neighbor model underperformed against sub-dataset \#4, testing subset \#1, earning an F1-score of only 0.3279. Time-wise, the model spent 106205148337753 ns ($\approx$29.5 hours) training against sub-dataset \#1 and 68650714770712 ns ($\approx$19.1 hours) training against sub-dataset \#4.

Table \ref{eval-4-1} highlights the performance of our 18 models against sub-dataset \#4, testing subset \#1. For comparison, refer to Table \ref{eval-1-1}, which showcases the performance of our 18 models against sub-dataset \#1, testing subset \#1.

Calculating the average F1-score for each sub-dataset yields the following values:

\begin{enumerate}
    \item \colorbox{Gray!30}{Sub-dataset \#1} \tab \colorbox{Gray!30}{\textbf{F1-score:} 0.45623659}
    \item \colorbox{Gray!5}{Sub-dataset \#2} \tab \colorbox{Gray!5}{\textbf{F1-score:} 0.49727649}
    \item \colorbox{Gray!30}{Sub-dataset \#3} \tab \colorbox{Gray!30}{\textbf{F1-score:} 0.52929375}
    \item \colorbox{Gray!5}{Sub-dataset \#4} \tab \colorbox{Gray!5}{\textbf{F1-score:} 0.53296775}
    \item \colorbox{Gray!30}{All sub-datasets} \tab \colorbox{Gray!30}{\textbf{F1-score:} 0.50389963}
\end{enumerate}

Sub-dataset \#1 appears to be somewhat more difficult than the others---the average F1-score of sub-dataset \#1 is $\approx$0.4562, somewhat lower than the overall average F1-score of $\approx$0.5039. Judging by F1-score, sub-datasets \#3 and \#4 are remarkably similar, while sub-dataset \#2 appears to be marginally more difficult than the overall dataset.

\textbf{Traditional machine learning vs. deep learning, supervised learning vs. unsupervised learning.} Our evaluation pitted (1) traditional machine learning models against deep learning models and (2) supervised learning models against unsupervised learning models. Our results suggest that that the type of model (logistic regression, isolation forest, BIRCH, multi-layer perceptron, etc.) is more important than traditional machine learning vs. deep learning or supervised learning vs. unsupervised learning. Among our supervised traditional machine learning models, several performed very well (e.g., logistic regression, isolation forest); others performed very poorly (e.g., support vector machine, extra trees). With unsupervised traditional machine learning, we saw exceptional performance from the BIRCH model, mediocre performance from mini-batch k-means clustering model, and inconsistent performance from the local outlier factor model. When it comes to deep learning, we have only two models to consider: (1) the supervised multi-layer perceptron and (2) the unsupervised restricted Boltzmann machine. The multi-layer perceptron performed exceptionally well, while the restricted Boltzmann machine performed exceptionally poorly. Therefore, we find both traditional machine learning and deep learning approaches to be well suited to the problem of automotive intrusion detection. Likewise, we find that both supervised and unsupervised models---if properly constructed and optimized---can function as automotive IDSs.

We have included detailed metrics---per model---in the appendix (see Section \ref{sec:appendix}).

In addition, the \href{https://bitbucket.org/brooke-lampe/can-benchmark/src/master/}{\texttt{can-benchmark}} repository contains the artifacts generated over the course of our benchmark evaluation. We also provide a Microsoft Excel spreadsheet, \texttt{benchmark.xlsx}, in which we meticulously organize the results of our experiments. The \texttt{can-benchmark} repository is available here: \url{https://bitbucket.org/brooke-lampe/can-benchmark/src/master/}.

\section{Future Work} \label{sec:future-work}

\begin{diag}
\begin{center}
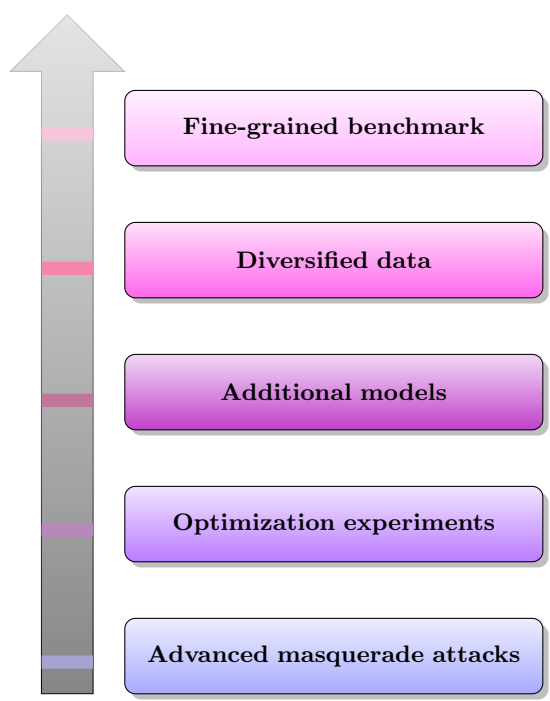

\smartdiagramset{
    set color list={Periwinkle!60,Purple!60,RedViolet!60,RubineRed!60,Lavender!60},
    border color=black,
}
\smartdiagram[priority descriptive diagram]{
    \textbf{Advanced masquerade attacks},
    \textbf{Optimization experiments},
    \textbf{Additional models},
    \textbf{Diversified data},
    \textbf{Fine-grained benchmark}
}
\captionof{figure}{Future work, prioritized}
\label{future-work}
\end{center}
\end{diag}

We have identified a number of opportunities for future work, which are highlighted in Figure \ref{future-work}. More specifically, we enumerate the following:

\begin{enumerate}
    \item Craft advanced masquerade attacks (i.e., no confliction)
    \item Dedicate experiments to optimization (i.e., finding the best parameters)
    \item Construct additional models (e.g., non-learning, traditional machine learning, deep learning)
    \item Collect diversified data (e.g., different vehicles, drivers, environments)
    \item Conduct a fine-grained benchmark (e.g., which attacks are easiest for the decision tree model to detect?)
\end{enumerate}

\textbf{Advanced masquerade attacks.} As mentioned in Section \ref{sec:background}, a sophisticated masquerade attack incorporates both spoofing and suppress attacks. Essentially, CAN frames from the legitimate ECU are \textit{suppressed,} so that there is no confliction when the attacker \textit{spoofs} CAN frames with the legitimate ECU's arbitration identifier. As such, the attacker \textit{masquerades} as the legitimate ECU. To construct advanced masquerade attacks, we plan to explore two options: (1) scrubbing legitimate CAN frames from a traffic capture during post-processing or (2) conducting actual suppress attacks. It would be easier to simply remove legitimate CAN frames from a traffic capture during post-processing in order to simulate an advanced masquerade attack, but we are concerned about loss of fidelity. If we were to conduct an actual suppress attack in a live vehicle, fidelity would be preserved, but the test vehicle might be adversely impacted. Successful suppress attacks on live vehicles generally involve ECU reprogramming \cite{advanced-can-injection-whitepaper, blackhat-advanced-can-injection, keen-tesla, keen-bmw}. If the ECU reprogramming process goes awry, we might not be able to restore the ECU to a functional state. Therefore, we will explore and evaluate our two options. Once we have suppressed the legitimate ECU's CAN frames---either live or during post-processing---we will conduct a spoofing attack. Because the legitimate ECU's CAN frames have been suppressed, there will be no confliction during the spoofing attack.

\textbf{Optimization experiments.} For our 18 machine learning models, we explored different parameter settings and conducted experiments with the goal of optimization. However, the focus of this work is our dataset and benchmark; as such, we did not rigorously optimize our models. In our future work, we plan to dedicate a number of experiments to optimization. We will meticulously record our parameters and results with the objective of finding optimal parameters for each model.

\textbf{Additional models.} In our future work, we plan to explore additional machine learning models---beyond \texttt{scikit-learn}'s offerings \cite{scikit-learn}. We will investigate both traditional machine learning and deep learning models, and we will experiment with both supervised and unsupervised learning paradigms. In particular, we are considering convolutional neural networks (CNNs) and transformers for our supervised deep learning models. For our unsupervised deep learning models, we are looking into autoencoders and generative adversarial networks (GANs) \cite{deep-learning-survey}. In addition, we intend to implement non-learning intrusion detection algorithms---e.g., interval-, frequency-, sequence-, and entropy-based techniques---to enrich our benchmark.

\textbf{Diversified data.} Automotive controller area network data varies widely by vehicle type, by manufacturer, and even by model. Though the 2011 Chevrolet Impala and the 2016 Chevrolet Silverado were both produced by General Motors and share the Chevrolet mark, they differ significantly when it comes to CAN frames. They share some arbitration identifiers but not others. If we look closely at the arbitration IDs they actually share, we can see that the data field is often longer for the Silverado's CAN frames than the Impala's. As such, it is essential to collect diversified data from different vehicle types, manufacturers, and models. A larger, more diversified dataset would provide a greater quantity of data---which is especially important for machine learning IDSs. In addition, we plan to recruit more test drivers. In particular, we will be looking at the experience level of our test drivers---inexperienced vs. experienced. We will also consider different test environments. Inter-city driving, for example, is drastically different from highway driving. Similarly, vehicles handle differently---and, thus, people drive differently---on minimum-maintenance gravel roads compared to paved asphalt roads. Lastly, we will look at additional driving conditions (e.g., weather) when enriching our dataset.

\textbf{Fine-grained benchmark.} Once we have cultivated additional attacks, additional---optimized---models, and additional data, we plan to conduct a fine-grained benchmark. In this work, we have evaluated the performance of our models against our four train/test sub-datasets and our four testing subsets. In a more fine-grained benchmark, we will also consider the performance of a given model against an attack of a given type. We will answer questions such as...

\begin{enumerate}
    \item Which attacks are easiest for the decision tree model to detect?
    \item Which model is best at detecting speed spoofing attacks?
    \item Across all models, which attacks are easiest to detect? Hardest?
\end{enumerate}

\section{Conclusion} \label{sec:conclusion}

The controller area network---CAN---bus has emerged as the de facto standard for in-vehicle networks (IVNs) around the globe. Safety-critical components (e.g., the brakes, the engine, the transmission) depend on the CAN bus for expedient, reliable communication. Unfortunately, while the CAN bus was designed to be resilient under harsh operating conditions, it was not designed to be resilient under \textit{adversarial} conditions. Standard security practices such as authentication, authorization, and encryption are completely lacking when it comes to the CAN bus. Researchers have since developed authentication, authorization, and encryption specifications for the CAN bus, but retroactive implementation of said security controls would be exorbitantly expensive---in terms of hardware, labor, engineering effort, and monetary cost. Therefore, the automotive intrusion detection system (IDS) has emerged in the literature as a low-cost, low-effort solution to the automotive [in]security problem. However, developing and evaluating an automotive IDS can be quite challenging---especially if researchers lack access to a test vehicle. Without a test vehicle, researchers are limited to publicly available CAN data, and existing CAN intrusion detection datasets come with various limitations. This lack of CAN data has become a barrier to entry into automotive intrusion detection research---and even automotive security research in general.

We seek to lower this barrier to entry by introducing a new CAN intrusion detection dataset, which facilitates the development and evaluation of automotive IDSs. Our dataset, \texttt{can-train-and-test}, offers real-world CAN traffic data from four different vehicles---a sedan, a compact SUV, a full-size SUV, and a pickup truck---produced by two different manufacturers. For each vehicle, we provide comparable attack captures, which enable researchers to assess a given IDS's ability to generalize to different vehicle types and models. Our dataset contains \texttt{.log} files for playback as well as labeled and unlabeled \texttt{.csv} files for supervised and unsupervised machine learning. As such, our dataset is well suited to a variety of different automotive intrusion detection and automotive security enterprises. In addition, \texttt{can-train-and-test} supplies nine unique attacks, ranging from denial of service (DoS) fuzzing to triple spoofing attacks. As such, researchers can select from a wide variety of attacks when partitioning the data into training and testing datasets. Alternatively, researchers can leverage our curated \texttt{can-train-and-test} repository, which is subdivided into four train/test sub-datasets and four testing subsets. As a benchmark, we pitted 18 machine learning models against the \texttt{can-train-and-test} repository. During our evaluation and analysis, we found that the multi-layer perceptron, gradient boosting, isolation forest, BIRCH, and logistic regression models consistently scored above 0.95 when it came to accuracy, precision, recall, and F1-score---regardless of the sub-dataset and testing subset. Across all experiments on all sub-datasets, we saw an average F1-score of $\approx$0.5039, indicating that our \texttt{can-train-and-test} dataset is indeed capable of distinguishing capable, well-trained IDSs from their less-than-capable counterparts. We present \texttt{can-train-and-test} as a contribution to the existing collection of open-access CAN intrusion detection datasets in hopes of filling in the gaps left by the existing collection.

\section{Acknowledgments} \label{sec:acknowledgments}
We would like to thank our test drivers, especially Christian Lampe, Julie Lampe, and Wayne \& Vickie Olsen.

\bibliographystyle{apalike}
\bibliography{dataset}

\newpage

\appendix

\section{Detailed Results} \label{sec:appendix}

Here, we provide the detailed results of our 18 traditional machine learning models. We conducted multiple experiments with different sets of parameters; we include only the parameters that resulted in the best performance (in terms of maximum F1-score). The results are tabulated by model and organized by learning style (i.e., traditional machine learning, deep learning, supervised learning, unsupervised learning).

Additionally, the \href{https://bitbucket.org/brooke-lampe/can-benchmark/src/master/}{\texttt{can-benchmark}} repository contains the artifacts generated over the course of our benchmark evaluation. We also provide a Microsoft Excel spreadsheet, \texttt{benchmark.xlsx}, in which we meticulously organize the results of our experiments. The \texttt{can-benchmark} repository is available here: \url{https://bitbucket.org/brooke-lampe/can-benchmark/src/master/}.

\subsection{Traditional Machine Learning, Supervised}

We provide the results of our supervised traditional machine learning models in Tables \ref{gaussian-naive-bayes}, \ref{k-nearest-neighbor}, \ref{linear-regression}, \ref{logistic-regression}, \ref{linear-support-vector-machine}, \ref{one-class-support-vector-machine}, \ref{support-vector-machine}, \ref{decision-tree}, \ref{extra-trees}, \ref{gradient-boosting}, \ref{isolation-forest}, and \ref{random-forest}.


\begin{table}[ht!]
\tiny

\begin{threeparttable}
\caption{Gaussian Naive Bayes\tnote{1}} \label{gaussian-naive-bayes}

\begin{tabular}{|p{0.45cm}|p{0.45cm}|p{0.6cm}|p{0.6cm}|p{0.6cm}|p{0.6cm}|p{0.5cm}|p{0.5cm}|p{0.5cm}|p{0.8cm}|p{0.8cm}|p{0.8cm}|p{0.8cm}|p{1.8cm}|p{1.6cm}|} \hline
    \rowcolor{Tan!80}
    \textbf{Data- set} & \textbf{Sub- set} &
    \textbf{Acc.} & \textbf{Pre.} & \textbf{Rec. (TPR)} & \textbf{F1-score} &
    \textbf{TNR} & \textbf{FPR} & \textbf{FNR} &
    \textbf{TP} & \textbf{TN} & \textbf{FP} & \textbf{FN} &
    \textbf{Training Time (ns)} & \textbf{Testing Time (ns)} \\ \hline \hline

    \rowcolor{Tan!40}
    \#1	&           \#1 &
    0.8820 &        0.9814 &        0.8820 &        0.9273 &
    0.8877 &        0.1123 &        0.1180 &
    23214 &         5006275 &       633115 &        40066 &
    2071507728 &                    617191930 \\ \hline

    \rowcolor{Tan!20}
    \#1	&           \#2	&
    0.8652 &        0.9479 &        0.8652 &        0.9044 &
    0.8868 &        0.1132 &        0.1348 &
    6227 &          5572327 &       711423 &        157940 &
    2071507728 &                    674632590 \\ \hline

    \rowcolor{Tan!40}
    \#1	&           \#3	&
    0.8839 &        0.9957 &        0.8839 &        0.9361 &
    0.8854 &        0.1146 &        0.1161 &
    5428 &          7626878 &       987572 &        14927 &
    2071507728 &                    1069338929 \\ \hline

    \rowcolor{Tan!20}
    \#1	&           \#4 &
    0.8906 &        0.9979 &        0.8906 &        0.9411 &
    0.8914 &        0.1086 &        0.1094 &
    919 &           11772352 &      1433959 &       12880 &
    2071507728 &                    1671649398 \\ \hline

    \rowcolor{Melon!40}
    \#2	&           \#1 &
    0.9901 &        0.9982 &        0.9901 &        0.9940 &
    0.9908 &        0.0092 &        0.0099 &
    3916 &          13085042 &      121269 &        9883 &
    2995791737 &                    1402378958 \\
\hline
\end{tabular}
\tiny
\begin{tablenotes}
    \item [1] \textbf{Parameters:} N/A
\end{tablenotes}
\end{threeparttable}

\end{table}


\begin{table}[ht!]
\tiny

\begin{threeparttable}
\caption{K-Nearest Neighbor\tnote{1}} \label{k-nearest-neighbor}

\begin{tabular}{|p{0.45cm}|p{0.45cm}|p{0.6cm}|p{0.6cm}|p{0.6cm}|p{0.6cm}|p{0.5cm}|p{0.5cm}|p{0.5cm}|p{0.8cm}|p{0.8cm}|p{0.8cm}|p{0.8cm}|p{1.8cm}|p{1.6cm}|} \hline
    \rowcolor{Tan!80}
    \textbf{Data- set} & \textbf{Sub- set} &
    \textbf{Acc.} & \textbf{Pre.} & \textbf{Rec. (TPR)} & \textbf{F1-score} &
    \textbf{TNR} & \textbf{FPR} & \textbf{FNR} &
    \textbf{TP} & \textbf{TN} & \textbf{FP} & \textbf{FN} &
    \textbf{Training Time (ns)} & \textbf{Testing Time (ns)} \\ \hline \hline

    \rowcolor{Tan!40}
    \#1	&           \#1 &
    0.9855 &        0.3376 &        0.3141 &        0.3254 &
    0.9931 &        0.0069 &        0.6859 &
    19877 &         5600381 &       39009 &         43403 &
    106205148337753 &               2698926122173 \\ \hline

    \rowcolor{Tan!20}
    \#1	&           \#2	&
    0.9714 &        0.0009 &        0.0001 &        0.0002 &
    0.9967 &        0.0033 &        0.9999 &
    19 &            6263171 &       20579 &         164148 &
    106205148337753 &               6628601627607 \\ \hline

    \rowcolor{Tan!40}
    \#1	&           \#3	&
    0.9976 &        0.4762 &        0.0005 &        0.0010 &
    1.0000 &        0.0000 &        0.9995 &
    10 &            8614439 &       11 &            20345 &
    106205148337753 &               3326301454816 \\ \hline

    \rowcolor{Tan!20}
    \#1	&           \#4 &   
    0.9982 &        0.0190 &        0.0148 &        0.0166 &
    0.9992 &        0.0008 &        0.9852 &
    204 &           13195779 &      10532 &         13595 &
    106205148337753 &               11972712642811 \\ \hline
 
    \rowcolor{Melon!40}
    \#2	&           \#1 &
    0.9988 &        0.4198 &        0.4774 &        0.4467 &
    0.9993 &        0.0007 &        0.5226 &
    6587 &          13197207 &      9104 &          7212 &
    218093481844902 &               4188777687199 \\
\hline
\end{tabular}
\tiny
\begin{tablenotes}
    \item [1] \textbf{Parameters:} n\_neighbors = 3
\end{tablenotes}
\end{threeparttable}

\end{table}


\begin{table}[ht!]
\tiny

\begin{threeparttable}
\caption{Linear Regression\tnote{1}} \label{linear-regression}

\begin{tabular}{|p{0.45cm}|p{0.45cm}|p{0.6cm}|p{0.6cm}|p{0.6cm}|p{0.6cm}|p{0.5cm}|p{0.5cm}|p{0.5cm}|p{0.8cm}|p{0.8cm}|p{0.8cm}|p{0.8cm}|p{1.8cm}|p{1.6cm}|} \hline
    \rowcolor{Tan!80}
    \textbf{Data- set} & \textbf{Sub- set} &
    \textbf{Acc.} & \textbf{Pre.} & \textbf{Rec. (TPR)} & \textbf{F1-score} &
    \textbf{TNR} & \textbf{FPR} & \textbf{FNR} &
    \textbf{TP} & \textbf{TN} & \textbf{FP} & \textbf{FN} &
    \textbf{Training Time (ns)} & \textbf{Testing Time (ns)} \\ \hline \hline

    \rowcolor{Tan!40}
    \#1	&           \#1 &
    0.3540 &        0.9680 &        0.3540 &        0.5042 &
    0.3467 &        0.6533 &        0.6460 &
    117735 &        2916636 &       5496650 &       41600 &
    3544294407 &                    141826800 \\ \hline

    \rowcolor{Tan!20}
    \#1	&           \#2	&
    0.3112 &        0.9668 &        0.3112 &        0.4499 &
    0.2989 &        0.7011 &        0.6888 &
    130470 &        2002190 &       4695855 &       23448 &
    3544294407 &                    127655570 \\ \hline

    \rowcolor{Tan!40}
    \#1	&           \#3	&
    0.3467 &        0.9939 &        0.3467 &        0.5136 &
    0.3469 &        0.6531 &        0.6533 &
    4432 &          3271141 &       6158658 &       13678 &
    3544294407 &                    159399960 \\ \hline

    \rowcolor{Tan!20}
    \#1	&           \#4 &
    0.3341 &        0.9758 &        0.3341 &        0.4757 &
    0.3204 &        0.6796 &        0.6659 &
    142195 &        2161436 &       4584635 &       7027 &
    3544294407 &                    137564940 \\ \hline

    \rowcolor{Melon!40}
    \#2	&           \#1 &
    0.3009 &        0.9447 &        0.3009 &        0.4440 &
    0.2961 &        0.7039 &        0.6991 &
    77230 &         1997231 &       4748840 &       71992 &
    2848875729 &                    113877510 \\
\hline
\end{tabular}
\tiny
\begin{tablenotes}
    \item [1] \textbf{Parameters:} threshold = 0.0
\end{tablenotes}
\end{threeparttable}

\end{table}


\begin{table}[ht!]
\tiny

\begin{threeparttable}
\caption{Logistic Regression\tnote{1}} \label{logistic-regression}

\begin{tabular}{|p{0.45cm}|p{0.45cm}|p{0.6cm}|p{0.6cm}|p{0.6cm}|p{0.6cm}|p{0.5cm}|p{0.5cm}|p{0.5cm}|p{0.8cm}|p{0.8cm}|p{0.8cm}|p{0.8cm}|p{1.8cm}|p{1.6cm}|} \hline
    \rowcolor{Tan!80}
    \textbf{Data- set} & \textbf{Sub- set} &
    \textbf{Acc.} & \textbf{Pre.} & \textbf{Rec. (TPR)} & \textbf{F1-score} &
    \textbf{TNR} & \textbf{FPR} & \textbf{FNR} &
    \textbf{TP} & \textbf{TN} & \textbf{FP} & \textbf{FN} &
    \textbf{Training Time (ns)} & \textbf{Testing Time (ns)} \\ \hline \hline

    \rowcolor{Tan!40}
    \#1	&           \#1 &
    0.9890 &        0.9871 &        0.9890 &        0.9878 &
    0.9964 &        0.0036 &        0.0110 &
    20723 &         5619213 &       20177 &         42557 &
    112252658417 &                  380374780 \\ \hline

    \rowcolor{Tan!20}
    \#1	&           \#2	&
    0.9565 &        0.9514 &        0.9565 &        0.9539 &
    0.9805 &        0.0195 &        0.0435 &
    6195 &          6161349 &       122401 &        157972 &
    112252658417 &                  199207470 \\ \hline

    \rowcolor{Tan!40}
    \#1	&           \#3	&
    0.9976 &        0.9954 &        0.9976 &        0.9964 &
    0.9999 &        0.0001 &        0.0024 &
    15 &            8613985 &       465 &           20340 &
    112252658417 &                  696710769 \\ \hline

    \rowcolor{Tan!20}
    \#1	&           \#4 &
    0.9976 &        0.9979 &        0.9976 &        0.9977 &
    0.9986 &        0.0014 &        0.0024 &
    5 &             13187768 &      18543 &         13794 &
    112252658417 &                  984385909 \\ \hline

    \rowcolor{Melon!40}
    \#2	&           \#1 &
    0.9983 &        0.9982 &        0.9983 &        0.9983 &
    0.9993 &        0.0007 &        0.0017 &
    1598 &          13196556 &      9755 &          12201 &
    54372904797 &                   342538899 \\
\hline
\end{tabular}
\tiny
\begin{tablenotes}
    \item [1] \textbf{Parameters:} N/A
\end{tablenotes}
\end{threeparttable}

\end{table}


\begin{table}[ht!]
\tiny

\begin{threeparttable}
\caption{Linear Support Vector Machine\tnote{1}} \label{linear-support-vector-machine}

\begin{tabular}{|p{0.45cm}|p{0.45cm}|p{0.6cm}|p{0.6cm}|p{0.6cm}|p{0.6cm}|p{0.5cm}|p{0.5cm}|p{0.5cm}|p{0.8cm}|p{0.8cm}|p{0.8cm}|p{0.8cm}|p{1.8cm}|p{1.6cm}|} \hline
    \rowcolor{Tan!80}
    \textbf{Data- set} & \textbf{Sub- set} &
    \textbf{Acc.} & \textbf{Pre.} & \textbf{Rec. (TPR)} & \textbf{F1-score} &
    \textbf{TNR} & \textbf{FPR} & \textbf{FNR} &
    \textbf{TP} & \textbf{TN} & \textbf{FP} & \textbf{FN} &
    \textbf{Training Time (ns)} & \textbf{Testing Time (ns)} \\ \hline \hline

    \rowcolor{Tan!40}
    \#1	&           \#1 &
    0.9841 &        0.3018 &        0.3275 &        0.3141 &
    0.9915 &        0.0085 &        0.6725 &
    20723 &         5591450 &       47940 &         42557 &
    35314797566 &                   344383409 \\ \hline

    \rowcolor{Tan!20}
    \#1	&           \#2	&
    0.9593 &        0.0538 &        0.0361 &        0.0432 &
    0.9834 &        0.0166 &        0.9639 &
    5923 &          6179525 &       104225 &        158244 &
    35314797566 &                   354491014 \\ \hline

    \rowcolor{Tan!40}
    \#1	&           \#3	&
    0.9975 &        0.0131 &        0.0009 &        0.0017 &
    0.9998 &        0.0002 &        0.9991 &
    19 &            8613017 &       1433 &          20336 &
    35314797566 &                   673546732 \\ \hline

    \rowcolor{Tan!20}
    \#1	&           \#4 &
    0.9942 &        0.0002 &        0.0009 &        0.0003 &
    0.9952 &        0.0048 &        0.9991 &
    12 &            13143123 &      63188 &         13787 &
    35314797566 &                   661366602 \\ \hline

    \rowcolor{Melon!40}
    \#2	&           \#1 &
    0.9990 &        0.0000 &        0.0000 &        0.0000 &
    1.0000 &        0.0000 &        1.0000 &
    0 &             13206311 &      0 &             13799 &
    57758151570 &                   664285259 \\
\hline
\end{tabular}
\tiny
\begin{tablenotes}
    \item [1] \textbf{Parameters:} loss = hinge
\end{tablenotes}
\end{threeparttable}

\end{table}


\begin{table}[ht!]
\tiny

\begin{threeparttable}
\caption{One-Class Support Vector Machine\tnote{1}\tnote{2}} \label{one-class-support-vector-machine}

\begin{tabular}{|p{0.45cm}|p{0.45cm}|p{0.6cm}|p{0.6cm}|p{0.6cm}|p{0.6cm}|p{0.5cm}|p{0.5cm}|p{0.5cm}|p{0.8cm}|p{0.8cm}|p{0.8cm}|p{0.8cm}|p{1.8cm}|p{1.6cm}|} \hline
    \rowcolor{Tan!80}
    \textbf{Data- set} & \textbf{Sub- set} &
    \textbf{Acc.} & \textbf{Pre.} & \textbf{Rec. (TPR)} & \textbf{F1-score} &
    \textbf{TNR} & \textbf{FPR} & \textbf{FNR} &
    \textbf{TP} & \textbf{TN} & \textbf{FP} & \textbf{FN} &
    \textbf{Training Time (ns)} & \textbf{Testing Time (ns)} \\ \hline \hline

    \rowcolor{Tan!40}
    \#1	&           \#1 &
    - &             - &             - &             - &
    - &             - &             - &
    - &             - &             - &             - &
    - &                             - \\ \hline

    \rowcolor{Tan!20}
    \#1	&           \#2	&
    - &             - &             - &             - &
    - &             - &             - &
    - &             - &             - &             - &
    - &                             - \\ \hline

    \rowcolor{Tan!40}
    \#1	&           \#3	&
    - &             - &             - &             - &
    - &             - &             - &
    - &             - &             - &             - &
    - &                             - \\ \hline

    \rowcolor{Tan!20}
    \#1	&           \#4 &
    - &             - &             - &             - &
    - &             - &             - &
    - &             - &             - &             - &
    - &                             - \\ \hline

    \rowcolor{Melon!40}
    \#2	&           \#1 &
    - &             - &             - &             - &
    - &             - &             - &
    - &             - &             - &             - &
    - &                             - \\ \hline

    \rowcolor{Melon!20}
    \#2	&           \#2 &
    - &             - &             - &             - &
    - &             - &             - &
    - &             - &             - &             - &
    - &                             - \\
\hline
\end{tabular}
\tiny
\begin{tablenotes}
    \item [1] \textbf{Parameters:} kernel = rbf
    \item [2] The one-class support vector machine exceeded our 6-day (518400000000000 ns) time limit
\end{tablenotes}
\end{threeparttable}

\end{table}


\begin{table}[ht!]
\tiny

\begin{threeparttable}
\caption{Support Vector Machine\tnote{1}} \label{support-vector-machine}

\begin{tabular}{|p{0.45cm}|p{0.45cm}|p{0.6cm}|p{0.6cm}|p{0.6cm}|p{0.6cm}|p{0.5cm}|p{0.5cm}|p{0.5cm}|p{0.8cm}|p{0.8cm}|p{0.8cm}|p{0.8cm}|p{1.8cm}|p{1.6cm}|} \hline
    \rowcolor{Tan!80}
    \textbf{Data- set} & \textbf{Sub- set} &
    \textbf{Acc.} & \textbf{Pre.} & \textbf{Rec. (TPR)} & \textbf{F1-score} &
    \textbf{TNR} & \textbf{FPR} & \textbf{FNR} &
    \textbf{TP} & \textbf{TN} & \textbf{FP} & \textbf{FN} &
    \textbf{Training Time (ns)} & \textbf{Testing Time (ns)} \\ \hline \hline

    \rowcolor{Tan!40}
    \#1	&           \#1 &
    0.9720 &        0.1505 &        0.3275 &        0.2062 &
    0.9793 &        0.0207 &        0.6725 &
    20723 &         5522435 &       116955 &        42557 &
    116557173711567 &               20169585048115 \\ \hline

    \rowcolor{Tan!20}
    \#1	&           \#2	&
    0.9618 &        0.0630 &        0.0361 &        0.0459 &
    0.9860 &        0.0140 &        0.9639 &
    5923 &          6195589 &       88161 &         158244 &
    116557173711567 &               22794629836354 \\ \hline

    \rowcolor{Tan!40}
    \#1	&           \#3	&
    0.9976 &        0.0260 &        0.0009 &        0.0018 &
    0.9999 &        0.0001 &        0.9991 &
    19 &            8613737 &       713 &           20336 &
    116557173711567 &               30536179241509 \\ \hline

    \rowcolor{Tan!20}
    \#1	&           \#4 &
    0.9981 &        0.0007 &        0.0006 &        0.0007 &
    0.9992 &        0.0008 &        0.9994 &
    8 &             13195569 &      10742 &         13791 &
    116557173711567 &               46633303631277 \\ \hline

    \rowcolor{Melon!40}
    \#2\tnote{2} &  \#1 &
    - &             - &             - &             - &
    - &             - &             - &
    - &             - &             - &             - &
    - &                             - \\
\hline
\end{tabular}
\tiny
\begin{tablenotes}
    \item [1] \textbf{Parameters:} kernel = rbf
    \item [2] This experiment exceeded our 6-day (518400000000000 ns) time limit
\end{tablenotes}
\end{threeparttable}

\end{table}


\begin{table}[ht!]
\tiny

\begin{threeparttable}
\caption{Decision Tree\tnote{1}} \label{decision-tree}

\begin{tabular}{|p{0.45cm}|p{0.45cm}|p{0.6cm}|p{0.6cm}|p{0.6cm}|p{0.6cm}|p{0.5cm}|p{0.5cm}|p{0.5cm}|p{0.8cm}|p{0.8cm}|p{0.8cm}|p{0.8cm}|p{1.8cm}|p{1.6cm}|} \hline
    \rowcolor{Tan!80}
    \textbf{Data- set} & \textbf{Sub- set} &
    \textbf{Acc.} & \textbf{Pre.} & \textbf{Rec. (TPR)} & \textbf{F1-score} &
    \textbf{TNR} & \textbf{FPR} & \textbf{FNR} &
    \textbf{TP} & \textbf{TN} & \textbf{FP} & \textbf{FN} &
    \textbf{Training Time (ns)} & \textbf{Testing Time (ns)} \\ \hline \hline

    \rowcolor{Tan!40}
    \#1	&           \#1 &
    0.9742 &        0.0724 &        0.1123 &        0.0880 &
    0.9838 &        0.0162 &        0.8877 &
    7109 &          5548288 &       91102 &         56171 &
    8271963209 &                    401618047 \\ \hline

    \rowcolor{Tan!20}
    \#1	&           \#2	&
    0.9738 &        0.0000 &        0.0000 &        0.0000 &
    0.9993 &        0.0007 &        1.0000 &
    0 &             6279265 &       4485 &          164167 &
    8271963209 &                    471393563 \\ \hline

    \rowcolor{Tan!40}
    \#1	&           \#3	&
    0.9821 &        0.0022 &        0.0148 &        0.0039 &
    0.9844 &        0.0156 &        0.9852 &
    301 &           8479721 &       134729 &        20054 &
    8271963209 &                    602445631 \\ \hline

    \rowcolor{Tan!20}
    \#1	&           \#4 &
    0.9827 &        0.0018 &        0.0276 &        0.0033 &
    0.9837 &        0.0163 &        0.9724 &
    381 &           12990683 &      215628 &        13418 &
    8271963209 &                    997735598 \\ \hline

    \rowcolor{Melon!40}
    \#2	&           \#1 &
    0.9982 &        0.3166 &        0.6010 &        0.4148 &
    0.9986 &        0.0014 &        0.3990 &
    8293 &          13188413 &      17898 &         5506 &
    43693779718 &                   1547987161 \\
\hline
\end{tabular}
\tiny
\begin{tablenotes}
    \item [1] \textbf{Parameters:} criterion = entropy, splitter = random, min\_samples\_split = 2
\end{tablenotes}
\end{threeparttable}

\end{table}


\begin{table}[ht!]
\tiny

\begin{threeparttable}
\caption{Extra Trees\tnote{1}} \label{extra-trees}

\begin{tabular}{|p{0.45cm}|p{0.45cm}|p{0.6cm}|p{0.6cm}|p{0.6cm}|p{0.6cm}|p{0.5cm}|p{0.5cm}|p{0.5cm}|p{0.8cm}|p{0.8cm}|p{0.8cm}|p{0.8cm}|p{1.8cm}|p{1.6cm}|} \hline
    \rowcolor{Tan!80}
    \textbf{Data- set} & \textbf{Sub- set} &
    \textbf{Acc.} & \textbf{Pre.} & \textbf{Rec. (TPR)} & \textbf{F1-score} &
    \textbf{TNR} & \textbf{FPR} & \textbf{FNR} &
    \textbf{TP} & \textbf{TN} & \textbf{FP} & \textbf{FN} &
    \textbf{Training Time (ns)} & \textbf{Testing Time (ns)} \\ \hline \hline

    \rowcolor{Tan!40}
    \#1	&           \#1 &
    0.9915 &        0.8199 &        0.3046 &        0.4442 &
    0.9992 &        0.0008 &        0.6954 &
    19275 &         5635157 &       4233 &          44005 &
    209531429363 &                  20667658310 \\ \hline

    \rowcolor{Tan!20}
    \#1	&           \#2	&
    0.9740 &        0.0000 &        0.0000 &        0.0000 &
    0.9994 &        0.0006 &        1.0000 &
    0 &             6280034 &       3716 &          164167 &
    209531429363 &                  24155191698 \\ \hline

    \rowcolor{Tan!40}
    \#1	&           \#3	&
    0.9976 &        1.0000 &        0.0001 &        0.0003 &
    1.0000 &        0.0000 &        0.9999 &
    3 &             8614450 &       0 &             20352 &
    209531429363 &  31514357019 \\ \hline

    \rowcolor{Tan!20}
    \#1	&           \#4 &
    0.9988 &        0.0000 &        0.0000 &        0.0000 &
    0.9998 &        0.0002 &        1.0000 &
    0 &             13203625 &      2686 &          13799 &
    209531429363 &                  49439825750 \\ \hline

    \rowcolor{Melon!40}
    \#2	&           \#1 &
    0.9990 &        0.5307 &        0.5210 &        0.5258 &
    0.9995 &        0.0005 &        0.4790 &
    7189 &          13199954 &      6357 &          6610 &
    1048485087501 &                 94353095263 \\
\hline
\end{tabular}
\tiny
\begin{tablenotes}
    \item [1] \textbf{Parameters:} n\_estimators = 50, criterion = gini, min\_samples\_split = 2
\end{tablenotes}
\end{threeparttable}

\end{table}


\begin{table}[ht!]
\tiny

\begin{threeparttable}
\caption{Gradient Boosting\tnote{1}} \label{gradient-boosting}

\begin{tabular}{|p{0.45cm}|p{0.45cm}|p{0.6cm}|p{0.6cm}|p{0.6cm}|p{0.6cm}|p{0.5cm}|p{0.5cm}|p{0.5cm}|p{0.8cm}|p{0.8cm}|p{0.8cm}|p{0.8cm}|p{1.8cm}|p{1.6cm}|} \hline
    \rowcolor{Tan!80}
    \textbf{Data- set} & \textbf{Sub- set} &
    \textbf{Acc.} & \textbf{Pre.} & \textbf{Rec. (TPR)} & \textbf{F1-score} &
    \textbf{TNR} & \textbf{FPR} & \textbf{FNR} &
    \textbf{TP} & \textbf{TN} & \textbf{FP} & \textbf{FN} &
    \textbf{Training Time (ns)} & \textbf{Testing Time (ns)} \\ \hline \hline

    \rowcolor{Tan!40}
    \#1	&           \#1 &
    0.6254 &        0.9892 &        0.6254 &        0.7585 &
    0.6212 &        0.3788 &        0.3746 &
    63280 &         3503305 &       2136085 &       0 &
    2451403662347 &                 8489738261 \\ \hline

    \rowcolor{Tan!20}
    \#1	&           \#2	&
    0.6626 &        0.9763 &        0.6626 &        0.7739 &
    0.6538 &        0.3462 &        0.3374 &
    164079 &        4108265 &       2175485 &       88 &
    2451403662347 &                 8834463042 \\ \hline

    \rowcolor{Tan!40}
    \#1	&           \#3	&
    0.9977 &        0.9977 &        0.9977 &        0.9966 &
    1.0000 &        0.0000 &        0.0023 &
    469 &           8614449 &       1 &             19886 &
    2451403662347 &                 11825961968 \\ \hline

    \rowcolor{Tan!20}
    \#1	&           \#4 &
    0.9984 &        0.9980 &        0.9984 &        0.9982 &
    0.9994 &        0.0006 &        0.0016 &
    644 &           13198117 &      8194 &          13155 &
    2451403662347 &                 18095133741 \\ \hline

    \rowcolor{Melon!40}
    \#2	&           \#1 &
    0.9992 &        0.9991 &        0.9992 &        0.9991 &
    0.9998 &        0.0002 &        0.0008 &
    5942 &          13203502 &      2809 &          7857 &
    4148960101197 &                 18509739315 \\
\hline
\end{tabular}
\tiny
\begin{tablenotes}
    \item [1] \textbf{Parameters:} loss = deviance, learning\_rate = 0.1, n\_estimators = 100, criterion = friedman\_mse, min\_samples\_split = 2
\end{tablenotes}
\end{threeparttable}

\end{table}


\begin{table}[ht!]
\tiny

\begin{threeparttable}
\caption{Isolation Forest\tnote{1}} \label{isolation-forest}

\begin{tabular}{|p{0.45cm}|p{0.45cm}|p{0.6cm}|p{0.6cm}|p{0.6cm}|p{0.6cm}|p{0.5cm}|p{0.5cm}|p{0.5cm}|p{0.8cm}|p{0.8cm}|p{0.8cm}|p{0.8cm}|p{1.8cm}|p{1.6cm}|} \hline
    \rowcolor{Tan!80}
    \textbf{Data- set} & \textbf{Sub- set} &
    \textbf{Acc.} & \textbf{Pre.} & \textbf{Rec. (TPR)} & \textbf{F1-score} &
    \textbf{TNR} & \textbf{FPR} & \textbf{FNR} &
    \textbf{TP} & \textbf{TN} & \textbf{FP} & \textbf{FN} &
    \textbf{Training Time (ns)} & \textbf{Testing Time (ns)} \\ \hline \hline

    \rowcolor{Tan!40}
    \#1	&           \#1 &
    0.9846 &        0.9779 &        0.9846 &        0.9812 &
    0.9956 &        0.0044 &        0.0154 &
    0 &             5614590 &       24800 &         63280 &
    197244480367 &                  88099567128 \\ \hline

    \rowcolor{Tan!20}
    \#1	&           \#2	&
    0.9550 &        0.9492 &        0.9550 &        0.9521 &
    0.9799 &        0.0201 &        0.0450 &
    0 &             6157449 &       126301 &        164167 &
    197244480367 &                  100337823607 \\ \hline

    \rowcolor{Tan!40}
    \#1	&           \#3	&
    0.9969 &        0.9957 &        0.9969 &        0.9963 &
    0.9992 &        0.0008 &        0.0031 &
    928 &           8607461 &       6989 &          19427 &
    197244480367 &                  134518109020 \\ \hline

    \rowcolor{Tan!20}
    \#1	&           \#4 &
    0.9881 &        0.9980 &        0.9881 &        0.9930 &
    0.9890 &        0.0110 &        0.0119 &
    1131 &          13061670 &      144641 &        12668 &
    197244480367 &                  202233230770 \\ \hline

    \rowcolor{Melon!40}
    \#2	&           \#1 &
    0.9990 &        0.9990 &        0.9990 &        0.9986 &
    1.0000 &        0.0000 &        0.0010 &
    695 &           13206294 &      17 &            13104 &
    324118674127 &                  206854622230 \\
\hline
\end{tabular}
\tiny
\begin{tablenotes}
    \item [1] \textbf{Parameters:} n\_estimators = 50, contamination = 0.001
\end{tablenotes}
\end{threeparttable}

\end{table}


\begin{table}[ht!]
\tiny

\begin{threeparttable}
\caption{Random Forest\tnote{1}} \label{random-forest}

\begin{tabular}{|p{0.45cm}|p{0.45cm}|p{0.6cm}|p{0.6cm}|p{0.6cm}|p{0.6cm}|p{0.5cm}|p{0.5cm}|p{0.5cm}|p{0.8cm}|p{0.8cm}|p{0.8cm}|p{0.8cm}|p{1.8cm}|p{1.6cm}|} \hline
    \rowcolor{Tan!80}
    \textbf{Data- set} & \textbf{Sub- set} &
    \textbf{Acc.} & \textbf{Pre.} & \textbf{Rec. (TPR)} & \textbf{F1-score} &
    \textbf{TNR} & \textbf{FPR} & \textbf{FNR} &
    \textbf{TP} & \textbf{TN} & \textbf{FP} & \textbf{FN} &
    \textbf{Training Time (ns)} & \textbf{Testing Time (ns)} \\ \hline \hline

    \rowcolor{Tan!40}
    \#1	&           \#1 &
    0.9879 &        0.0088 &        0.0008 &        0.0015 &
    0.9990 &        0.0010 &        0.9992 &
    52 &            5633511 &       5879 &          63228 &
    763511969876 &                  38249637250 \\ \hline

    \rowcolor{Tan!20}
    \#1	&           \#2	&
    0.9720 &        0.0000 &        0.0000 &        0.0000 &
    0.9974 &        0.0026 &        1.0000 &
    0 &             6267567 &       16183 &         164167 &
    763511969876 &                  42700298556 \\ \hline

    \rowcolor{Tan!40}
    \#1	&           \#3	&
    0.9976 &        1.0000 &        0.0004 &        0.0009 &
    1.0000 &        0.0000 &        0.9996 &
    9 &             8614450 &       0 &             20346 &
    763511969876 &                  56868215083 \\ \hline

    \rowcolor{Tan!20}
    \#1	&           \#4 &
    0.9988 &        0.0011 &        0.0001 &        0.0003 &
    0.9999 &        0.0001 &        0.9999 &
    2 &             13204519 &      1792 &          13797 &
    763511969876 &                  88901090110 \\ \hline

    \rowcolor{Melon!40}
    \#2	&           \#1 &
    0.9991 &        0.6009 &        0.5464 &        0.5724 &
    0.9996 &        0.0004 &        0.4536 &
    7540 &          13201303 &      5008 &          6259 &
    3114043843818 &                 136611348151 \\
\hline
\end{tabular}
\tiny
\begin{tablenotes}
    \item [1] \textbf{Parameters:} n\_estimators = 100, criterion = gini, min\_samples\_split = 2
\end{tablenotes}
\end{threeparttable}

\end{table}

\subsection{Traditional Machine Learning, Unsupervised}

We provide the results of our unsupervised traditional machine learning models in Tables \ref{k-means-clustering}, \ref{mini-batch-k-means-clustering}, \ref{birch}, and \ref{local-outlier-factor}.


\begin{table}[ht!]
\tiny

\begin{threeparttable}
\caption{K-Means Clustering\tnote{1}} \label{k-means-clustering}

\begin{tabular}{|p{0.45cm}|p{0.45cm}|p{0.6cm}|p{0.6cm}|p{0.6cm}|p{0.6cm}|p{0.5cm}|p{0.5cm}|p{0.5cm}|p{0.8cm}|p{0.8cm}|p{0.8cm}|p{0.8cm}|p{1.8cm}|p{1.6cm}|} \hline
    \rowcolor{Tan!80}
    \textbf{Data- set} & \textbf{Sub- set} &
    \textbf{Acc.} & \textbf{Pre.} & \textbf{Rec. (TPR)} & \textbf{F1-score} &
    \textbf{TNR} & \textbf{FPR} & \textbf{FNR} &
    \textbf{TP} & \textbf{TN} & \textbf{FP} & \textbf{FN} &
    \textbf{Training Time (ns)} & \textbf{Testing Time (ns)} \\ \hline \hline

    \rowcolor{Tan!40}
    \#1	&           \#1 &
    0.8564 &        0.9763 &        0.8564 &        0.9124 &
    0.8660 &        0.1340 &        0.1436 &
    0 &             4883881 &       755509 &        63280 &
    97416359356 &                   638434302 \\ \hline

    \rowcolor{Tan!20}
    \#1	&           \#2	&
    0.8654 &        0.9467 &        0.8654 &        0.9042 &
    0.8880 &        0.1120 &        0.1346 &
    0 &             5580250 &       703500 &        164167 &
    97416359356 &                   709549141 \\ \hline

    \rowcolor{Tan!40}
    \#1	&           \#3	&
    0.8658 &        0.9970 &        0.8658 &        0.9258 &
    0.8660 &        0.1340 &        0.1342 &
    15172 &         7460471 &       1153979 &       5183 &
    97416359356 &                   963140938 \\ \hline

    \rowcolor{Tan!20}
    \#1	&           \#4 &
    0.8872 &        0.9985 &        0.8872 &        0.9392 &
    0.8875 &        0.1125 &        0.1128 &
    8883 &          11720655 &      1485656 &       4916 &
    97416359356 &                   1463804724 \\ \hline

    \rowcolor{Melon!40}
    \#2	&           \#1 &
    0.8872 &        0.9985 &        0.8872 &        0.9392 &
    0.8875 &        0.1125 &        0.1128 &
    8811 &          11720655 &      1485656 &       4988 &
    130622830521 &                  1447033632 \\
\hline
\end{tabular}
\tiny
\begin{tablenotes}
    \item [1] \textbf{Parameters:} n\_clusters = 2, algorithm = full
\end{tablenotes}
\end{threeparttable}

\end{table}


\begin{table}[ht!]
\tiny

\begin{threeparttable}
\caption{Mini-Batch K-Means Clustering\tnote{1}} \label{mini-batch-k-means-clustering}

\begin{tabular}{|p{0.45cm}|p{0.45cm}|p{0.6cm}|p{0.6cm}|p{0.6cm}|p{0.6cm}|p{0.5cm}|p{0.5cm}|p{0.5cm}|p{0.8cm}|p{0.8cm}|p{0.8cm}|p{0.8cm}|p{1.8cm}|p{1.6cm}|} \hline
    \rowcolor{Tan!80}
    \textbf{Data- set} & \textbf{Sub- set} &
    \textbf{Acc.} & \textbf{Pre.} & \textbf{Rec. (TPR)} & \textbf{F1-score} &
    \textbf{TNR} & \textbf{FPR} & \textbf{FNR} &
    \textbf{TP} & \textbf{TN} & \textbf{FP} & \textbf{FN} &
    \textbf{Training Time (ns)} & \textbf{Testing Time (ns)} \\ \hline \hline

    \rowcolor{Tan!40}
    \#1	&           \#1 &
    0.4681 &        0.9669 &        0.4681 &        0.6304 &
    0.4729 &        0.5271 &        0.5319 &
    2491 &          2666944 &       2972446 &       60789 &
    7678285764 &                    3796294688 \\ \hline

    \rowcolor{Tan!20}
    \#1	&           \#2	&
    0.4969 &        0.9271 &        0.4969 &        0.6470 &
    0.5099 &        0.4901 &        0.5031 &
    88 &            3204152 &       3079598 &       164079 &
    7678285764 &                    4126732218 \\ \hline

    \rowcolor{Tan!40}
    \#1	&           \#3	&
    0.4741 &        0.9969 &        0.4741 &        0.6408 &
    0.4732 &        0.5268 &        0.5259 &
    17101 &         4076409 &       4538041 &       3254 &
    7678285764 &                    5800975603 \\ \hline

    \rowcolor{Tan!20}
    \#1	&           \#4 &
    0.4748 &        0.9984 &        0.4748 &        0.6429 &
    0.4745 &        0.5255 &        0.5252 &
    10415 &         6267005 &       6939306 &       3384 &
    7678285764 &                    9274790788 \\ \hline

    \rowcolor{Melon!40}
    \#2	&           \#1 &
    0.8872 &        0.9985 &        0.8872 &        0.9392 &
    0.8875 &        0.1125 &        0.1128 &
    8808 &          11720655 &      1485656 &       4991 &
    13176242457 &                   8605857753 \\
\hline
\end{tabular}
\tiny
\begin{tablenotes}
    \item [1] \textbf{Parameters:} n\_clusters = 2, batch\_size = 512
\end{tablenotes}
\end{threeparttable}

\end{table}


\begin{table}[ht!]
\tiny

\begin{threeparttable}
\caption{BIRCH\tnote{1}} \label{birch}

\begin{tabular}{|p{0.45cm}|p{0.45cm}|p{0.6cm}|p{0.6cm}|p{0.6cm}|p{0.6cm}|p{0.5cm}|p{0.5cm}|p{0.5cm}|p{0.8cm}|p{0.8cm}|p{0.8cm}|p{0.8cm}|p{1.8cm}|p{1.6cm}|} \hline
    \rowcolor{Tan!80}
    \textbf{Data- set} & \textbf{Sub- set} &
    \textbf{Acc.} & \textbf{Pre.} & \textbf{Rec. (TPR)} & \textbf{F1-score} &
    \textbf{TNR} & \textbf{FPR} & \textbf{FNR} &
    \textbf{TP} & \textbf{TN} & \textbf{FP} & \textbf{FN} &
    \textbf{Training Time (ns)} & \textbf{Testing Time (ns)} \\ \hline \hline

    \rowcolor{Tan!40}
    \#1	&           \#1 &
    0.9889 &        0.9779 &        0.9889 &        0.9834 &
    1.0000 &        0.0000 &        0.0111 &
    0 &             5639390 &       0 &             63280 &
    214197454658 &                  605665184 \\ \hline

    \rowcolor{Tan!20}
    \#1	&           \#2	&
    0.9745 &        0.9497 &        0.9745 &        0.9620 &
    1.0000 &        0.0000 &        0.0255 &
    0 &             6283750 &       0 &             164167 &
    214197454658 &                  662717530 \\ \hline

    \rowcolor{Tan!40}
    \#1	&           \#3	&
    0.9976 &        0.9953 &        0.9976 &        0.9965 &
    1.0000 &        0.0000 &        0.0024 &
    0 &             8614450 &       0 &             20355 &
    214197454658 &                  1004912418 \\ \hline

    \rowcolor{Tan!20}
    \#1	&           \#4 &
    0.9990 &        0.9979 &        0.9990 &        0.9984 &
    1.0000 &        0.0000 &        0.0010 &
    0 &             13206311 &      0 &             13799 &
    214197454658 &                  1171693676 \\ \hline

    \rowcolor{Melon!40}
    \#2	&           \#1 &
    0.9990 &        0.9979 &        0.9990 &        0.9984 &
    1.0000 &        0.0000 &        0.0010 &
    0 &             13206311 &      0 &             13799 &
    297512429069 &                  1163219087 \\
\hline
\end{tabular}
\tiny
\begin{tablenotes}
    \item [1] \textbf{Parameters:} threshold = 0.5, branching\_factor = 25, n\_clusters = 2
\end{tablenotes}
\end{threeparttable}

\end{table}


\begin{table}[ht!]
\tiny

\begin{threeparttable}
\caption{Local Outlier Factor\tnote{1}} \label{local-outlier-factor}

\begin{tabular}{|p{0.45cm}|p{0.45cm}|p{0.6cm}|p{0.6cm}|p{0.6cm}|p{0.6cm}|p{0.5cm}|p{0.5cm}|p{0.5cm}|p{0.8cm}|p{0.8cm}|p{0.8cm}|p{0.8cm}|p{1.8cm}|p{1.6cm}|} \hline
    \rowcolor{Tan!80}
    \textbf{Data- set} & \textbf{Sub- set} &
    \textbf{Acc.} & \textbf{Pre.} & \textbf{Rec. (TPR)} & \textbf{F1-score} &
    \textbf{TNR} & \textbf{FPR} & \textbf{FNR} &
    \textbf{TP} & \textbf{TN} & \textbf{FP} & \textbf{FN} &
    \textbf{Training Time (ns)} & \textbf{Testing Time (ns)} \\ \hline \hline

    \rowcolor{Tan!40}
    \#1	&           \#1 &
    0.9254 &        0.9728 &        0.9254 &        0.9465 &
    0.9343 &        0.0657 &        0.0746 &
    72077 &         7860912 &       552374 &        87258 &
    53788202512612 &                2439824879120 \\ \hline

    \rowcolor{Tan!20}
    \#1	&           \#2	&
    0.1714 &        0.9557 &        0.1714 &        0.2640 &
    0.1560 &        0.8440 &        0.8286 &
    129536 &        1044762 &       5653283 &       24382 &
    53788202512612 &                7698684672802 \\ \hline

    \rowcolor{Tan!40}
    \#1	&           \#3	&
    0.9386 &        0.9965 &        0.9386 &        0.9665 &
    0.9400 &        0.0600 &        0.0614 &
    4080 &          8863608 &       566191 &        14030 &
    53788202512612 &                2504114371275 \\ \hline

    \rowcolor{Tan!20}
    \#1	&           \#4 &
    0.1661 &        0.8882 &        0.1661 &        0.2726 &
    0.1644 &        0.8356 &        0.8339 &
    36488 &         1108865 &       5637206 &       112734 &
    53788202512612 &                7995192332965 \\ \hline

    \rowcolor{Melon!40}
    \#2	&           \#1 &
    0.9081 &        0.9559 &        0.9081 &        0.9314 &
    0.9280 &        0.0720 &        0.0919 &
    1753 &          6260040 &       486031 &        147469 &
    59034611309148 &                1175546111752 \\
\hline
\end{tabular}
\tiny
\begin{tablenotes}
    \item [1] \textbf{Parameters:} n\_neighbors = 20, algorithm = kd\_tree, contamination = auto
\end{tablenotes}
\end{threeparttable}

\end{table}

\subsection{Deep Learning, Supervised}

We provide the results of our supervised deep model---the multi-layer perceptron---in Table \ref{multi-layer-perceptron}.


\begin{table}[ht!]
\tiny

\begin{threeparttable}
\caption{Multi-Layer Perceptron\tnote{1}} \label{multi-layer-perceptron}

\begin{tabular}{|p{0.45cm}|p{0.45cm}|p{0.6cm}|p{0.6cm}|p{0.6cm}|p{0.6cm}|p{0.5cm}|p{0.5cm}|p{0.5cm}|p{0.8cm}|p{0.8cm}|p{0.8cm}|p{0.8cm}|p{1.8cm}|p{1.6cm}|} \hline
    \rowcolor{Tan!80}
    \textbf{Data- set} & \textbf{Sub- set} &
    \textbf{Acc.} & \textbf{Pre.} & \textbf{Rec. (TPR)} & \textbf{F1-score} &
    \textbf{TNR} & \textbf{FPR} & \textbf{FNR} &
    \textbf{TP} & \textbf{TN} & \textbf{FP} & \textbf{FN} &
    \textbf{Training Time (ns)} & \textbf{Testing Time (ns)} \\ \hline \hline

    \rowcolor{Tan!40}
    \#1	&           \#1 &
    0.9788 &        0.9837 &        0.9788 &        0.9811 &
    0.9861 &        0.0139 &        0.0212 &
    20723 &         5560852 &       78538 &         42557 &
    3088090156514 &                 11708139937 \\ \hline

    \rowcolor{Tan!20}
    \#1	&           \#2	&
    0.9666 &        0.9529 &        0.9666 &        0.9594 &
    0.9908 &        0.0092 &        0.0334 &
    6227 &          6226232 &       57518 &         157940 &
    3088090156514 &                 13276672102 \\ \hline

    \rowcolor{Tan!40}
    \#1	&           \#3	&
    0.9974 &        0.9953 &        0.9974 &        0.9964 &
    0.9998 &        0.0002 &        0.0026 &
    11 &            8612758 &       1692 &          20344 &
    3088090156514 &                 17550210412 \\ \hline

    \rowcolor{Tan!20}
    \#1	&           \#4 &
    0.9973 &        0.9979 &        0.9973 &        0.9976 &
    0.9983 &        0.0017 &        0.0027 &
    204 &           13184381 &      21930 &         13595 &
    3088090156514 &                 27715472855 \\ \hline

    \rowcolor{Melon!40}
    \#2	&           \#1 &
    0.9992 &        0.9990 &        0.9992 &        0.9991 &
    0.9997 &        0.0003 &        0.0008 &
    6272 &          13202688 &      3623 &          7527 &
    6061045048456 &                 27148087661 \\
\hline
\end{tabular}
\tiny
\begin{tablenotes}
    \item [1] \textbf{Parameters:} activation = logistic, solver = adam, alpha = 0.0001, learning\_rate = constant
\end{tablenotes}
\end{threeparttable}

\end{table}

\subsection{Deep Learning, Unsupervised}

We provide the results of our unsupervised deep model---the restricted Boltzmann machine---in Table \ref{restricted-boltzmann-machine}.


\begin{table}[ht!]
\tiny

\begin{threeparttable}
\caption{Restricted Boltzmann Machine\tnote{1}} \label{restricted-boltzmann-machine}

\begin{tabular}{|p{0.45cm}|p{0.45cm}|p{0.6cm}|p{0.6cm}|p{0.6cm}|p{0.6cm}|p{0.5cm}|p{0.5cm}|p{0.5cm}|p{0.8cm}|p{0.8cm}|p{0.8cm}|p{0.8cm}|p{1.8cm}|p{1.6cm}|} \hline
    \rowcolor{Tan!80}
    \textbf{Data- set} & \textbf{Sub- set} &
    \textbf{Acc.} & \textbf{Pre.} & \textbf{Rec. (TPR)} & \textbf{F1-score} &
    \textbf{TNR} & \textbf{FPR} & \textbf{FNR} &
    \textbf{TP} & \textbf{TN} & \textbf{FP} & \textbf{FN} &
    \textbf{Training Time (ns)} & \textbf{Testing Time (ns)} \\ \hline \hline

    \rowcolor{Tan!40}
    \#1	&           \#1 &
    0.0111 &        0.0001 &        0.0111 &        0.0002 &
    0.0000 &        1.0000 &        0.9889 &
    63280 &         0 &             5639390 &       0 &
    916661700320 &                  9736619899 \\ \hline

    \rowcolor{Tan!20}
    \#1	&           \#2	&
    0.0255 &        0.0006 &        0.0255 &        0.0013 &
    0.0000 &        1.0000 &        0.9745 &
    164167 &        0 &             6283750 &       0 &
    916661700320 &                  10970949396 \\ \hline

    \rowcolor{Tan!40}
    \#1	&           \#3	&
    0.0024 &        0.0000 &        0.0024 &        0.0000 &
    0.0000 &        1.0000 &        0.9976 &
    20355 &         0 &             8614450 &       0 &
    916661700320 &                  14603996262 \\ \hline

    \rowcolor{Tan!20}
    \#1	&           \#4 &
    0.0010 &        0.0000 &        0.0010 &        0.0000 &
    0.0000 &        1.0000 &        0.9990 &
    13799 &         0 &             13206311 &      0 &
    916661700320 &                  22041127093 \\ \hline

    \rowcolor{Melon!40}
    \#2	&           \#1 &
    0.0010 &        0.0000 &        0.0010 &        0.0000 &
    0.0000 &        1.0000 &        0.9990 &
    13799 &         0 &             13206311 &      0 &
    1354503991934 &                 24294537797 \\
\hline
\end{tabular}
\tiny
\begin{tablenotes}
    \item [1] \textbf{Parameters:} n\_components = 16, learning\_rate = 0.1, batch\_size = 20, threshold = -0.0001
\end{tablenotes}
\end{threeparttable}

\end{table}

\end{document}